\documentclass[sigconf,review=false, anonymous=false]{acmart}

\input{datalabmacros}

\setcopyright{none}

\renewcommand\footnotetextcopyrightpermission[1]{} %

\pgfplotsset{compat=1.17}

\begin{document}
\newcommand{\ourTitle}{\diagrams: a pattern-preserving diagrammatic representation of non-disjunctive Relational Queries}
\title[\diagrams]{\ourTitle}

\author{Wolfgang	Gatterbauer}
\orcid{0000-0002-9614-0504}
\affiliation{%
    \orcidicon{0000-0002-9614-0504}
	Northeastern University\country{USA}}
\email{w.gatterbauer@northeastern.edu}

\author{Cody Dunne}
\orcid{0000-0002-1609-9776}%
\affiliation{%
    \orcidicon{0000-0002-1609-9776}
	Northeastern University\country{USA}}
\email{c.dunne@northeastern.edu}

\author{Mirek	Riedewald}
\orcid{0000-0002-6102-7472}
\affiliation{
    \orcidicon{0000-0002-6102-7472}
	Northeastern University\country{USA}}
\email{m.riedewald@northeastern.edu}

\begin{abstract}

Analyzing relational languages by their \emph{logical expressiveness} is well understood.
Something not well understood or even formalized is the vague concept of \emph{relational query patterns}.
What are query patterns?
And how can we reason about query patterns across different relational languages, irrespective of their syntax and their procedural or declarative nature?
In this paper, we formalize the concept of query patterns with a variant of \emph{pattern-preserving mappings} between the relational tables of queries.
This formalism allows us to analyze the relative pattern expressiveness of relational query languages 
and to create a hierarchy of languages that have equal logical expressiveness yet different pattern expressiveness.
We show that relational calculus can express a greater class of patterns
than the basic operators of relational algebra.
And we propose a complete and sound diagrammatic representation of safe relational calculus that is not only relationally complete, but can also express all query patterns for the large and useful fragment of non-disjunctive relational calculus.
Among all diagrammatic representations for relational queries that we are aware of, 
our \diagrams are the only one that is relationally complete and that can represent all query patterns in the non-disjunctive fragment.

\end{abstract}

\maketitle

\section{Introduction}

When designing and comparing query languages, we are usually concerned about \emph{logical expressiveness}: 
can a language express a particular query we want?
For relational languages, questions of expressiveness 
have been studied for decades, and 
formalisms for comparing expressiveness are well developed and understood. 
Logical expressiveness basically boils down to the notion of logical equivalence: 
does a particular query in one language have a logically-equivalent representation in another?

We do not have the same sophisticated machinery to reason about the up-to-now vague notion of \emph{relational query patterns}:
can a given language express a particular query pattern from another language?
We posit that identifying patterns in queries may have several advantages, 
akin to how formalizing best practices in software \emph{design patterns} has aided software engineers \cite{Gamma:1995ys}.
For one, general and reusable query patterns could assist in teaching students how to write complicated queries.
Queries written using common patterns could then potentially be easier to interpret quickly.
But how would we actually define a \emph{relational query pattern}?
And what would it mean for a given target language to be able to express a particular pattern?
Importantly, a formalisation of relational patterns should be applicable across identical fragments of the four important languages Datalog, Relational Algebra (RA), Relational Calculus (RC), and SQL,
and thus be orthogonal to questions of syntax
and language design.
Furthermore, it should allow us to answer whether such patterns can be expressed to the same extent in all relational query languages, or whether there is a hierarchy of pattern expressiveness among popular languages.

\begin{figure}[t]
\centering
\begin{subfigure}[b]{.32\linewidth}
	\begin{align*}
		I(x,y)	& \datarule R(x,\_),  S(y). 		\\[-1mm]
		Q(x,y) 	& \datarule R(x,y), \neg I(x,y).
	\end{align*}
	\vspace{-4mm}
    \caption{}
	\label{Fig_RA_vs_Datalog_a}
\end{subfigure}
\hspace{-0.6mm}
\begin{subfigure}[b]{.24\linewidth}
	\begin{align*}
		R - \big( \pi_A R \times S \big)
	\end{align*}
	\vspace{-4mm}
    \caption{}
	\label{Fig_RA_vs_Datalog_b}
\end{subfigure}
\hspace{0mm}
\begin{subfigure}[b]{.33\linewidth}
	\vspace{1mm}
    \includegraphics[scale=0.36]{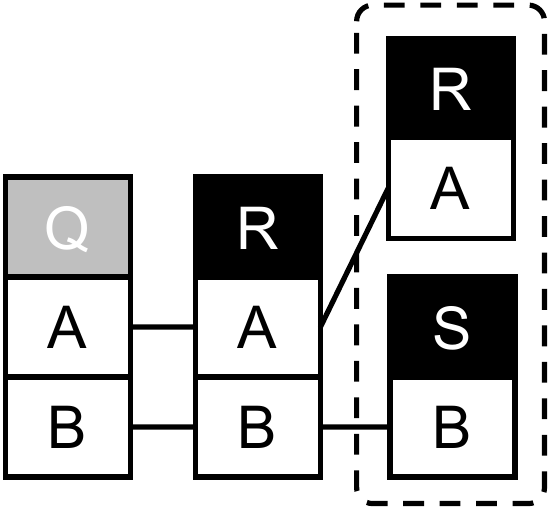}
	\vspace{-2mm}
    \caption{}
	\label{Fig_RA_vs_Datalog_c}
\end{subfigure}
\begin{subfigure}[b]{.32\linewidth}
	\begin{align*}
		I(y)	& \datarule R(\_,y), \neg S(y). 		\\[-1mm]
		Q(x,y) 	& \datarule R(x,z), I(y).
	\end{align*}
	\vspace{-4mm}
    \caption{}
	\label{Fig_RA_vs_Datalog_d}
\end{subfigure}
\hspace{2mm}
\begin{subfigure}[b]{.24\linewidth}
	\begin{align*}
		R \Join \big( \pi_B R - S \big)
	\end{align*}
	\vspace{-4mm}
    \caption{}
	\label{Fig_RA_vs_Datalog_e}
\end{subfigure}
\hspace{0mm}
\begin{subfigure}[b]{.33\linewidth}
    \includegraphics[scale=0.36]{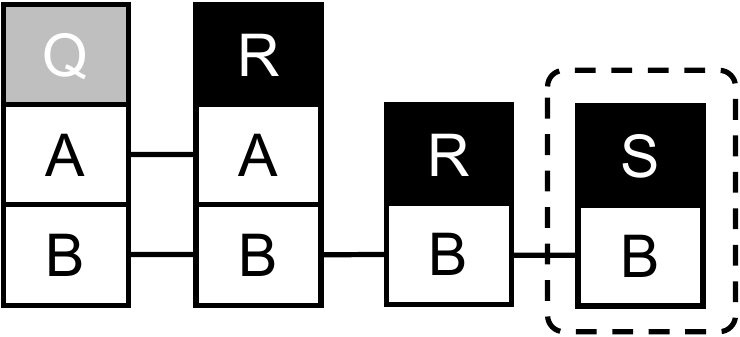}
	\vspace{-2mm}
    \caption{}
	\label{Fig_RA_vs_Datalog_f}
\end{subfigure}
\begin{subfigure}[b]{.32\linewidth}
	\begin{align*}
		Q(x,y) 	& \datarule R(x,y), \neg S(y).
	\end{align*}
	\vspace{-4mm}
    \caption{}
	\label{Fig_RA_vs_Datalog_g}
\end{subfigure}
\hspace{1mm}
\begin{subfigure}[b]{.26\linewidth}
\centering
	\h{\frownie{}}
	\vspace{1mm}
    \caption{}
	\label{Fig_RA_vs_Datalog_h}
\end{subfigure}
\hspace{-0.6mm}
\begin{subfigure}[b]{.33\linewidth}
    \includegraphics[scale=0.36]{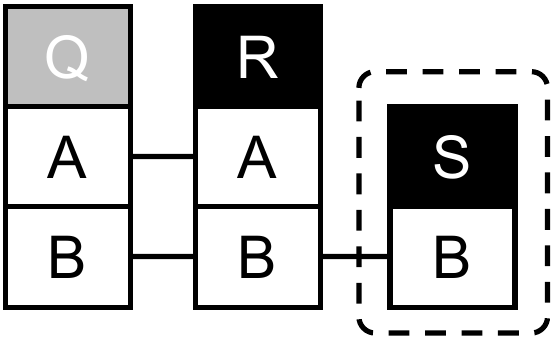}
	\vspace{-2mm}
    \caption{}
	\label{Fig_RA_vs_Datalog_i}
\end{subfigure}
\caption{The first column (a, d, g) shows three logically-equivalent $\DatalogN$ queries that use three different ``query patterns.'' 
The first two (a, d) can also be expressed in Relational Algebra (RA) (b, e), whereas the third pattern (g) 
cannot
be expressed in RA,
i.e. it is not possible
to write a logically equivalent query in basic RA that uses only one occurence of $R$ and $S$ each.
The third column (c, f, i) shows \diagrams that 
use the same ``query patterns''
as the $\DatalogN$ queries in the first column
(we call them later \emph{pattern-isomorph}).
To the best of our knowledge, our proposed \diagrams are the first diagrammatic representation of relational queries that are relationally-complete and able to represent
the full range of query patterns for non-disjunctive relational calculus.
}
\label{Fig_RA_vs_Datalog}
\end{figure}

\introparagraph{Our 1$^{\mathrm{st}}$ contribution: \emph{query patterns}} 
We develop a language-independent notion of relational query patterns
that allows us to compare the abilities of relational query languages to express query patterns 
present in other languages, and thus to reason
about their relative pattern expressiveness.
This approach allows us to contribute a novel hierarchy of pattern-expressiveness among the 
non-disjunctive fragment of above mentioned four languages.
The intuition of our language-independent formalism 
is to reason in terms of mapping between the extensional relations used in two queries.
However, it is not trivial to turn this intuition into a corresponding algorithm that 
that can be applied to any relational
query, no matter the language used for expressing it 
(we include examples to show that more intuitive mappings would fail on queries). 

\begin{example}[RA vs.\ Datalog]
\label{ex:intro}
Consider the $\DatalogN$ query in \cref{Fig_RA_vs_Datalog_g}, which returns all tuples in $R(A,B)$ 
whose attribute $B$ does not appear in the unary table $S(B)$.
The query uses each of the input tables $R$ and $S$ exactly once. 
As we will later prove, there is no way to express this query in basic Relational Algebra ($\RA$) by using each of the tables $R$ and $S$ only once. 
\Cref{Fig_RA_vs_Datalog_b,Fig_RA_vs_Datalog_e} show
two logically-equivalent queries in $\RA$,
each of which uses the table $R(A,B)$ twice
(intuitively, \cref{Fig_RA_vs_Datalog_b} adds a column before applying a negation, 
whereas \cref{Fig_RA_vs_Datalog_e} adds the column after the negation).
We also added equivalent $\DatalogN$ queries, which for those two $\RA$ expression use the exact same ``logical pattern'' (a concept we will formalize later). 
Intuitively (and we prove this later more formally), $\DatalogN$ can express strictly more query patterns than RA;
it has a higher ``pattern-expressiveness'' despite having the same logical expressiveness.
We believe that any diagrammatic language for illustrating and reasoning about query patterns used in queries should be able to express the full range of possible patterns across existing
relational query languages (such as the one in \cref{Fig_RA_vs_Datalog_g}).
It follows that any diagrammatic representation of relational queries that relies on a one-to-one mapping with the operators of RA 
\emph{cannot} represent the full spectrum of query patterns of relational queries.
\end{example}

\introparagraph{Our 2$^{\mathrm{nd}}$ contribution: \diagrams} 
Motivated by prior user studies 
\cite{Reisner1975:HumanFactors,DBLP:journals/csur/Reisner81,DBLP:conf/sigmod/LeventidisZDGJR20}
showing that diagrammatic representations of queries can help users understand them faster,
we first discuss the basic limits of diagrammatic languages in contrast to textual languages to represent such patterns.
We then design an arguably simple and intuitive diagrammatic representation of relational queries called \diagrams 
that ($i$) is relationally complete and ($ii$) preserves the structure of query patterns in the fragment of non-disjunctive relational calculus.

\begin{example}
In the rightmost column of \cref{Fig_RA_vs_Datalog}, we show \diagrams (that we will formalize later) that use the same query patterns 
as the associated queries in $\DatalogN$ and RA to their left.
\end{example}

\textbf{Outline of the paper.}
\Cref{sec:nondisjunctivefragment} defines the \emph{non-disjunctive fragment} of relational query languages for 
Datalog, Relational Algebra ($\RA$), Tuple Relational Calculus (\TRC), and $\SQL$,
and proves that they have equivalent logical expressiveness.

\Cref{sec:structureisomorphism}
develops our formal approach for comparing the relative \emph{pattern expressiveness} among relational query languages.
We apply the approach and contribute a novel hierarchy of pattern expressiveness among the above four languages.

\Cref{sec:QV} 
shows that the non-disjunctive fragment allows for a very intuitive diagrammatic representation system we term \diagrams.
We give the formal translation from the non-disjunctive fragment of $\TRC$ to \diagrams and back, 
and define their formal validity.
We also prove that, for this fragment, \diagrams have the same pattern expressiveness as \TRC.

\Cref{sec:completeness}
adds a single visual element (a union at the root) to make \diagrams relationally complete.\footnote{Although disjunctions can be composed from conjunction and negation using De Morgan's law ($A \vee B = \neg(\neg A \wedge \neg B)$),
this additional visual symbol is necessary: for safe relational queries, DeMorgan is not enough, as
there is no way to write a safe Tuple Relational Calculus ($\TRC$) expression ``\emph{Return all entries that appear in either R or S}'' 
that avoids a union operator.
This argument is part of the textbook argument for the union operator  being an essential, non-redundant operator for relational algebra.
}

\Cref{sec:sentences}
makes a minor modification to the definition of \diagrams that allows them to also represent logical sentences 
(or, equivalently, Boolean queries or logical constraints).
This extension allows us to compare our diagrammatic formalism against a long history of diagrams for 
representing logical sentences.

\Cref{sec:relatedWork}
contrasts our formalism with selected related work. 
In particular, we discuss the connection to Peirce's existential graphs \cite{peirce:1933,Roberts:1992,Shin:2002} and 
show that our formalism is more general and solves interpretational problems of Peirce's graphs, which have been the focus of intense research for over a century.

Due to space constraints, we had to move all proofs
and several intuitive illustrating examples
\iflabelexists{sec:sec:appendix:nomenclature}
{to the appendix.}
{to an optional appendix \cite{relationaldiagrams-TR}.}

\section{The non-disjunctive fragment of relational query languages}
\label{sec:nondisjunctivefragment}
\label{SEC:NONDISJUNCTIVEFRAGMENT}

In this section, we define 
the \emph{non-disjunctive fragment} of 
relational query languages.
We focus here on non-Boolean queries and discuss Boolean queries (logical sentences) later in 
\cref{sec:sentences}.

We assume the reader to be familiar with 
$\DatalogN$ (Non-recursive Datalog with negation),
$\RA$ (Relational Algebra),
$\TRC$ (safe Tuple Relational Calculus),
$\SQL$ (Structured Query Language),
and the necessary safety conditions for $\TRC$ and $\DatalogN$ to be equivalent in logical expressiveness to $\RA$.
We also assume familiarity with concepts such as relations, predicates, atoms, and the named and unnamed perspective of relational algebra.
Currently, the most comprehensive exposition of these topics we know of is Ullman's 1988 textbook \cite{Ullman1988PrinceplesOfDatabase},
together with resources for translating between $\SQL$ and relational calculus 
\cite{DBLP:journals/tse/CeriG85, VanDenBusscheVansummeren:2009:SQL}.
These connections are also discussed in most database textbooks 
\cite{Elmasri:dq,RamakrishnanGehrke:DBMS2000, Garcia-MolinaUW2009:DBSystems, DBLP:books/mg/SKS20},
though in less detail.
To save space, we build upon those well-established results 
and focus on the novel concepts introduced in this paper.
We do not cover safe Domain Relational Calculus (DRC) as we will show later in \cref{sec:Peirce} that $\TRC$
has a more natural translation into diagrams.

\subsection{Why disjunctions are harder to represent}

Assume Alice calls Bob and tells him ``I see a blue car that has a flat tire.''
What is the mental image that Bob has from this information? 
It is a car with two conditions:
it is blue, and it has a flat tire as in \cref{Fig_blue_car_a}.
Next, instead assume that Alice instead tells Bob
``I see a car that is either blue or that has a flat tire.''
What is the mental image that Bob has from that information? 

There is no single mental image that could capture that situation. 
Bob needs to imagine two \emph{different} images.
If Bob sees one image with two different cars (one blue, the other with a flat tire), 
then he actually sees \emph{two separate cars}.
Bob needs to add some additional visual symbol representing the logic that those are two \emph{different overlapping possible worlds}.
In other words, any situation (think of a concrete arrangement of items) can only display conjunctive information~\cite{Shin:2002}.
This diagrammatic representation problem is not as apparent in text:
``\sql{Car.color = `blue' OR Car.tire = `flat'}''.\footnote{To provide some additional intuition,
recall that \emph{conjunctions} of selections can be simply modeled as concatenation of simple selections,
e.g. $\sigma_{C_1 \wedge C_2}(R)$ is the same as $\sigma_{C_1}(\sigma_{C_2}(R))$. Thus conjunctions are an inherently more natural logical connective 
than disjunctions; disjunction cannot be represented without additional visual symbols.}

Next, we define non-disjunctive fragments of four query languages.
We will show in \cref{sec:QV} that these languages lend themselves to a rather natural diagrammatic representation that we term \diagrams, which have nice structure-preserving properties (\cref{sec:structureisomorphism}).
Further on in \cref{sec:completeness}, we achieve relational completeness by adding an additional visual construct.
In the following subsections, we use the SQL query in \cref{fig:disjunctionSQLor} over a schema $R(A), S(A), T(A)$ as running example for a query that \emph{cannot} be expressed in the non-disjunctive fragment.

\subsection{Non-recursive Datalog with negation}
We start with $\Datalog$ since the definition is most straightforward.
$\Datalog$ expresses disjunction (or union) by repeating an Intentional Database predicate (IDB) in the head of multiple rules.
For example, \cref{fig:disjunctionSQLor} expressed in $\Datalog$ becomes:
\begin{align*}
	Q(x) &\datarule R(x), S(x), T(\_).	\\
	Q(x) &\datarule R(x), S(\_), T(x).
\end{align*}	
This query cannot be expressed without defining at least one IDB at least twice,
in our case the result table $Q(x)$.
This leads to a natural definition of the non-disjunctive fragment of $\DatalogN$:

\begin{definition}[$\DatalogND$]
	\emph{Non-disjunctive non-recursive Datalog with negation} ($\DatalogND$)
	is the non-recursive fragment of $\DatalogN$ in which every IDB appears in the head of one single rule.
\end{definition}

\subsection{Relational Algebra (RA)}
We focus on the fragment of basic $\RA$ that contains no union operator $\cup$ 
and in which all selection conditions are simple (i.e., they do not use the disjunction operator $\vee$).
A simple condition is $C=(X \theta Y)$ where $X$ is an attribute, 
$Y$ is either an attribute or a constant, 
and $\theta$ is a comparison operator from $\{=, \neq, >, \geq, <, \leq\}$.
Notice that conjunctions of selections can be modeled as concatenation of selections,
e.g., $\sigma_{C_1 \wedge C_2}(R)$ is the same as $\sigma_{C_1}(\sigma_{C_2}(R))$.
\cref{fig:disjunctionSQLor} cannot be expressed in that fragment and requires either the disjunction operator $\vee$ as in:
\begin{align*}
	\pi_{A} \big(\sigma_{A=B \vee A=C} \big( R \times \rho_{A\rightarrow B} (S) \times \rho_{A \rightarrow C}(T)\big) \big)
\end{align*}
or the union operator $\cup$ as in:
\begin{align*}
	\pi_{A} \big(R \Join S \times \rho_{A \rightarrow C}(T)\big) 
	\cup
	\pi_{A} \big(R \Join T \times \rho_{A \rightarrow B}(S)\big) 
\end{align*}

\begin{definition}
	[$\NDRA$]
	The non-disjunctive fragment of 
	Relational Algebra ($\NDRA$) results from disallowing the union operators $\cup$ 
	and by restricting selections to conjunctions of simple predicates.
\end{definition}

\begin{figure}[t]
\begin{subfigure}[b]{1\linewidth}
    \includegraphics[scale=0.25]{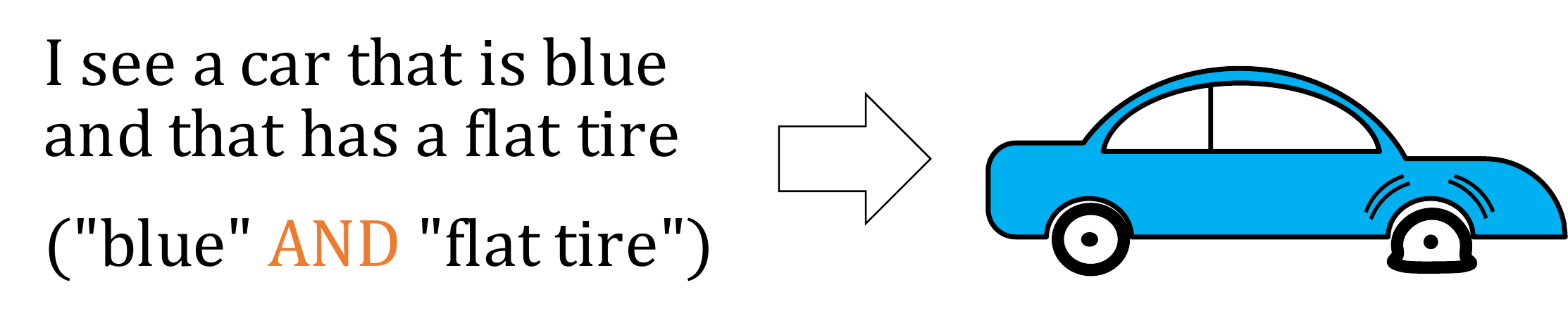}
	\vspace{-3mm}
    \caption{}
	\label{Fig_blue_car_a}	
\end{subfigure}
\begin{subfigure}[b]{1\linewidth}
    \includegraphics[scale=0.25]{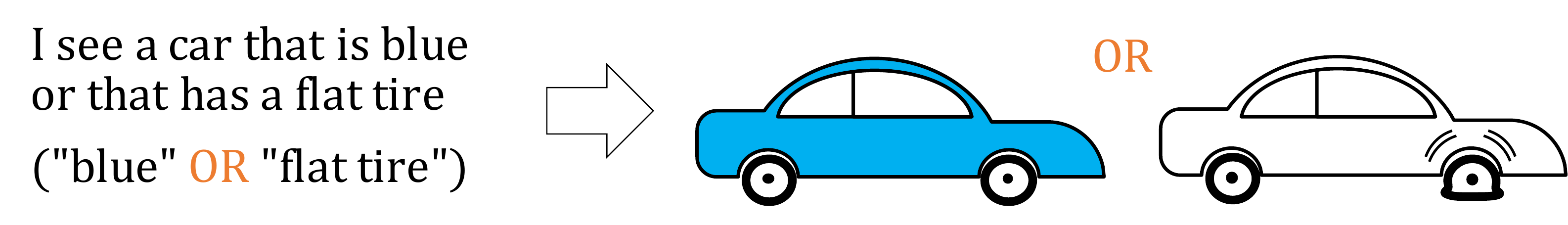}
	\vspace{-3mm}
    \caption{}
	\label{Fig_blue_car_b}	
\end{subfigure}
\caption{Showing a car that has a blue color or a flat tire.
}
\label{Fig_blue_car}  
\end{figure}

\subsection{Tuple Relational Calculus (TRC)}
\label{sec:TRC}
Recall that safe $\TRC$ only allows existential quantification (and not universal quantification)
\cite{Ullman1988PrinceplesOfDatabase}.
Predicates are either join predicates 
``$r.A \,\theta\, s.B$'' or selection predicates 
``$r.A \,\theta\, v$'', with $r, s$ being table variables and $v$ a domain value.
WLOG, every existential quantifier can be pulled out as early as to either be at the start of the query, or directly following a negation operator. 
For example, instead of
$\neg(\exists r \in R [r.A = 0 \wedge \exists s \in S[s.B = r.B]])$
we rather write this sentence canonically as
$\neg(\exists r \in R, s \in S [r.A = 0 \wedge s.B = r.B])$.
This canonical representation implies that a set of existential quantifiers is always predated by the negation operator, 
except for the table variables outside any scope of negation operators.

We will define an additional requirement that each predicate contains 
a local (or what we refer to as \emph{anchored}) attribute
whose table is quantified within the scope of the last negation.
For example, we do not allow 
$\neg(\exists r \in R[\neg(r.A=0)])$ 
because the table variable $r$ is defined outside the scope of the most inner negation around the predicate $r.A=0$.
However, we allow the logically-equivalent 
$\neg(\exists r \in R[r.A \neq 0)])$ where the table variable $r$ is existentially quantified within the same scope as the attribute $r.A \neq 0$.

\begin{definition}[Anchored predicate]\label{def:anchor}
	A predicate is anchored if it contains at least one attribute of a table that is existentially quantified 
	inside the same negation scope as that predicate.
\end{definition}

Intuitively, this anchoring condition guarantees that the predicates can be applied in the same logical scope
where a table is defined.
This requirement also avoids a hidden disjunction.
To illustrate, consider the following $\TRC$ query:
\begin{align*}
	\{ q(A) \mid \exists r \in R [q.A = r.A \wedge 
	\neg (\exists s \in S[r.A = 0 \wedge s.B = r.B])] \}	
\end{align*}
This query contains no apparent disjunction, however the predicate ``$r.A = 0$''
could be pulled outside the negation,
and after applying De Morgan's law on the expression we get a disjunction:
\begin{align*}
	\{ q(A) \mid \exists r \in R [q.A = r.A \wedge 
	\big(
	r.A \neq 0 
	\vee
	\neg (\exists s \in S[s.B = r.B])
	\big) ]\}		
\end{align*}

To avoid disjunctions and ``hidden disjunctions'' the non-disjunctive fragment 
avoids disjunctions entirely and
also requires that predicates are pulled up in the nesting hierarchy as much as possible.

\begin{definition}
	[$\NDTRC$]	
	\label{def:NDTRC}
	The non-disjunctive fragment of safe $\TRC$ 
	restricts predicates to conjunctions of anchored predicates.
\end{definition}

In order to express \cref{fig:disjunctionSQLor}, we need the disjunction operator.
Two possible translations are:
\begin{align*}
	\{ q(A) \mid \exists r \in R,  s \in S,  t \in T 
	[q.A = r.A \wedge (r.A = s.A \vee r.A = t.A)]\}		
\end{align*}
and:
\begin{align*}
	\{ q(A) 
	\mid \;
	& \exists r \in R,  s \in S,  t \in T 
	[q.A = r.A \wedge r.A = s.A] 
	\vee  \\
	& \exists r \in R,  s \in S,  t \in T 
	[q.A = r.A \wedge r.A = s.A]
	\}		
\end{align*}

\subsection{SQL under set semantics}

Structured Query Language (SQL) uses bag instead of set semantics and uses a ternary logic with NULL values.
In order to treat $\SQL$ as a logical query language,
we assume binary logic and no NULL values in the input database. 
It has been pointed out that 
``SQL's logic of nulls confuses people'' and even programmers tend to think in terms of the familiar two-valued logic~\cite{DBLP:conf/pods/ConsoleGLT20}.
Our focus here is devising a general formalism to capture logical query patterns across relational languages,
not on devising a visual representation for SQL's idiosyncrasies.
To emphasize the set semantic interpretation, we write the DISTINCT operator in all our SQL statements.

We define the non-disjunctive fragment of $\SQL$ as the syntactic shown in \cref{table:supported_grammar},
interpreted under set semantics (no duplicates by using \sql{DISTINCT}) 
and under binary logic (no null values allowed in the input tables).
Notice we also have the same syntactic restriction as for $\NDTRC$: 
every predicate needs to anchored (\cref{def:anchor}), i.e.\ reference at least one table within the scope of the last NOT.

\begin{definition}
	\emph{$\NDSQL$}:
	Non-disjunctive SQL under set semantics ($\NDSQL$) is the syntactic restriction of $\SQL$ 
	under binary logic (no NULL values in the input tables)
	to the grammar defined in \cref{table:supported_grammar}, and additionally
	requiring that every predicate is anchored.
\end{definition}

Each such query can be brought into a canonical form that shows a straightforward one-to-one correspondance with $\NDTRC$, 
which will simply our later discussion.
The idea is to replace membership and quantified subqueries with existential subqueries (\cref{table:supported_grammar})
and then unnest any existential quantifiers, i.e., to only use ``\sql{NOT EXISTS}''.
This pulling up quantification as early as possible is logically identical to the way we defined the canonical form of $\NDTRC$.

\subsection{Logical expressiveness of the fragment}

We show that the 4 previously defined non-disjunctive fragments are equivalent in their logical expressiveness.
The proof is available 
\iflabelexists{sec:sec:appendix:nomenclature}
{to the appendix}
{in an online appendix~\cite{relationaldiagrams-TR}}
and
is an adaptation of the standard proofs of equal expressiveness as found, for example, in
\cite{Ullman1988PrinceplesOfDatabase}.
However, the translations also 
need to pay attention to the restricted fragment (e.g., we cannot use union to define an active domain).
The translations between the languages attempt to keep the numbers of extensional database atoms the same if possible.
This detail will be important later in \cref{sec:structureisomorphism}, where we show that those 4 fragments differ in the types of query patterns they can express.

\begin{theorem}\label{th:equivalence}[Logical expressiveness]
	$\DatalogND$, 
	$\NDRA$, 
	$\NDTRC$, and $\NDSQL$ 
	have the same logical expressiveness.
\end{theorem}

\begin{figure}[t]
\centering
\renewcommand{\arraystretch}{0.85}
	\textup{
	\textsf{
	\footnotesize
	\setlength{\tabcolsep}{0.6mm}
	\begin{tabular}{@{} r @{\hspace{2mm}} l @{\hspace{4mm}} l @{\hspace{6mm}}l l}
		Q::=    	& SELECT DISTINCT	C \{, C\}  		& main query\\
		  	  		& FROM 		R \{, R\} 					\\
		  	  		& [WHERE  	P]  								\\[0.5mm]
		S::=    	& SELECT (C \{, C\} $\mid \ ^{*}$)  	& subquery\\
		  	  		& FROM 		R \{, R\} 					\\
		 			& [WHERE  	P]  							 	\\[0.5mm]
		C::=    	& [T.]A     							& column or attribute	\\[0.5mm]
		R::=    	& T [[AS] T]  							& table (table alias)	\\[0.5mm]
		P::=    	& P \{AND P\}     						& conjunction of predicates\\
		$\mid\,$  	& C O C      							& \hspace{3mm}join predicate\\
		$\mid\,$  	& C O V     							& \hspace{3mm}selection predicate\\
		$\mid\,$  	& NOT `('P`)'			 				& \hspace{3mm}negation\\
		$\mid\,$  	& [NOT] EXISTS `('S`)' 					& \hspace{3mm}existential subquery\\
		$\mid\,$  	& C [NOT] IN `('S`)'  					& \hspace{3mm}membership subquery\\
		$\mid\,$  	& C O (ALL `('S`)' $\mid$ ANY `('S`)')  & \hspace{3mm}quantified subquery\\[0.5mm]		
		O::=    	& $< \ \mid \ \leq \ \mid \ = \ \mid \ \geq \ \mid \ > \ \mid \ <>$ 
															& comparison operator\\[0.5mm]
		T::=    	&           					& table identifier\\[0.5mm]
		A::=    	&           					& attribute identifier\\[0.5mm]
		V::=    	&           					& string or number
	\end{tabular}
	}
	}
\vspace{2mm}
\caption{Grammar of $\NDSQL$.
Statements enclosed in [\hspace{1mm}] are optional; 
statements separated by \textbf{$\mid$} indicate 
a choice between alternatives.
In addition, all predicates (joins and selections) need to be anchored.
}
\label{table:supported_grammar}  
\end{figure}

\section{Pattern-preserving mappings between queries across languages}
\label{sec:structureisomorphism}
\label{SEC:STRUCTUREISOMORPHISM}

Our goal is to establish a formalism that allows us to reason about the so-far vague notion of a \emph{relational query pattern}.
We wish to use this formalism to compare relational query languages by their relative abilities to express the query patterns.
Thus, our definitions need to be applicable to all relational query languages, irrespective of their syntax and language-dependent peculiarities.

\subsection{Mapping patterns across queries}

\introparagraph{Intuition}
Our idea is to formalize patterns based on the only common symbols in queries across languages: 
the \emph{input relations} from the database.
Intuitively, we will define two queries to be \emph{pattern-isomorph} if there is a 1-to-1 mapping between the relational input tables across the queries such that appropriately synchronized changes to the input tables (e.g., inserting another tuple)
will keep the two queries logically equivalent.\footnote{Recall that an \emph{isomorphism} is a structure-preserving mapping between two structures that can be reversed.
For it to be reversible, it needs to be
surjective
(each element in the target is mapped to)
 and injective
(different elements in the source need to map to different elements in the target)
\cite{Gallier:2011ul}.
We decided to use the term pattern-isomorphisms in this paper since our focus is on particular structures we refer to as patterns.
}
We call any such 1-to-1 correspondence between queries a \emph{pattern-preserving mapping}.

For a definition of pattern-preserving mapping
to correctly identify patterns instead of logical equivalence, 
we need to be able to 
treat repeated instances of the same input table  (also called self-joins)
in a query as \emph{independent}.
As example, consider the relational query 
$R - \big( \pi_A R \times S \big)$
from \cref{Fig_RA_vs_Datalog_b}.
From a logical point of view, the query is a function $q(R, S)$ that maps input relations $R$ and $S$ to a binary output table.
In that perspective of queries as functions mapping relations to an output relation, 
we call the \emph{signature} of this query
its relational input $(R, S)$.
We then consider what we call the \emph{query representation}
$\overline{q}(R, R, S)$
which represents the function mapping three explicitly-referenced tables to an output.
We then call the query 
${q}'(R_1, R_2, S) = R_1 - \big( \pi_A R_2 \times S \big)$ that treats all explicit tables as distinct
as \emph{shattered query} of $q$.
And we define the relational query pattern to be that shattered query.

\introparagraph{Formalisation}
To make these intuitions precise across the varying syntax of relational query languages, 
we need a way to refer to the individual appearance of \emph{extensional tables}
in a query, irrespective of the language.

\begin{definition}[Extensional table]
	We call an \emph{extensional table} any existentially-quantified and language-specific appearance of a relational input table
	in a query.
\end{definition}

\begin{definition}[Query representation] 
	Given a query $q(\mathcal{T})$ 
	that maps a set of database tables $\mathcal{T}$ to an output table
	using a signature $\overline{\mathcal{T}}$ of extensional tables,
	we call \emph{the query representation} $\overline q(\overline{\mathcal{T}})$ of query $q$ 
	the language-specific representation of the query.
\end{definition}

The intuition is that every relational query language has language-specific ways 
to reference tables, which are then used to define the actual query. 
These two definitions allows us to refer to the
existentially-quantified
input relations 
while abstracting away details of the language.
Recall that a query 
can quantify the same input relation more than once.
For example, the $\Datalog$ query $Q(x)\datarule R(x,y), R(\_,y)$ has two extensional tables, 
both of which refer to the same input table $R$.
It thus represents a relational function $q(R)$ that maps a any valid instance of table $R$ to an output table.
Its query representation however is $\overline Q(R, R)$ since it uses two occurrences of $R$, i.e.\ 
two extensional tables.\footnote{Without loss of generality, the order of the extensional tables in the signature of a query representation is the order of quantification of those tables in the actual query.}
We can now analyze the query defined by a query representation, treating each extensional table as different.
Thus, from $\overline Q(R, R)$, we consider what we call its \emph{shattered query} $Q'(R_1, R_2)$ defined as
$Q'(x)\datarule R_1(x,y), R_2(\_,y)$.
Notice that if we replace the signature $(R_1, R_2)$ of the shattered query $Q'$
with the signature $(R, R)$ of the query representation $\overline Q$,
we get a logically equivalent formulation of the original query $Q$.

\begin{definition}[Shattered query]
	Given a query  $q(\mathcal{T})$ with representation  $\overline{q}(\overline{\mathcal{T}})$, 
	we call $q'(\overline{\mathcal{T}}')$ its shattered query 
	iff
	$q'(\overline{\mathcal{T}}) \equiv q(\mathcal{T})$.
\end{definition}

In the above example, $Q'(R_1, R_2)$ 
with signature $\overline{\mathcal{T}}' = (R_1, R_2)$
is the shattered query of $Q(R)$
because
$Q'(R, R) \equiv Q(R)$. Both are $Q(x)\datarule R(x,y), R(\_,y)$.
The intuition behind our formalisms is that the shattered query
defines a function that maps a set of \emph{extensional tables} (not just a set of tables) to an output table.
Thus \emph{the shattered query
is a semantic definition}
of a relational query pattern across different relational query languages.
Two queries the use the same query pattern if their shattered queries are logically equivalent,
up to renaming and reordering of the input tables.
We first give the formal definition and then illustrate with more examples.

\begin{definition}[Query pattern]
	The semantics of a query pattern for a relational query $q(\mathcal{T})$ is defined by its shattered query $q'(\mathcal{T}')$.
\end{definition}

\begin{definition}[Pattern isomorphism]
	\label{def:isomorphism}
	Given two queries $q_1$ and $q_2$ with shattered queries
	$q_1'(\mathcal{T}')$
	and
	$q_2'(\mathcal{T}'')$.
	The queries are pattern-isomorph iff there is a bijective homomorphism 
	$h: \mathcal{T'} \rightarrow \mathcal{T''}$
	such that
	$q_1'(\mathcal{T'}) \equiv q_2'(h(\mathcal{T'}))$.	
\end{definition}

\begin{example}[Different patterns]
\label{ex:querystructureiso1}
We give an example of logically-equivalent queries that use arguably different query patterns.
Consider
table $R(A,B)$ and the two queries
$Q_1(R)$ and $Q_2(R)$ with
\begin{align*}
	Q_1(x) 	&\datarule \h{R}(x,\_), \h{R}(x,\_). \\
	Q_2(x)	&\datarule \h{R}(x,y),\h{R}(\_,y).
\end{align*}
Both queries are logically equivalent to $Q(x) \datarule \h{R}(x, \_)$, 
and thus also logically equivalent to each other.
However, $Q_1$ and $Q_2$ represent arguably different patterns:
$Q_1$ never uses the second attribute of $R$ whereas $Q_2$ uses it to join both occurrences of $R$.
This difference becomes even more apparent when writing the two queries with the same join pattern in SQL:
\cref{Fig_isomorphism_sql_1}
would even work if $R$ was unary, whereas \cref{Fig_isomorphism_sql_2} requires $R$ to be at least binary.

We next show that 
shattering allows us to formally distinguish the two patterns in the queries,
i.e.\ they are \emph{not pattern-isomorph}.
The shattered queries are 
$Q_1'(R_1, R_2)$
and
$Q_2'(R_3, R_4)$
with
\begin{align*}
	Q_1'(x) &\datarule \h{R_1}(x,\_),\h{R_2}(x,\_). \\
	Q_2'(x)	&\datarule \h{R_3}(x,y), \h{R_4}(\_,y).
\end{align*}
Neither of the two possible mappings between the shattered queries' extensional tables,
$h_1 = \{(R_1, R_3), (R_2, R_4)\}$
nor
$h_2 = \{(R_1, R_4),$ $(R_2, R_3)\}$,
preserves 
logical equivalence for the shattered queries.

However, $Q_1$ is pattern-isomorph to the $\TRC$ query $q_3(R)$ with
\begin{align*}
	&\{ q_3(A) \mid \exists r_1 \in \h{R}, r_2 \in \h{R} 
	[q.A = r_1.A \wedge r_1.A = r_2.A] \}	\\
\intertext{
To see that, notice that its shattered query $q_3'(R_5, R_6)$ with
}
	&\{ q_3'(A) \mid \exists r_1 \in \h{R_5}, r_2 \in \h{R_6}
	[q.A = r_1.A \wedge r_1.A = r_2.A] \}	
\end{align*}
allows the reversible mapping 
$h_3 = \{(R_1, R_5), (R_2, R_6)\}$
from $Q_1'$ to $q_3'$
that preserves logical equivalence.

By the same arguments, $Q_1$ is pattern-isomorph to the query
in 
\cref{Fig_isomorphism_sql_1},
and $Q_2$ is pattern-isomorph to the query
in
\cref{Fig_isomorphism_sql_2}.

\end{example}

Our formalism is similar in spirit to \emph{edge-preserving graph homomorphisms}
that map two nodes in graph $G_1$ linked by an edge to two nodes in graph $G_2$ that are also linked by an edge.
In our pattern-preserving isomorphisms between queries,
the role of nodes is played by the extensional tables in the queries
and the queries themselves play the role of the edges.
Notice the difference to well-known homomorphisms between conjunctive queries for determining query containment~\cite{Chandra:1977:OIC:800105.803397}: 
in that formalism, the role of nodes is played by variables (and constants)
and the relational atoms play the role of edges.
Also notice from
\cref{ex:querystructureiso1}
that a simpler mapping between the relational symbols (instead of the repeated extensional tables)
between two queries alone would not work.

Notice that---by design---our definition does not classify different join orders as different query patterns.
Also---by design---our definition does not include any notion of views or intermedia tables. 
This is achieved by excluding Intensional Database Predicates (IDBs) (as in $\Datalog$) from the definition of extensional tables.
We again illustrate the intuition for that design with examples.

\begin{example}[Join orders and views do not affect patterns]
\label{ex:querystructureiso2}
Assume that the edges of a directed graph are stored in a binary relation $E(A,B)$.
Consider a query  
returning nodes which form the starting point of a length 3 directed path.
We write the query in two different ways in unnamed RA where indices replace attribute names \cite{DBLP:books/aw/AbiteboulHV95}.
The first query applies projections as late as possible, 
whereas the second query applies the projections as early as possible:
\begin{align*}
	q_1(E) &= \pi_{1}\sigma_{2=3 \wedge 4=5}(\h{E} \times \h{E} \times  \h{E}) \\
	q_2(E) &= \pi_{1}\sigma_{2=3}(\h{E} \times \pi_{1}\sigma_{2=3}(\h{E} \times  \pi_{1}\h{E})) 	
\end{align*}
We also write these two queries in the more familiar named perspective of RA. 
These queries encode the same algebraic operations but are more verbose since 
the named perspective of RA requires a rename operator $\rho$ to unambiguously express the identical queries:
\begin{align*}
	q_1(E) &= \pi_{E.A}\sigma_{E.B=F.A \wedge F.B=G.A}(\h{E} \times \rho_{E \rightarrow F} 
				\h{E} \times  \rho_{E \rightarrow G} \h{E}) \\
	q_2(E) &= \pi_{E.A}\sigma_{E.B=F.A}(\h{E} \times \pi_{F.A}\sigma_{F.B=G.A}(
				\rho_{E \rightarrow F} \h{E} \times \rho_{\h{E} \rightarrow G} \pi_{A} \h{E})) 	
\end{align*}

Both RA queries use the same relational pattern according to our definition,
and we think the distinction between query patterns and join orders is important:
If join orders determined relational query pattern, then relational query patterns would be inherently tied to relational algebra;
concepts of join order and early projections are not meaningful in the context of 
declarative logical query languages.
To see that consider the logically-equivalent query in $\Datalog$:
\begin{align*}
	Q_3(x) 	&\datarule \h{E}(x,y), \h{E}(y,z), \h{E}(z,w).
\end{align*}
Query $Q_3$ logically specifies on what attributes the three tables need to be joined, 
but it does not specify any order or joins nor when projections happen.
Furthermore notice that RA query $q_1$ does not even specify a join order between the three extensional tables.

For a similar reason, temporary tables such as Intensional Database Predicates (IDBs) in $\Datalog$ do not count as extensional tables.
Thus, the following $\Datalog$ query uses
the same logical pattern (find three edges that join and keep the starting node),
even though it defines the intermediate intensional database predicate $I$:
\begin{align*}
	I(y)  	&\datarule \h{E}(y,z), \h{E}(z,w). \\
	Q_4(x) 	&\datarule \h{E}(x,y), I(y).
\end{align*}

\end{example}

\begin{figure}[t]
\centering
\begin{subfigure}[b]{.3\linewidth}
\begin{lstlisting}
SELECT DISTINCT R1.A
FROM R R1, R R2
WHERE R1.A = R2.A
\end{lstlisting}
\vspace{-7mm}
\caption{}
\label{Fig_isomorphism_sql_1}
\end{subfigure}
\hspace{5mm}
\begin{subfigure}[b]{.3\linewidth}
	\centering
	\includegraphics[scale=0.35]{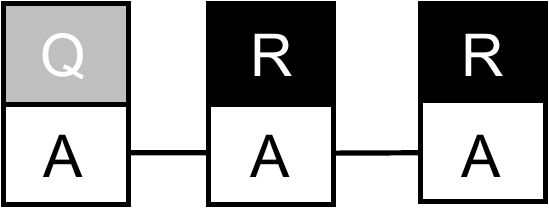}
	\vspace{-1mm}
	\caption{}
	\label{Fig_isomorphism_def_1}
\end{subfigure}
\hspace{15mm}
\begin{subfigure}[b]{.3\linewidth}
\begin{lstlisting}
SELECT DISTINCT R1.A
FROM R R1, R R2
WHERE R1.B = R2.B
\end{lstlisting}
\vspace{-7mm}
\caption{}
\label{Fig_isomorphism_sql_2}
\end{subfigure}
\hspace{5mm}
\begin{subfigure}[b]{.3\linewidth}
	\centering
	\includegraphics[scale=0.35]{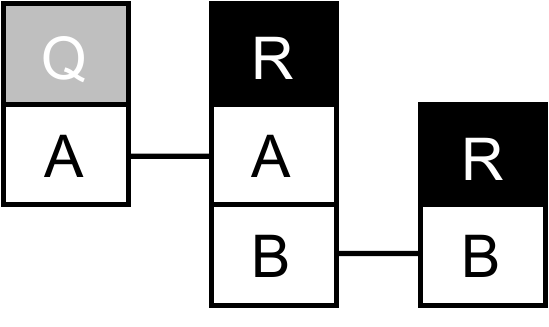}
	\vspace{-1mm}
	\caption{}
	\label{Fig_isomorphism_def_2}
\end{subfigure}
\caption{
\Cref{ex:querystructureiso1}: 
Two queries (a) and (c) that are logically-equivalent, but not pattern-isomorph.
Their associated \diagrams are shown in (b) and (d), respectively.}
\label{Fig_isomorphism_def}
\end{figure}

Notice that it follows immediately that two pattern-isomorph queries 
need to have the same number of extensional tables.
We believe that such a pattern-preserving mapping between queries is important if we want to help readers understand the 
\emph{exact logical pattern}
behind a relational query, irrespective of the language it is written in.
In particular, if we want to help users understand the logic of an existing relational query (recall that we focus on set semantics and binary logic), 
the diagrammatic representation needs to preserve this 1-to-1 correspondence with the query.

\subsection{Comparing relational languages in terms of ``pattern-expressiveness''}

We next add the final definition needed to formally compare relational query languages based on their relative abilities to represent query patterns.

\begin{definition}[Representation equivalence]
	We say that a query language $\L_2$ can \emph{pattern-represent} a query language $\L_1$
	(written as $\L_1 \subseteq^\rep \L_2$) 
	iff 
	for every legal query  $q_1 \in \L_1$ 
	there is a pattern-isomorphic query $q_2 \in \L_2$.
	We call a query languages $\L_2$ \emph{pattern-dominating} another language $\L_1$ 
	(written as $\L_1 \subsetneq^\rep \L_2$)
	iff
	$\L_1 \subseteq^\rep \L_2$
	but
	$\L_1 \not \supseteq^\rep \L_2$.
	We call $\L_1, \L_2$ \emph{representation equivalent}
	(written as $\L_1 \equiv^\rep \L_2$)
	iff 
	$\L_1 \subseteq^\rep \L_2$
	and
	$\L_1 \supseteq^\rep \L_2$,
	i.e.\ if both language can represent the same set of patterns.	
\end{definition}

We are now ready to state our result on the hierarchy of pattern expressiveness 
of these four relational query languages.
Recall that we are only considering the non-disjunctive fragment of the languages that we defined earlier in \cref{sec:nondisjunctivefragment}.

\begin{theorem}
	\label{th:representations}
	$\NDRA \subsetneq^\rep \DatalogND \subsetneq^\rep \NDTRC \equiv^\rep \NDSQL$.
\end{theorem}

Thus, we prove that \emph{relational calculus has relational patterns that cannot be expressed in relational algebra}.
In more detail, our proofs 
\iflabelexists{sec:sec:appendix:nomenclature}
{available in the appendix}
{available in the online appendix \cite{relationaldiagrams-TR}}
show that
there exists a query for which 
$\NDRA$
needs $50\%$ more extensional tables than $\DatalogND$,
and
there exists a query for which 
$\DatalogND$
needs $33.\dot{3}\%$ more extensional tables than $\NDTRC$ or $\NDSQL$.
The important consequence is that $\NDRA$, $\DatalogND$ or any diagrammatic language modeled after them 
would not be a suitable target language for helping users understand all existing relational query patterns
(including those used by $\NDSQL$).

We will later see that most existing visual query representations are modeled after relational algebra in that they model data flowing between relational operators, which implies they cannot truthfully represent 
all relational query patterns from $\TRC$ or $\NDSQL$.

\section[Relational diagrams]{Relational diagrams}
\label{SEC:QV}
\label{sec:QV}

This section introduces the basic visual elements of 
\diagrams
(\cref{sec:visualelements}).
It gives the formal translation 
from $\NDTRC$ 
(\cref{sec:fromTRCtoRD})
and back (\cref{sec:fromRDtoTRC}),
showing that there is a one-to-one correspondence between $\NDTRC$ expressions and \diagrams
and proving their validity (\cref{sec:RD_validity}).

\begin{figure}[t]
\centering
\begin{subfigure}[b]{.35\linewidth}
	\centering
    \includegraphics[scale=0.35]{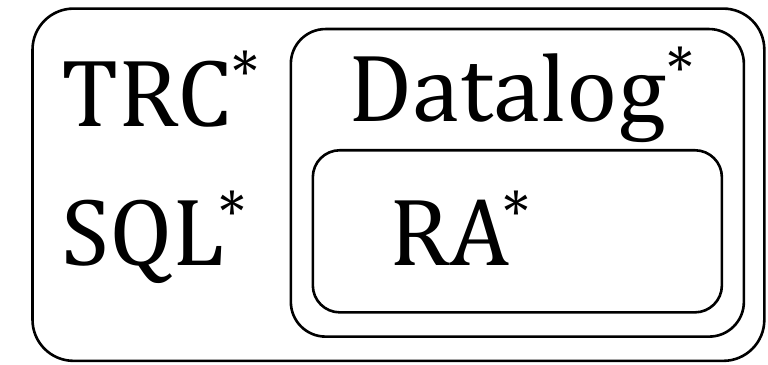}
	\vspace{-2mm}
    \caption{}
    \label{Fig_Representation_Hierarchy_main}
\end{subfigure}	
\hspace{1mm}
\begin{subfigure}[b]{.58\linewidth}
	\centering
    \includegraphics[scale=0.34]{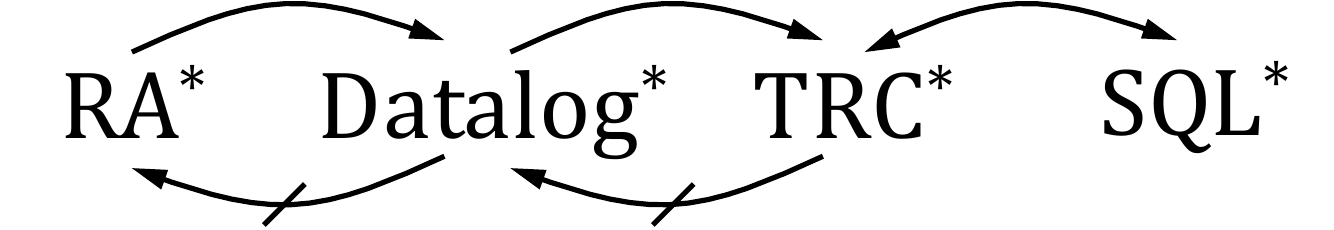}
	\vspace{1mm}
    \caption{}
    \label{Fig_Representation_main}
\end{subfigure}	
\caption{\cref{th:representations}: (a) Representation hierarchy between the non-disjunctive fragments of four query languages.
(b) Directions of pattern-preservation (and non-preservation) used in the proof.}
\end{figure}

\subsection[Visual elements]{Visual elements}
\label{sec:visualelements}

In designing our diagrammatic representation, 
we started from existing widely-used visual metaphors and then added the minimum necessary visual elements to 
obtain expressiveness for full $\NDTRC$.
In the following 5 points, we discuss both 
($i$) necessary specifications for \diagrams, and
($ii$) concrete design choices that are not formally required but justified based on HCI and visualization guidelines and best practices.
We use the term \emph{canvas} to refer to the plane in which the \diagram is displayed.
\Cref{Fig_TRC_vs_RD_b} displays several of the examples discussed next.

\emph{(1) Tables and attributes}:
We use the set-of-mappings definition of relations~\cite{Ullman1988PrinceplesOfDatabase}
in which a tuple is a mapping from attributes' names to values---in contrast to the set-of-lists representation in which order of presentation matters and which more closely matches the typical vector representation.
Thus a table is represented by any visual grouping of its attributes.
We use the typical UML convention of representing tables as rectangular boxes
with table name on top and attribute names below in separate rows. 
For maximum legibility, we ensure that text has maximal luminance contrast with its background \cite{munzner2015visualization}.
Table names are shown with white text on a black background and, to differentiate them, attributes use black text on a white background.
For example, table \tableBox{R} with attribute \attributeBox{A}.
To simplify query interpretation, we do not use table aliases, similar to $\Datalog$ and $\RA$ (and different from $\SQL$ and $\TRC$).
We also reduce visual complexity by only showing attributes that are used in the particular query, similar to $\SQL$ and $\TRC$ (and different from $\Datalog$).
Database users are commonly familiar with relational schema diagrams.
Thus, we argue that a simple conjunctive query should be visualized similarly to
a typical database schema representation,
as used prominently in standard introductory database textbooks \cite{Elmasri:dq,DBLP:books/mg/SKS20}.
We illustrate with \cref{Fig_TRC_vs_RD_b}.

\emph{(2) Selection predicates}:
Selection predicates are filters and are shown ``in place.'' 
For example, an attribute ``$r_2.C > 1$'' is shown as \selPredicateBox{C$>$1}
in the corresponding instance of table $R$.
An attribute participating in multiple selection predicates is repeated at least as many times as there are selections
(e.g., to display ``$r_2.C > 1 \wedge r_2.C < 3$'', we would repeat $\sql{R.C}$ twice as \selPredicateBox{C$>$1} and \selPredicateBox{C$<$3}).
An attribute participating in $k$ selection predicates, then we repeat it $k$ times.

\emph{(3) Join predicates}:
Equi-join predicates (e.g., ``$s_2.A = t_2.A$'') , which arguably are the most common type of join in practice, are represented
by lines connecting the joined attributes.
For the other less-frequent theta join operators 
$\{\neq, <, \leq, \geq, >\}$, 
we additionally place the operator as a label on the line
and use an arrowhead to indicate the reading order and correct application of the operator \emph{in the direction of the arrow}.
For example, for a predicate
``$r_1.A > r_2.B$'', the label is $>$ and the arrow points from 
attribute $\sql{A}$ of the first $\sql{R}$ occurrence to $\sql{B}$ of the second:
$\sql{A} {\scriptscriptstyle \xrightarrow{>}} \sql{B}$.
Notice that the direction of arrows can be flipped, along with flipping the operator, while maintaining the identical meaning:
$\sql{A} {\scriptscriptstyle \xleftarrow{<}} \sql{B}$. 
To avoid ambiguity with the standard left-to-right reading convention for operators, we normalize arrows to never point from right to left.
An attribute participating in multiple join predicates needs to be shown only once and has several lines connecting it to other attributes.
An attribute participating in one or more join predicates and $k$ selection predicates, is shown $k+1$ times.\footnote{In practice, one can reduce the size of a \diagram by reusing an existing selection predicate also for joins. This comes at the conceptual complication that the exact graph topology of the \diagram (which attributes are connected) is not uniquely determined (though it still allows only one correct interpretation).
In our example \cref{Fig_TRC_vs_RD_b}, one could remove the attribute \sql{R.C} of $r_2$ and connect \sql{Q.D} to either \selPredicateBox{C$>$1} or \selPredicateBox{C$<$3} instead.}

\emph{(4) Negation boxes}:
In $\NDTRC$, negations are either avoided (e.g., $\neg (R.A = S.B)$ is identical to $R.A \neq S.B$) or 
placed before the existential quantifiers.
We represent a negation with a closed line that partitions the canvas into a subcanvas that is negated (inside the bounding box)
and everything else that is not (outside of the bounding box).
As convention, we use dashed rounded rectangles.\footnote{Rectangles allow better use of space than ellipses, and rounded corners together with dashed lines distinguish those negation boxes clearly from the rectangles with solid edges and right angles used for tables and attributes.}
Recursive partitioning of the canvas allows us to represent a tree-based nesting order that corresponds to
the nested scopes of quantified tuple variables in $\TRC$
(and also the nesting order of subqueries in $\SQL$).
We call \emph{the main canvas} the root of that nesting hierarchy and each node a \emph{partition of the canvas}.

\emph{(5) Output table}:
We display an output table to emphasize the compositional nature of relational queries: a relational query uses several tables as input, and returns one new table as output. 
We use the same symbol for that output table as the $\TRC$ expression, for which we most commonly use $Q$.
We use a gray background \selectBox{} to make this table visually distinct from input tables.
As the title is always Q, the reduced luminance contrast of white on gray is acceptable as it little impacts query readability.

\subsection{From TRC to Relational Diagrams}
\label{sec:fromTRCtoRD}

We next describe the 5-step translation from any valid $\NDTRC$ expression to a \diagram.
We also illustrate by translating the $\NDTRC$ expression from \cref{Fig_TRC_vs_RD_a} into the \diagram from \cref{Fig_TRC_vs_RD_b}.
Notice that the translation critically leverages 3 conditions fulfilled by the input:
(1) Safe $\TRC$ (and thus also $\NDTRC$) only allows existential and not universal quantification \cite{Ullman1988PrinceplesOfDatabase},
(2) $\NDTRC$ only allows conjunction between predicates, and
(3) all predicates in $\NDTRC$ are anchored (recall \cref{def:anchor}).

\emph{(1) Creating canvas partitions}:
The scopes of the negations in a $\TRC$ are nested by definition.
We translate this hierarchy of the scopes for each negation (the \textit{negation hierarchy}) into a nested partition of the canvas.
\cref{Fig_TRC_vs_RD_d} illustrates the nested partitions as derived from the 
negation hierarchy \cref{Fig_TRC_vs_RD_c} of the original $\NDTRC$ expression.
Notice that the double negation ``$\neg(\neg(\ldots))$'' results in the scope $q_1$ of the negation hierarchy to be empty.

\emph{(2) Placing tables}:
Each table variable defines a table that gets placed into the canvas partition 
that corresponds to the respective negation scope. For example, the tables corresponding to the table variables $r_1$, $r_2$, and $s_1$ are all outside any negation scope and are thus placed in the root partition $q_0$.
Notice that similar to \Datalog\ and \RA\ (and in contrast to $\TRC$ and $\SQL$), \diagrams do not need table aliases.

\begin{figure}[t]
\centering
\begin{subfigure}[b]{1\linewidth}
	\centering	
    \includegraphics[scale=0.42]{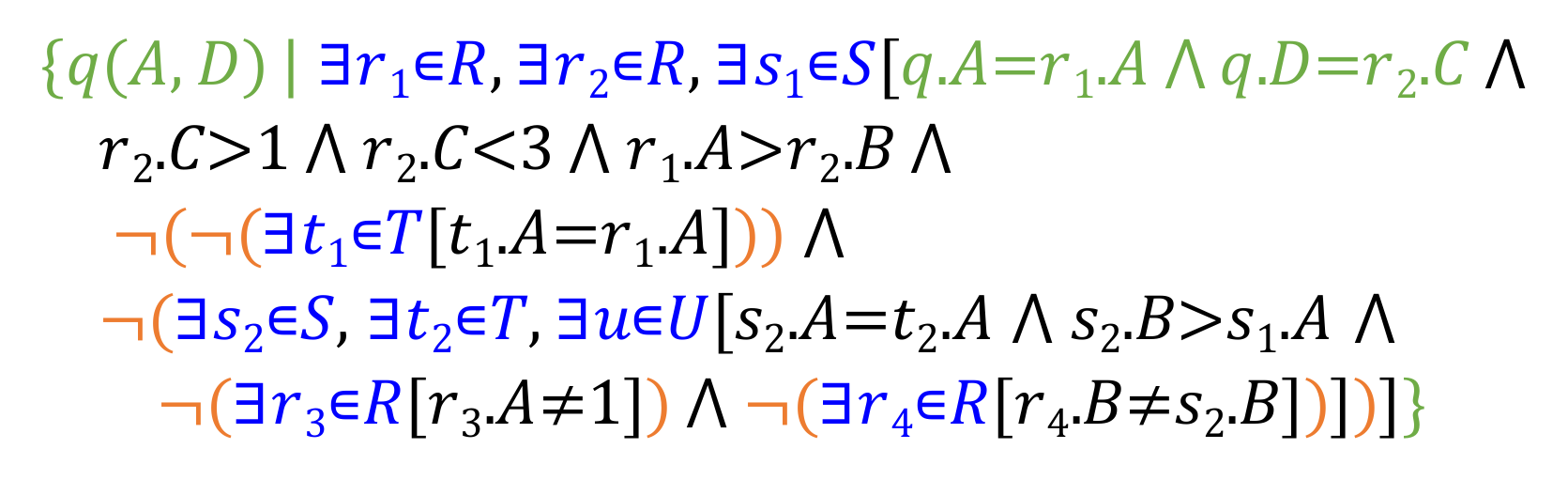}
	\vspace{-3mm}
    \caption{$\NDTRC$}
    \label{Fig_TRC_vs_RD_a}
\end{subfigure}
\begin{subfigure}[b]{0.45\linewidth}
	\centering
    \includegraphics[scale=0.42]{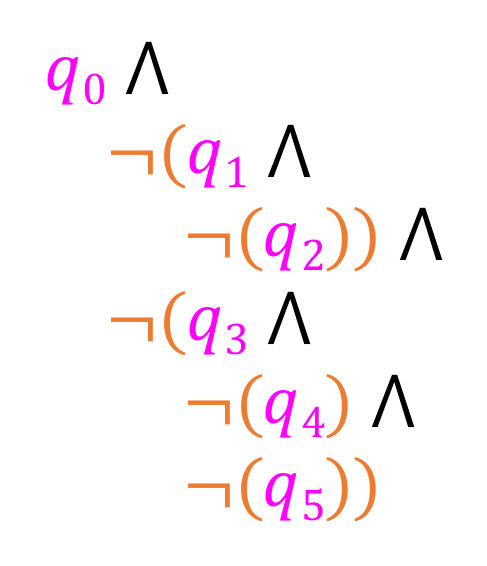}
	\vspace{-3mm}	
    \caption{Negation hierarchy}
    \label{Fig_TRC_vs_RD_c}
\end{subfigure}
\hspace{5mm}
\begin{subfigure}[b]{0.45\linewidth}
	\centering
    \includegraphics[scale=0.42]{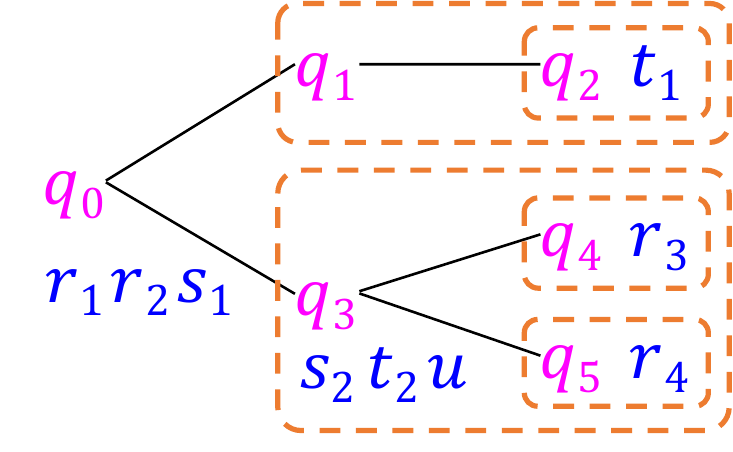}
    \caption{Canvas partitions}
    \label{Fig_TRC_vs_RD_d}
\end{subfigure}
\begin{subfigure}[b]{1\linewidth}
	\centering	
	\vspace{2mm}		
    \includegraphics[scale=0.42]{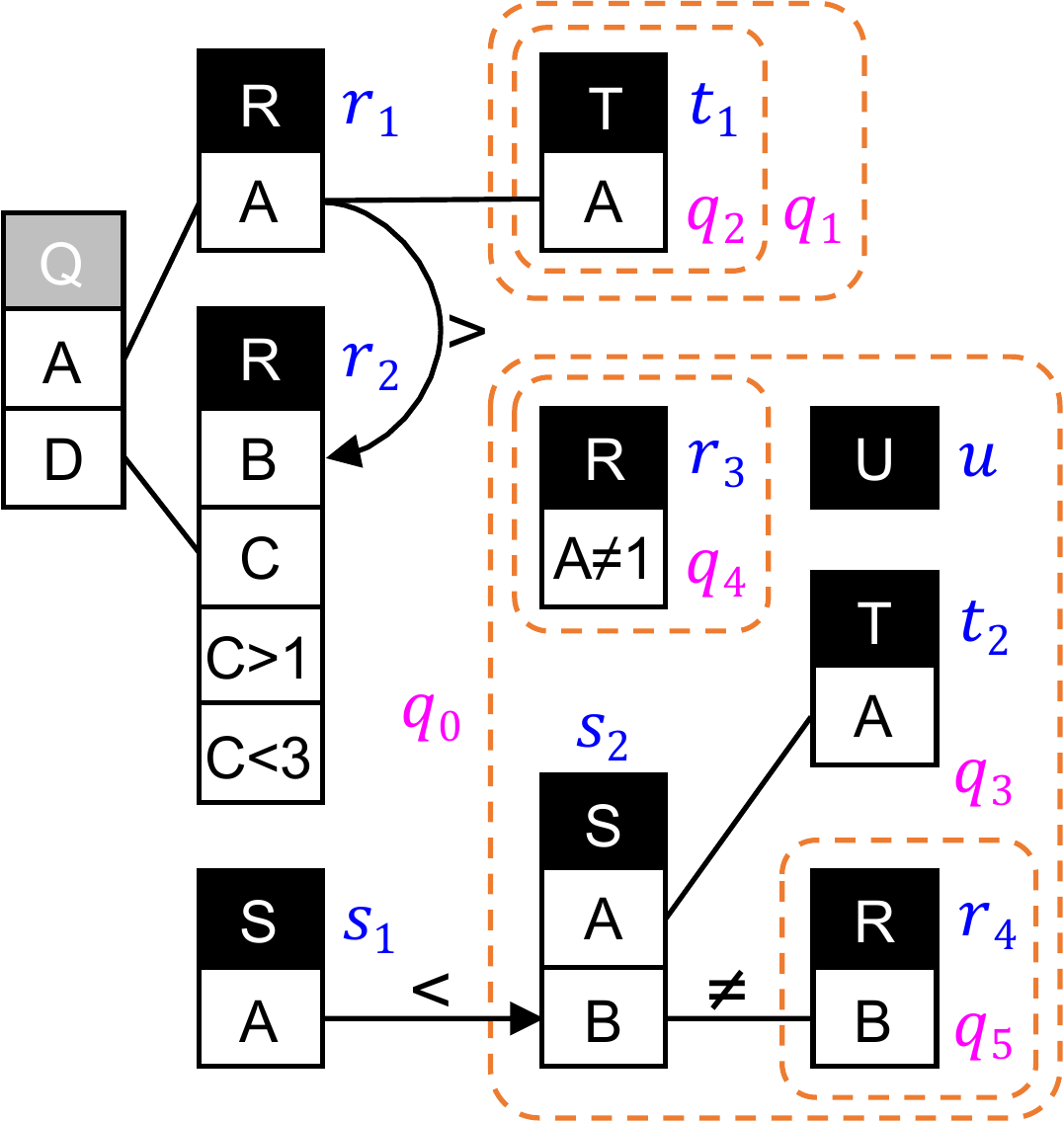}
    \caption{\diagram}
    \label{Fig_TRC_vs_RD_b}
\end{subfigure}
\caption{Example $\NDTRC$ expression (a), its corresponding \diagram (d), and derivation of the nesting hierarchy (b, c).
Colored partitions $q_i$ (purple) and table variables $r_i$ (blue) are not part of \diagrams and displayed to discuss the correspondence.
}
\label{Fig_TRC_vs_RD}
\vspace{-2mm}
\end{figure}

\emph{(3) Placing selection predicates}:
The predicates within each scope are combined via conjunction and are thus added one after the other.
Since all selection predicates are anchored, the selection predicates can be placed in the same partition as their respective table, which allows correct interpretation (see \cref{sec:fromRDtoTRC}).
For example, for 
$\neg(\exists r_3 \in R[r_3.A \neq 1])$, 
the predicate ``$A \neq 1$'' is placed directly below $R$ in $q_4$.
An example of a predicate that is not anchored would be $\exists r_3 \in R[\neg(r_3.A=1)]$: the scope of the negation contains a predicate of a table that is not existentially quantified in that scope. 

\emph{(4) Placing join predicates}:
For each each join predicate, we add the two attributes (if not already present) and connect them via an edge with any comparison operator drawn at the middle.
An attribute participating in multiple join predicates needs to be shown only once. 
Equi-joins are the standard and no operator is shown. 
Asymmetric joins include an arrowhead at one end of the edge (see \cref{sec:visualelements}).
Since for anchored join predicates at least one of the two attributes is in the partition of a local table, 
the negation can be correctly interpreted.
An example of a predicate that is not anchored would be
$\neg(r_4.B=s_2.B)$. 
What is possible is the logically-equivalent 
$r_4.B \neq s_2.B$ (as long as one of the two attributes is in the local scope of the last negation.
In our example, this is the case in 
$\neg(\exists r_4 \in R[r_4.B \neq s_2.B])$.

\emph{(5) Place and connect output table}:
The safety conditions 
for $\TRC$ \cite{Ullman1988PrinceplesOfDatabase} 
imply that the output predicates can only be chosen from tables outside of all negations,
thus in the root scope or partition $q_0$.
We place an additional table with a new name (we commonly use $Q$ for query)
and use a unique gray background \selectBox{} to imply the difference from extensional tables.

\introparagraph{Completeness}
Notice that this 5-step translation guarantees uniqueness of the following aspects:
(1) nesting hierarchy (corresponding to the negation hierarchy),
(2) where tables are placed (canvas partitions corresponding to the negation scope),
(3) which attributes have selection predicates, and
(4) which attributes participate in joins and how.
The following aspects are not uniquely defined
(without impact on the later interpretation):
(1) the order of attributes below each table;
(2) the direction of arrows can be flipped with simultaneous label flip
e.g., 
$s_1.A {\scriptscriptstyle \xleftarrow{>}} s_2.B$
and
$s_1.A {\scriptscriptstyle \xrightarrow{<}} s_2.B$
are identical (by convention we avoid arrows from right-to-left, but allow them up-to-down and down-to-up);
(3) the size of visual elements and their relative arrangement; and
(4) any optional changes in style (e.g., other than dashed negation boxes, distinct visual appearance between tables and attributes).

\subsection{From a Relational Diagram to TRC}
\label{sec:fromRDtoTRC}

We next describe the reverse 5-step translation from any valid \diagram to a valid and unique $\NDTRC$ expression. 
At the end, we summarize the conditions of a \diagram to be valid,
which are the set of requirements listed for each of the 5 steps. 
We again illustrate with the examples from \cref{Fig_TRC_vs_RD}.

\emph{(1) Determine the nested scopes of negation}:
From the nested canvas partitions (\cref{Fig_TRC_vs_RD_d}), 
create the nested scopes of the negation operators of the later $\NDTRC$ expression (\cref{Fig_TRC_vs_RD_c}).

\emph{(2) Quantification of table variables}:
Each table in a partition corresponds to an existentially-quantified table variable. 
WLOG, we use a small letter indexed by number of occurrence for repeated tables.
We add those quantified table variables in the respective scope of the negation hierarchy (\cref{Fig_TRC_vs_RD_d}).
For example, table $T$ in $q_2$ becomes
$\exists t_1 \in T[\ldots]$ and replaces $q_2$ in \cref{Fig_TRC_vs_RD_intermediate}.
Notice that partition $q_1$ is empty and the resulting negation scope does not contain any expression other than another negation scope.
We require that the leaves of the partition are not empty and contain at least one table.
Otherwise, expressions $\wedge \neg()$ and $\wedge \neg(\neg())$
would both have to be true, leaving the meaning of an empty leaf partition ambiguous.
This also implies that an empty canvas (there is only one partition, in which root and leaf are empty) is not a valid \diagram.

\emph{(3) Selection predicates}:
Selection attributes are placed into the scope in which its table is defined.
For example, the predicate $R.A \neq 1$ in partition $q_4$ leads to
$\neg(\exists r_3 \in R[r_3.A \neq 1])$.

\emph{(4) Join predicates}:
For join predicates (lines connecting attributes in \diagrams with optional direction and operator), we have a validity condition 
that they can only connect attributes of tables that are in same partition or different partitions that are in a direct-descendant relationship.
In our example, $T.A$ in $q_2$ connects to $R.A$ in $q_0$ (here $q_0$ is the root and grandparent of $q_2$.)
However, we could not connect any attribute in $q_5$ with any attribute in $q_4$ (which are siblings in the nesting hierarchy).
This requirement is the topological equivalent of scopes for quantified variables in $\TRC$ and
guarantees that only already-defined table variables are referenced.
Each such predicate is placed in the scope of the lower of the two partitions in the hierarchy, which guarantees the predicate to be ``anchored." 
For example, the inequality join connecting $S.B$ in $q_3$ and $R.B$ in $q_5$ is placed in the scope of $q_5$.

\emph{(5) Output table}:
The validity condition for the output table is that each of its one or more attributes is connected to exactly one attribute from a table in the root partition $q_0$. This corresponds to the standard safety condition of safe $\TRC$.
This steps adds the set parentheses, the output tables, and its attribute and output predicates shown in green in \cref{Fig_TRC_vs_RD_a}.

\introparagraph{Soundness}
Notice that this 5-step translation guarantees that the resulting $\NDTRC$ is uniquely determined up to 
(1) renaming of the tuple variables;
(2) reordering the predicates in conjunctions, and
(3) flipping the left/right positions of attributes in each predicate.
It follows that \diagram are sound, 
and their logical interpretation \emph{unambiguous}.

\subsection{Valid \diagrams}
\label{sec:RD_validity}

\begin{figure}[t]
\centering
\includegraphics[scale=0.42]{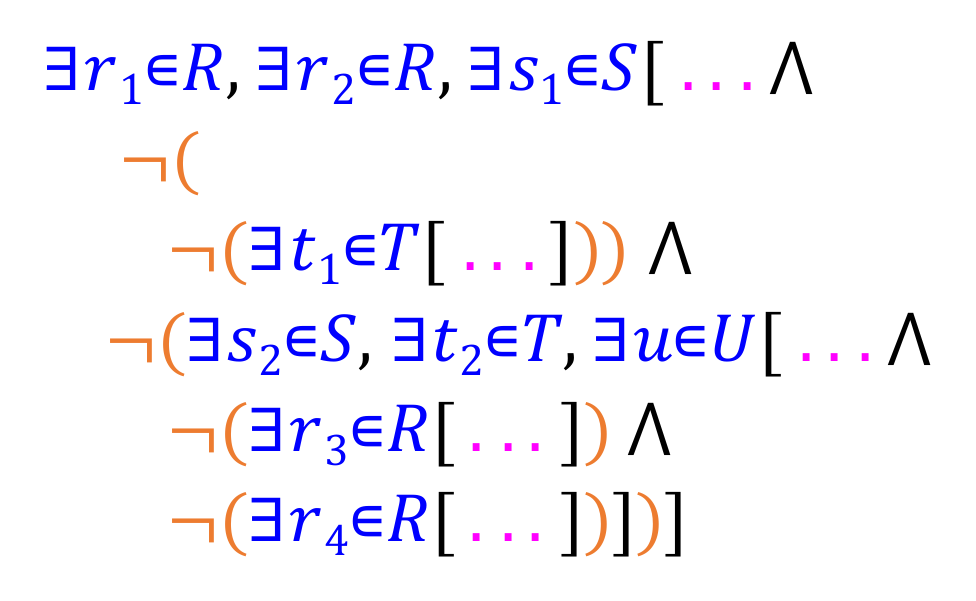}
\caption{\Cref{sec:fromRDtoTRC}: $\NDTRC$ stub after step 2 of the translation.}
\label{Fig_TRC_vs_RD_intermediate}
\vspace{-2mm}
\end{figure}

In order for a \diagram
to be valid we require that
each of the conditions for the 5-step translation process are fulfilled.

\begin{definition}[Validity]
\label{def:validRelationalDiagram}
A \diagram is valid iff
\begin{enumerate}
\item
The nested hierarchy of optional negation boxes partitions the canvas (any two dashed boxes are either disjoint or one is completely contained within the other).

\item
Each table, its attributes, and its selection predicates are discernible and reside in exactly one canvas partition.

\item
Each leaf in the canvas partition contains at least one table.

\item
Joins only happen between attributes of tables in partitions that are descendants (not siblings or their descendants).
Join predicates with asymmetric operators such as $<$ and $>$ require a line with directionality (e.g., an arrow head).

\item
The output table has at least one attribute, and each attribute connects to exactly one attribute in the root partition $q_0$ 
(safety condition of $\TRC$).
\end{enumerate}
\end{definition}

\begin{figure*}[t]
\centering
\begin{minipage}{0.14\linewidth}
\begin{subfigure}[b]{\linewidth}		
\begin{lstlisting}
SELECT DISTINCT R.A
FROM R
WHERE not exists
 (SELECT *
 FROM S
 WHERE not exists
  (SELECT *
  FROM R AS R2
  WHERE (R2.B = S.B
  OR R2.C = S.C)
  AND R2.A = R.A))
\end{lstlisting}
\vspace{3.5mm}
    \caption{}
    \label{fig:disjunctionSQL}
\end{subfigure}
\end{minipage}
\hspace{4mm}
\begin{minipage}{0.14\linewidth}
\begin{subfigure}[b]{\linewidth}		
\begin{lstlisting}
SELECT DISTINCT R.A
FROM R
WHERE not exists
 (SELECT *
 FROM S
 WHERE not exists
  (SELECT *
  FROM R AS R2
  WHERE R2.B = S.B)
  AND R2.A = R.A)
 AND not exists
  (SELECT *
  FROM R AS R2
  WHERE R2.C = S.C
  AND R2.A = R.A))
\end{lstlisting}
\vspace{-6mm}
    \caption{}
    \label{fig:disjunctionSQL2}
\end{subfigure}	
\end{minipage}
\hspace{2mm}	
\begin{minipage}{0.25\linewidth}
\begin{subfigure}[b]{\linewidth}
	\vspace{2mm}
    \includegraphics[scale=0.42]{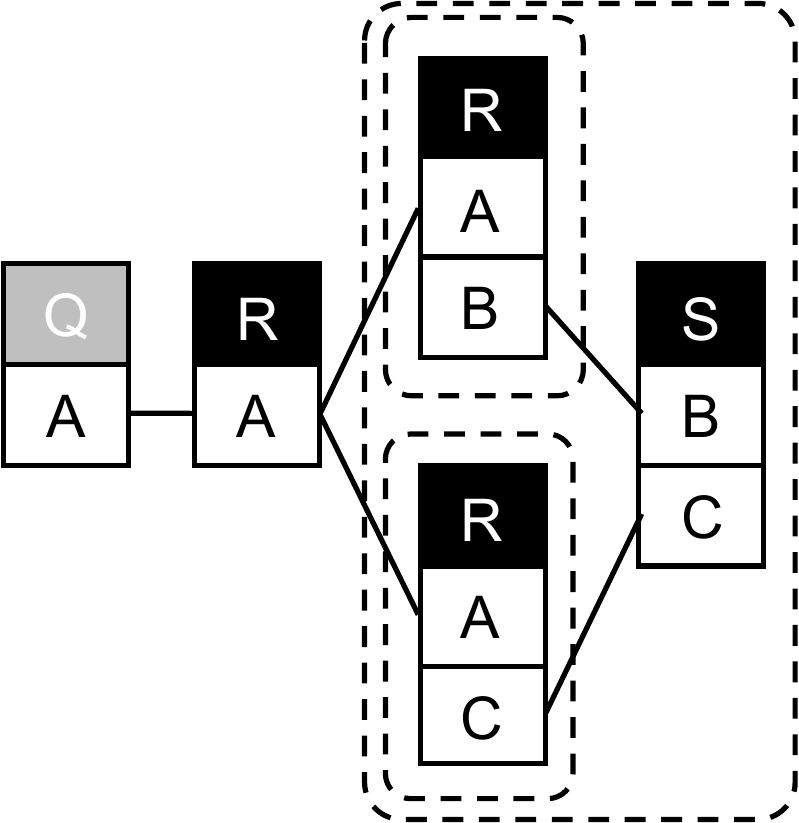}
	\vspace{2mm}
    \caption{}
    \label{Fig_disjunction_QV}
\end{subfigure}	
\end{minipage}
\hspace{3mm}	
\begin{minipage}{0.145\linewidth}
\begin{subfigure}[b]{\linewidth}		
\begin{lstlisting}
SELECT DISTINCT R.A
FROM R, S, T
WHERE R.A=S.A
OR R.A=T.A
\end{lstlisting}
\vspace{-7mm}
    \caption{}
    \label{fig:disjunctionSQLor}
\end{subfigure}
\begin{subfigure}[t]{\linewidth}		
\vspace{-1mm}
\begin{lstlisting}
(SELECT DISTINCT R.A
FROM R, S, T
WHERE R.A=S.A)
UNION
(SELECT DISTINCT R.A
FROM R, S, T
WHERE R.A=T.A)
\end{lstlisting}
\vspace{-6mm}
    \caption{}
    \label{fig:disjunctionSQLor2}
\end{subfigure}	
\end{minipage}
\hspace{2mm}
\begin{minipage}{0.18\linewidth}
\begin{subfigure}[b]{\linewidth}
	\vspace{8mm}
    \includegraphics[scale=0.39]{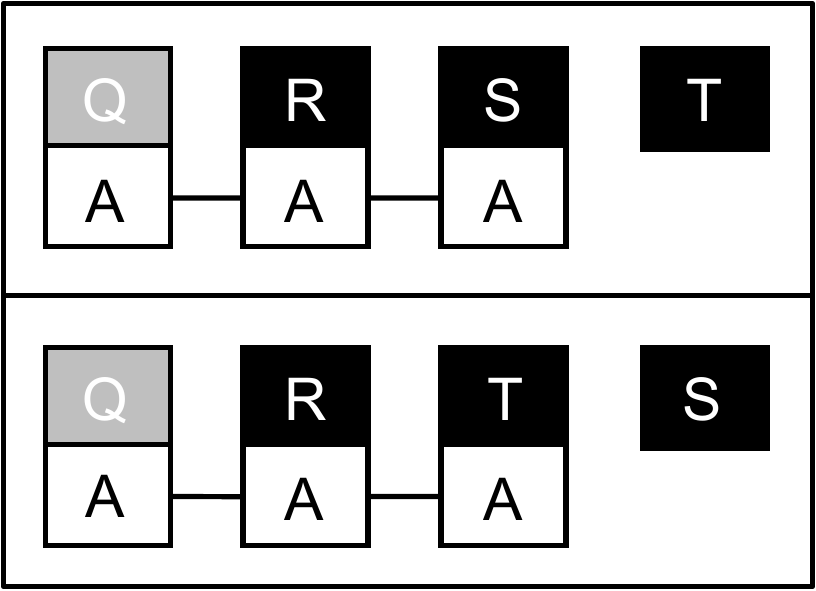}
	\vspace{3mm}
    \caption{}
    \label{Fig_disjunction_or_QV}
\end{subfigure}	
\end{minipage}
\caption{Illustrations for \Cref{ex:disjunction1} on replacing disjunctions:
(a) $\SQL$ with disjunctions,
(b) logically-equivalent (yet not representation-equivalent) $\NDSQL$ statement, and
(c) \diagrams.
Illustrations for \Cref{ex:disjunction2} on creating the union of queries:
(d) $\SQL$ with disjunctions,
(e) logically-equivalent union of $\NDSQL$ statement, and
(f) \diagrams with union cells.}
\label{fig:disjunction}
\end{figure*}

\begin{theorem}[Unambiguous \diagrams]
\label{unambiguous_RDs}
Every valid \diagram has an unambiguous interpretation in $\NDTRC$.
\end{theorem}

The constructive translation from \cref{sec:fromRDtoTRC} forms the proof.

\begin{theorem}[Representation-equivalence]
\label{representation-equivalence}
$\NDTRC$ and 
\diagrams
are re\-pre\-sentation-equivalent. 
\end{theorem}

Hence both languages can represent the same set of query patterns.
The 
translations from
\cref{sec:fromTRCtoRD,sec:fromRDtoTRC} 
keep a 1-to-1 correspondence between extensional tables
and thus form the proof.

\section{Relational completeness}
\label{sec:completeness}
\label{SEC:COMPLETENESS}

Recall from our discussion in \cref{sec:nondisjunctivefragment}
that any single situation can display only conjunctive information.
Shin goes further and claims that ``Any diagrammatic system that seeks to represent disjunctive information needs to bring in an artificial syntactic device with its own convention.'' \cite{shin_1995}.
The syntactic device we use is 
inspired by the representation of disjunction in $\Datalog$:
we allow placing several \diagrams on the same canvas, each in a separate \emph{union cell}.
Each cell of the canvas then contains one \diagram displaying only conjunctive information, 
yet the relation among the different cells is disjunctive.
We also allow logical transformations that are not pattern-preserving 
(thus we focus on \emph{logical equivalence} only, 
and not our more strict criterion of \emph{representation equivalence}).
This is arguably a loss of \emph{representation power}, but in line with the current understanding 
of the limits of diagrams to express disjunctive information~\cite{shin_1995,Shin:2002}.

These two \textit{extensions} together make \diagrams
\emph{relationally complete}: 
every query expressible in full $\RA$, safe $\TRC$, $\DatalogN$, or 
our prior $\NDSQL$ fragment extended by 
disjunctions of predicates\footnote{Extend the grammar from
\cref{table:supported_grammar} 
with one additional rule: 
\textsf{P::= `('P OR P`)'}.
}
can then be represented 
as \diagram.

We illustrate with two examples. The first shows how to avoid disjunctions if they are not at the root level.
The second shows how to replace disjunctions in the root by unions of queries.

\begin{example}[Replacing disjunctions]\label{ex:disjunction1}
Consider the $\SQL$ query from \cref{fig:disjunctionSQL}
which contains a disjunction and is not in $\NDSQL$. 
Using De Morgan's Law $\neg(A \vee B) = \neg A \wedge \neg B$, we can first reformulate the conditions including disjunction as DNF, 
and then distribute the quantifier over the conjuncts. 
This leads to a query without disjunctions, yet comes at the cost of having to repeat relation $R$:
\begin{align*}
\phantom{=} &\{ q(A) \mid 
	\exists r \in R [q.A \equal r.A \wedge \neg (\exists s \in S\\
	&\hspace{10.5mm}[\h{\neg (\exists r_2 \in R [ }(r_2.B \equal s.B \;\h{\vee}\; r_2.C \equal s.C) \wedge r_2.A \equal r.A \h{] )} ])] \}
\\
= &\{ q(A) \mid 
  \exists r \in R [q.A \equal r.A \wedge \neg (\exists s \in S\\
	&\hspace{10.5mm}[\h{\neg (\exists r_2 \in R [}r_2.B \equal s.B \wedge r_2.A \equal r.A \h{])} \;\h{\wedge}\;	\\
	&\hspace{11.7mm}\h{\neg (\exists r_3 \in R [}r_3.C \equal s.C \wedge r_3.A \equal r.A \h{])} ]] \}	
\end{align*}
\Cref{fig:disjunctionSQL2} shows this query as representation-equivalent $\NDSQL$ query,
and \cref{Fig_disjunction_QV} as 
\diagram.
\end{example}

\begin{example}[Union of queries]
	\label{ex:disjunction2}
Consider three unary tables $R(A)$, $S(A)$, and $T(A)$ and the query from \cref{fig:disjunctionSQLor}
that we had already used in \cref{sec:nondisjunctivefragment}.
We can replace disjunction by pulling it to the root and replacing the query with a union of disjunction-free $\NDSQL$ queries:
\begin{align*}
\phantom{=} &\h{\{} q(A) \mid
	\exists r \in R, \exists s \in S, \exists t \in T 
	[q.A \equal r.A \wedge (r.A \equal s.A \;\h{\vee}\; 
	r.A \equal t.A)] \h{\}}	\\
= &\h{\{} q(A) \mid 
	\exists r \in R, \exists s \in S, \exists t \in T 
	[q.A \equal r.A \wedge r.A \equal s.A] \h{\}}  \;\h{\cup}\; \\
  &\h{\{} q(A) \mid 
	\exists r \in R, \exists s \in S, \exists t \in T 
	[q.A \equal r.A \wedge r.A \equal t.A] \h{\}}		
\end{align*}
\Cref{fig:disjunctionSQLor2} shows the representation-equivalent $\NDSQL$ query.
\Cref{Fig_disjunction_or_QV} shows this union of two queries as two separate queries,
each in a separate cell, and each with the same attribute signature in the output table.
Notice that we cannot leave away the non-connected tables in the individual queries; if any of the tables are empty, 
then the query returns an empty result.
\end{example}

The additional validity criterion for multiple canvas cells follows the conditions of union or disjunction in the \emph{named perspective}~\cite{DBLP:books/aw/AbiteboulHV95}
of query languages:
for disjunction in $\TRC$, 
each operand needs to have the same arity, and the mapping between them is achieved by reusing the same variables.

\begin{definition}[Validity (\cref{def:validRelationalDiagram} continued)]
(6) The output tables in multiple cells for the same query need to have the same name and same set of attributes.
\end{definition}

\begin{theorem}[Completeness]
	\label{th:completeness}
\diagrams extended with union cells are relationally complete.
\end{theorem}

\section{From Queries to Sentences}
\label{sec:sentences}
\label{SEC:SENTENCES}

This section discusses a simple generalization
of \diagrams
from \emph{relational queries} to \emph{relational sentences} (or Boolean queries).
This generalization allows us to express constraints (sentences that need to be true),
an extension that is not available in relational algebra.
It also allows us to compare our formalism against a long history of formalisms for logical statements.

The idea is simple: we leave away the output table.
An additional freedom with sentences is that the safety conditions of relational calculus fall away. 
Thus, we can express statements that do not have any existentially-quantified relations in the main canvas.

For the Boolean fragment of $\NDTRC$, we thus allow the first negation before the first existentially-quantified table.
In the translation back and forth from \diagrams (\cref{sec:fromTRCtoRD,sec:fromRDtoTRC}), we leave away point 5.
For $\NDSQL$, we need to make sure that it always returns either true or false (and not only true or the empty set). 
This can be achieved by adding to the rule for $Q$ in \cref{table:supported_grammar}
two alternative derivations
\textsf{Q::= SELECT NOT (P)}
and
\textsf{Q::= SELECT [NOT] EXISTS (S)}.
We illustrate with an intuitive example here.
More illustrative examples are included in our
online appendix \cite{relationaldiagrams-TR}.

\begin{figure}[t]
\centering	
\begin{subfigure}[b]{.38\linewidth}		
\begin{lstlisting}
SELECT not exists
 (SELECT *
 FROM Sailor s
 WHERE not exists
  (SELECT b.bid
  FROM Boat b, Reserves r
  WHERE b.color = 'red' 
  AND r.bid = b.bid
  AND r.sid = s.sid))
\end{lstlisting}
\vspace{-6mm}
\caption{}
\label{sql:Sailor_sentences_universal}
\end{subfigure}	
\hspace{3mm}
\begin{subfigure}[b]{.45\linewidth}
    \includegraphics[scale=0.4]{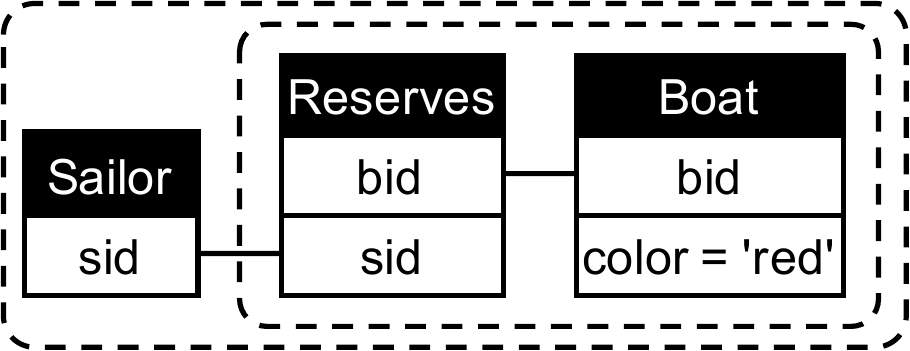}
\vspace{4mm}
\caption{}
\label{Fig_Sailor_all_Sailors_some_red_boat}
\end{subfigure}	
\caption{\Cref{ex:allsailors}: 
All sailors reserve some red boat.}
\label{fig:Sailor_sentences_universal}
\end{figure}

\begin{example}
Consider 
the statement: \emph{``All sailors reserve a red boat.''}\label{ex:allsailors}
\begin{align}
\begin{aligned}
\neg(\exists s \in \mathit{Sailor} 
	&[\neg (\exists b \in \mathit{Boat}, r \in \mathit{Reserves} [b.\mathit{color} = \mathit{'red'} \wedge 	\\
	&r.\mathit{bid} = b.\mathit{bid} \wedge r.\mathit{sid} = s.\mathit{sid}])])
\end{aligned}
\label{trc:allsailoredboats}
\end{align}

The first 4 steps of the translation in \cref{sec:fromTRCtoRD} still work:
the root canvas $q_0$
does not contain any relation
(\cref{Fig_Sailor_all_Sailors_some_red_boat}).
Similarly, the equivalent canonical $\NDSQL$ statement contains no \sql{FROM} clause before the first \sql{NOT}.
Notice that 
\cref{def:isomorphism} of query pattern isomorphism still works as
it is defined based the relational tables.
\end{example}

\section{Related Work}
\label{sec:relatedWork}
\label{SEC:RELATEDWORK}

\subsection{Peirce's beta existential graphs}
\label{sec:Peirce}

\diagrams represent nested quantifiers
in a similar way as the influential and widely-studied \emph{Existential Graphs} 
by Charles Sanders Peirce~\cite{peirce:1933,Roberts:1992,Shin:2002} for expressing logical statements
(or, equivalently, Boolean queries).
Peirce's graphs come in two variants called alpha and beta.
Alpha graphs correspond to propositional logic, whereas beta graphs correspond to first-order logic (FOL). 
Both variants use so-called \emph{cuts} to express negation (similar to our nesting boxes),
and beta graphs use a syntactical element called the \emph{Line of Identity} (LI) to denote both the existence of objects and the identity between objects.

\begin{example}[Nested negation] 
\Cref{fig:EG_comparison}
\label{ex:similarities}%
shows 2 beta graphs: 
\begin{align*}
&\textrm{\cref{Fig_Sailor_some_Sailors_some_red_boat_EG}: 	\emph{There exists a sailor who reserved a red boat.}}			\\
&\textrm{\cref{Fig_all_red_boats_some_Sailors_EG}: 			\emph{All red boats were reserved by some sailor.}}
\end{align*}
Beta graphs cannot represent constants and thus need to replace a selection of boats that are red with a dedicated new predicate
``is a red boat.''
Their respective translations into $\DRC$ are:
\begin{align*}
	\textrm{\cref{Fig_Sailor_some_Sailors_some_red_boat_EG}: }
	&\exists x, y[\sql{Sailor}(x) \wedge \sql{RedBoat}(y) \wedge \sql{Reserves}(x,y)]	\\
	\textrm{\cref{Fig_all_red_boats_some_Sailors_EG}: }
	&\neg (\exists y[\sql{RedBoat}(y) \wedge \neg(\exists x[\sql{Sailor}(x) \wedge \sql{Reserves}(x,y)])])
\end{align*}
Contrast the beta graphs with their respective \diagrams
and $\TRC$:
\begin{align}
&\begin{aligned}
\textrm{\cref{Fig_Sailor_some_Sailors_some_red_boat}: }
\exists s \in \sql{Sailor},
	&b \in \sql{Boat}, r \in \sql{Reserves} [r.\sql{bid} = b.\sql{bid} \wedge \\
	&r.\sql{sid} = s.\sql{sid} \wedge b.\sql{color} = \sql{`red'}]
\end{aligned}
\\
&
\textrm{\cref{Fig_all_red_boats_some_Sailors}: }
\neg(\exists b \in \sql{Boat}
	[B.\sql{color}=\sql{`red'} \wedge \neg (\exists s \in \sql{Sailor}, r \in \sql{Reserves}	\notag\\
	&\hspace{24mm}[r.\sql{bid} = b.\sql{bid} \wedge r.\sql{sid} = s.\sql{sid}])])
\end{align}
\end{example}

\begin{figure}[t]
\centering	
\begin{subfigure}[b]{.45\linewidth}
    \includegraphics[scale=0.4]{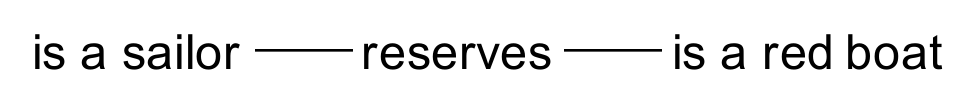}
	\vspace{-2.3mm}
    \caption{}
    \label{Fig_Sailor_some_Sailors_some_red_boat_EG}
\end{subfigure}	
\hspace{3mm}
\begin{subfigure}[b]{.45\linewidth}
    \includegraphics[scale=0.4]{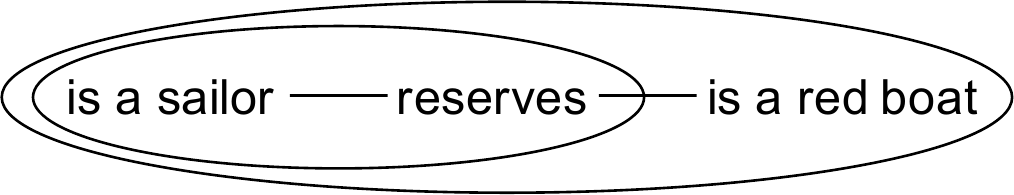}
	\vspace{-4mm}
    \caption{}
    \label{Fig_all_red_boats_some_Sailors_EG}
\end{subfigure}	
\hspace{5mm}
\begin{subfigure}[b]{.45\linewidth}
    \includegraphics[scale=0.4]{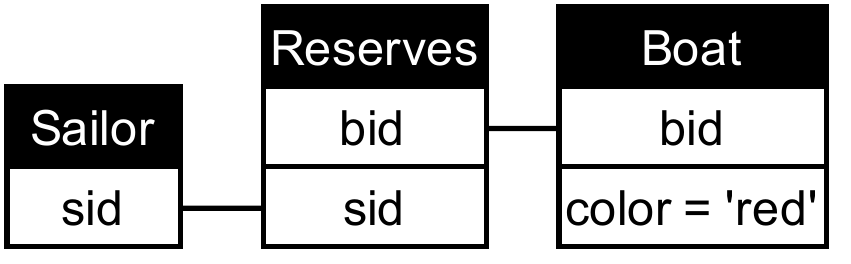}
\vspace{-1mm}
\caption{}
\label{Fig_Sailor_some_Sailors_some_red_boat}
\end{subfigure}	
\hspace{5mm}
\begin{subfigure}[b]{.45\linewidth}
    \includegraphics[scale=0.4]{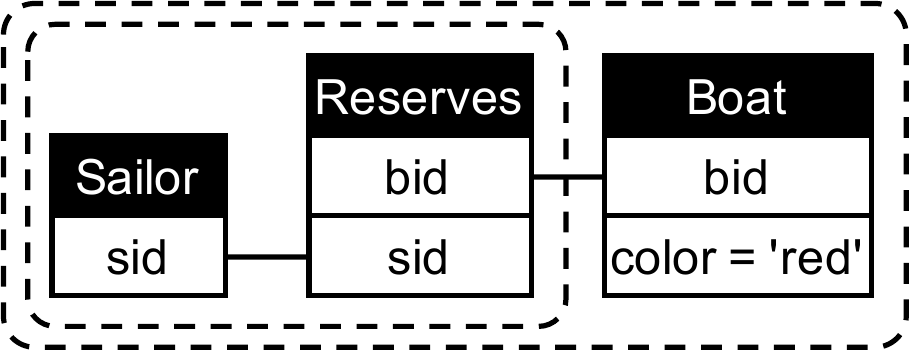}
\vspace{-2mm}
\caption{}
\label{Fig_all_red_boats_some_Sailors}
\end{subfigure}	
\caption{Diagrams for \Cref{ex:similarities} comparing the representations of negation in beta graphs (top) and \diagrams (bottom). 
}
\label{fig:EG_comparison}
\end{figure}

\introparagraph{Differences}
The 4 key differences of beta graphs 
vs.\ \diagrams
are:
(1) beta graphs can only represent sentences and not queries;
(2) beta graphs cannot represent constants, so selections cannot be modeled and instead require dedicated predicates;
(3) beta graphs can only represent identity predicates (and no comparisons); and
(4) Lines of Identity (LIs) in beta graphs have multiple meanings (existential quantification and identity)
and are a primary symbol.\footnote{Every beta graph has lines and graphs with lines but no predicates have meanings.
See, e.g., the recursive definition in \cite[p.\ 41]{Shin:2002}.}
This can make reading the graphs ambiguous.
We, in contrast, have predicates inspired from $\TRC$. Lines only connect two attributes and have no loose ends. 
Interpreting a graph as a $\TRC$ formula is straightforward and can be summarized in a simple set of rules (recall \cref{sec:QV}).
We now discuss this last point in more detail.

\introparagraph{Problems from abusing lines in beta graphs}
While over 100 years old, 
Peirce's beta system has led 
to multiple misinterpretations and 
ongoing discussions about how to interpret a valid beta graph correctly.
The literature contains many attempts to provide formal ``interpretations'' and provide consistent readings of these graphs.
How can it be that something that should be unambiguous still gives so much margin of error?
In our opinion, beta graphs have one important design problem leading to those misunderstandings:
it is the \emph{overloading of the meaning of the Lines of Identity (LI)},
and thus the abuse of lines as symbol.
As mentioned before, LI's are used to denote two different concepts:
(1) the \emph{existence} of objects (intuitively an existential quantification of a variable in $\DRC$ such as $\exists x$), and
(2) the \emph{identity} between objects (intuitively, R.A=S.A in $\TRC$).
This non-separation of concerns
leads to unfixable ambiguities.

\begin{example}[A red boat]
\label{ex:aredboat}%
Consider
the sentence ``There is a red boat''
shown in
\cref{Fig_redboat_EG} as a beta graph.
As beta graphs cannot represent constants, the graph requires a special unary predicate ``red boat.''
The LI represents both ``there exists something'' and ``that something is equal to a red boat.''
Thus a line (which arguably suggests two items being connected or joined) is meant as a quantified variable,
and the beta graph can be interpreted in $\DRC$ as:
\begin{align*}
	\exists x[\sql{RedBoat}(x)]
\end{align*}

\Cref{Fig_redboat} shows the same sentence as a \diagram.
Notice that ``there exists something'' is represented by just placing this something (a predicate) on the canvas.
There is no need for an existential line.
Also notice how the modern UML diagram allows predicates (relational atoms) with several attributes, and one of those attributes 
can be set equal to a constant (here \selPredicateBox{\textup{color = `red'}}). 
It can be read rather naturally like a $\TRC$ statement:
\begin{align*}
	\exists b \in \sql{Boat}[\sql{B.color=`red'}]
\end{align*}
\end{example}

\begin{figure}[t]
\centering	
\hspace{10mm}
\begin{subfigure}[b]{.2\linewidth}
    \includegraphics[scale=0.4]{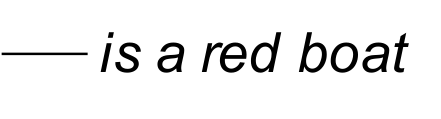}
\vspace{-4mm}
\caption{}
\label{Fig_redboat_EG}
\end{subfigure}	
\hspace{20mm}
\begin{subfigure}[b]{.15\linewidth}
    \includegraphics[scale=0.4]{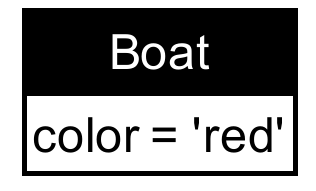}
\vspace{-5mm}
\caption{}
\label{Fig_redboat}
\end{subfigure}	
\hspace{10mm}
\caption{\Cref{ex:aredboat} illustrating that in LI's in beta graphs are needed as a symbol for quantification, not just equivalence.
}
\vspace{-2mm}
\end{figure}

The interpretation of beta graphs where one LI represents one existentially-quantified variable
can at times be intuitive and simply correspond to a modern $\DRC$ interpretation (recall \cref{ex:similarities}).
However, such a  simple interpretation is not always possible.

\begin{example}[Exactly one red boat]
	\label{ex:exactlyoneredboat}
Consider the sentence ``There exists exactly one red boat.''
\Cref{Fig_exactly_one_red_boat_EGa,Fig_exactly_one_red_boat_EGb}
show two beta graphs with \emph{different cut nestings} that can both be read as
\begin{align*}
	\exists x[\sql{RedBoat}(x) \wedge \neg(\exists y[\sql{RedBoat}(y) \wedge x \neq y]
\end{align*}
Now a single LI needs to represent \emph{two} existentially-quantified variables,
and two \emph{different nestings} of the cuts can represent the same statement.
Contrast this with \cref{Fig_exactly_one_red_boat}, read in $\TRC$ as:
\begin{align*}
\exists b \in \sql{Boat}
	&[\sql{B.color=`red'} \wedge \neg(\exists b_2 \in \sql{Boat} \\
	&[b_2.\sql{color}=\sql{`red'} \wedge b.\sql{bid} \neq b_2.\sql{bid}])]
\end{align*}
Notice here that the inequality is simply represented by a \emph{label} of a join between two predicates. 
Two tuple variables are represented by two different atoms and the interpretation is unambiguous:
\emph{There exists a boat whose color is red, and there does not exist another boat whose color is red and whose $\sql{bid}$ is different.}
\end{example}

The fact that one LI can branch into multiple endings (also called \emph{ligatures}), and may have \emph{loose endings}, and may represent multiple existentially-quantified variables, together with cuts being applied to such LI's can quickly lead to hard-to-interpret diagrams (see e.g., the increasingly-unreadable figures in \cite[pp. 42-49]{Shin:2002}).
This led to several attempts in the literature to provide ``reading algorithms'' of those graphs (e.g., \cite{Zeman:1964,Roberts:1973,Shin:2002}) and rather complicated proofs of the expressiveness of beta graphs~\cite{Zeman:1964}, assuming a correct reading.
As example, the paper by Dau \cite{Dau:2006} points out an error in Shin's reading algorithm \cite{Shin:2002}.
However, Dau's correction to Shin~\cite{Dau:2006} itself also has errors
(e.g., the interpretation of the right-most diagram in \cite[Fig 2]{Dau:2006} misses one equality, 
see \cite{relationaldiagrams-TR} for details).
\introparagraph{Why \diagrams avoid the problem}
\diagrams use the line only for connecting two attributes. 
The type of connection is unambiguously represented by a label.
Quantification is represented by predicates themselves.
Thus, on a more philosophical level, 
we think that our visual formalism solves those problems based on a more modern interpretation of first order logic:
$\TRC$ was created by Edgar Codd in the 1960s and 1970s
in order to provide a declarative database-query language for data manipulation in the relational data model~\cite{Codd:1970}.
In contrast, beta graphs were proposed even before first-order logic (FOL), which was only clearly articulated some years after Peirce's death in the 1928 first edition of David Hilbert and Wilhelm Ackermann's ``Grundzüge der theoretischen Logik''~\cite{HilbertAckerman:1928}.
Zeman, in his 1964 PhD thesis~\cite{Zeman:1964}, was the first to note that beta graphs are isomorphic to first-order logic with equality. However, the secondary literature, especially Roberts~\cite{Roberts:1973} and Shin~\cite{Shin:2002}, does not agree on just how this is so~\cite{wiki:existentialgraph}. 
We did not start from Peirce's beta graphs and attempted
to fix the issues that have been occupying a whole community for years.
Rather, we started from 
the modern UML reading of relational schemas and an understanding of $\TRC$, 
and tried to achieve a minimal visual extension to provide relational completeness and pattern-isomorphism to $\TRC$,
which happens to provide a natural solution of interpretation problems of beta graphs.
We believe that \diagrams provide a clean, unambiguous, and, in hindsight, simple abstraction of query patterns.

\begin{figure}[t]
\centering
\begin{minipage}{0.45\linewidth}
\begin{subfigure}[b]{\linewidth}
    \includegraphics[scale=0.4]{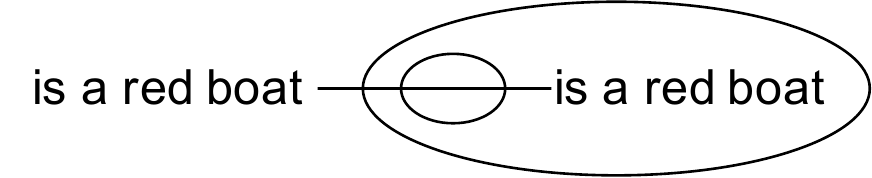}
\vspace{-1mm}
\caption{}
\label{Fig_exactly_one_red_boat_EGa}
\end{subfigure}
\begin{subfigure}[b]{\linewidth}
    \includegraphics[scale=0.4]{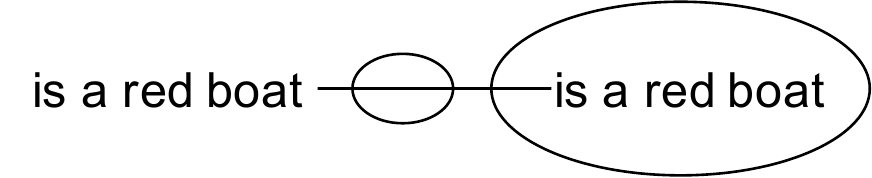}
\vspace{-1mm}
\caption{}
\label{Fig_exactly_one_red_boat_EGb}
\end{subfigure}
\end{minipage}
\hspace{6mm}
\begin{minipage}{0.38\linewidth}
\vspace{4.5mm}
\begin{subfigure}[b]{\linewidth}
    \includegraphics[scale=0.4]{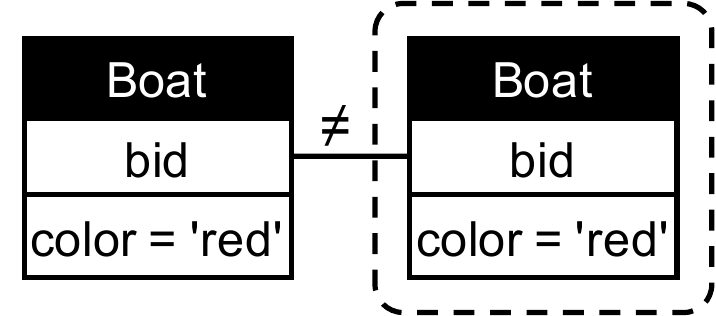}
\vspace{0mm}
\caption{}
\label{Fig_exactly_one_red_boat}
\end{subfigure}
\end{minipage}
\caption{\Cref{ex:exactlyoneredboat}: The combination of LI's and nesting symbols (called ``cuts'' in beta graphs) provide ambiguous ways to nest cuts.}
\vspace{-2mm}
\end{figure}

\subsection{$\queryviz$}
\label{sec:queryvis1_vs_2}

Some of our design decisions are similar
to our earlier query representation called 
$\queryviz$ \cite{DanaparamitaG2011:QueryViz,DBLP:conf/sigmod/LeventidisZDGJR20,gatterbauer2011databases}.
In $\queryviz$ diagrams, grouping boxes are used to group all tables within a local scope, 
i.e., \emph{for each individual query block}. 
Those boxes thus cannot show their respective nesting, and an additional symbol of directed arrows is needed to ``encode'' the nesting.
The high-level consequence of those design decisions is that 
(1) $\queryviz$ does not guarantee to unambiguously visualize nested queries with nesting depth $\geq 4$ 
(please see our online appendix \cite{relationaldiagrams-TR} for a minimum example),
(2) each grouping box needs to contain at least one relation 
(thus $\queryviz$ cannot represent the query in \cref{Fig_TRC_vs_RD}),
and (3) $\queryviz$ cannot represent general Boolean sentences
(e.g., the sentence ``All sailors have reserved some red boat'').
Thus $\queryviz$ is not sound and not relationally complete, 
even for the disjunctive fragment.

\introparagraph{Problems from abusing lines in $\queryviz$}
Similar to beta graphs, we think, in hindsight, that the design of $\queryviz$
abuses the line symbol by using it for two purposes: 
for ($i$) joining atoms and 
($ii$) for \emph{representing the negation hierarchy}.
In contrast, \diagrams use the line only for connecting two attributes
and represent the negation hierarchy explicitly by nesting negation boxes.
\diagrams fix those the completeness and soundness issues, 
and, in addition, can show logical sentences and queries or sentences lacking tables in one or more of the negation scopes of nested queries.

\subsection{Query-By-Example (QBE)}

The development of QBE~\cite{DBLP:journals/ibmsj/Zloof77} was strongly influenced by $\DRC$.
However, QBE can express relational division only by using COUNT
or by 
breaking the query into two logical steps and using a temporary relation~\cite[Ch. 6.9]{RamakrishnanGehrke:DBMS2000}.
But in doing so, QBE
uses the query pattern from $\RA$ and $\DatalogN$ of implementing relational division (or universal quantification)
in a dataflow-type, sequential manner, 
requiring two occurrences of the Sailor table.

\begin{figure}[t]
\centering
    \begin{subfigure}[b]{1\linewidth}
		\hspace{5mm}
        \includegraphics[scale=0.35]{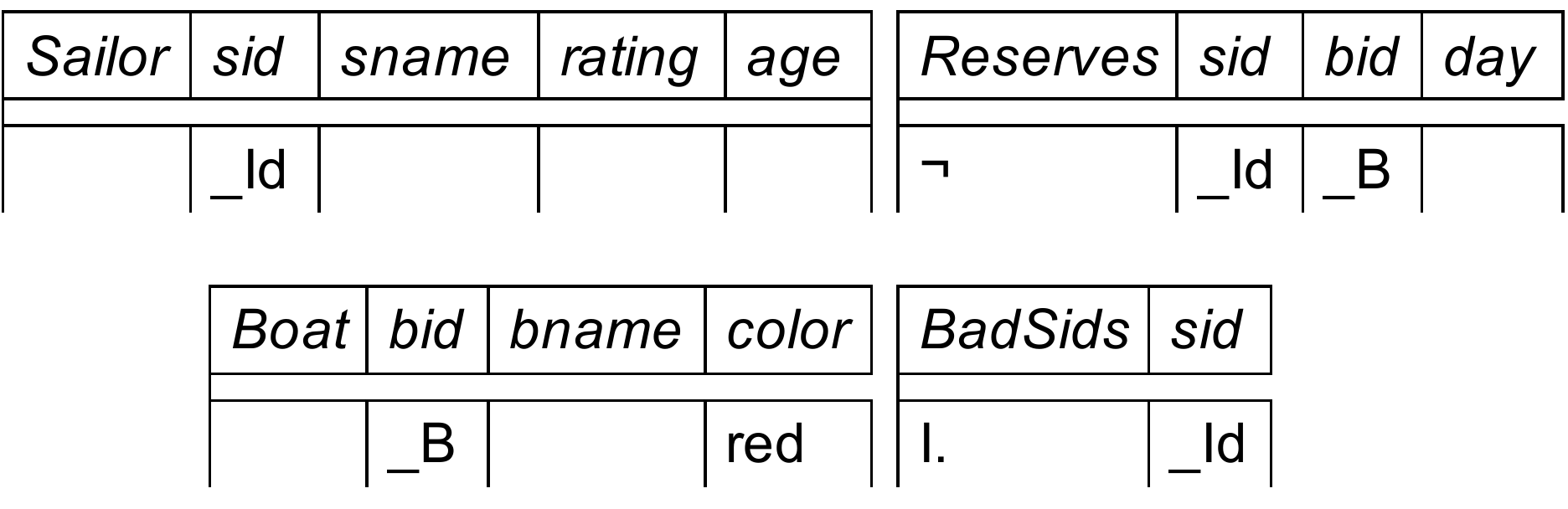}
        \caption{Temporary relation \sql{BadSids(sid)} (``I.'' stands for insert)}
        \label{Fig_QBE_red_boats_a}
    \end{subfigure}	
    \begin{subfigure}[b]{1\linewidth}
		\hspace{5mm}		
        \includegraphics[scale=0.35]{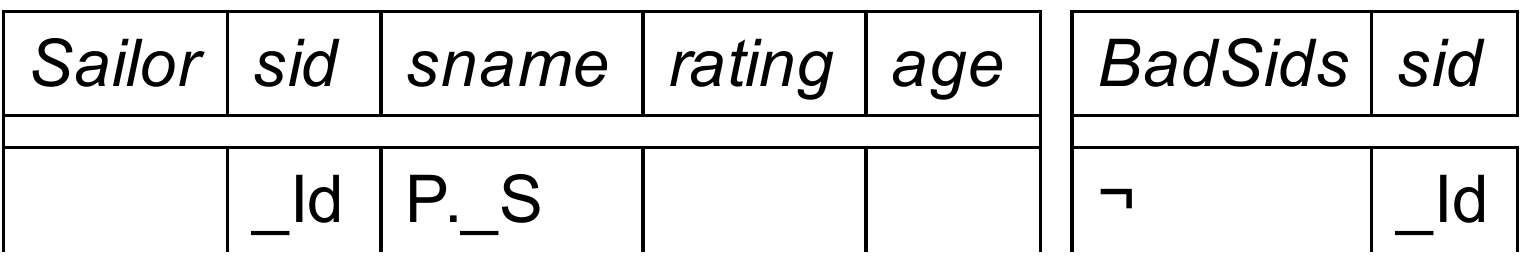}
        \caption{Actual answer \sql{Q(sid)} (``P.'' stands for print)}
        \label{Fig_QBE_red_boats_b}
    \end{subfigure}	
\caption{\Cref{QBE_redboats}: QBE needs to create a temporary relation \sql{BadSids} in order to express relational division.
It does follow the query pattern of relational algebra and not relational calculus.}
\label{Fig_QBE_red_boats}
\vspace{-2mm}
\end{figure}

\begin{example}[Sailors reserving all red boats in QBE]
\label{QBE_redboats}
	Consider the query 
	``\emph{Find sailors who have reserved all red boats}'' 
	QBE needs to first create an intermediate relation 
	\sql{I}
	that stores all Sailors for whom there is a red boat that is not reserved by the sailor
	(\cref{Fig_QBE_red_boats_a}),
	and then finds all the other Sailors
	(\cref{Fig_QBE_red_boats_b}).
	The pattern of this query in QBE thus matches exactly the one of $\DatalogN$ 
	(it requires two occurrences of the relational \sql{Sailor} instead of one as in calculus),
	which is arguably a more dataflow (one relation accessed after the other) than logical query language pattern:
	\begin{align}
	\begin{aligned}
		I_1(x,y)	& \datarule \sql{Reserves}(x,y,\_). 		 \\
		I_2(x)	& \datarule \sql{Sailor}(x,\_,\_,\_), \sql{Boat}(y,\_,\sql{'red'}), \neg I_1(x,y). 		 \\
		Q(y)	& \datarule \sql{Sailor}(x,y,\_,\_), \neg I_2(x). 
	\end{aligned}		
	\label{eq:QBE_equivalent_Datalog}
	\end{align}	
	\end{example}	

More formally, the QBE query from \cref{Fig_QBE_red_boats} is pattern-isomorph to the $\DatalogN$ query in
\cref{eq:QBE_equivalent_Datalog}.
Furthermore, the the following logically-equivalent $\TRC$ query
has no pattern-isomorph representation in QBE:
(i.e.\ with one single occurence of the \sql{Sailor} relation).
\begin{align}
\begin{aligned}
\!\!\!\!\!\!\{q(\sql{sname}) \mid 
	& \exists s \in \sql{Sailor} [q.\sql{sname}=s.\sql{sname} \ \wedge \\
	&\neg (\exists b \in \sql{Boat} [b.\sql{color} = \sql{`red'} \wedge \\
	&\neg (\exists r \in \sql{Reserves} [r.\sql{bid} = b.\sql{bid} \wedge r.\sql{sid} = s.\sql{sid}])]) \}
\end{aligned}
\label{q:trc:sailorsallredboats}
\end{align}

\subsection{Other relationally-complete formalisms}

Our online appendix \cite{relationaldiagrams-TR} 
compares \diagrams to other related visualizations like
DFQL (Dataflow Query Language) \cite{DBLP:journals/iam/ClarkW94,DBLP:journals/vlc/CatarciCLB97}.
On a high-level, all visual formalisms
that we are aware of and that are proven to be relationally complete
(including those listed in
\cite{DBLP:journals/vlc/CatarciCLB97})
are at their core visualizations of relational algebra operators.
This applies even to the more abstract \emph{graph data structures (GDS)} from \cite{DBLP:conf/vdb/Catarci91} and the later \emph{graph model (GM)} from \cite{DBLP:journals/is/CatarciSA93}, which are related to our concept of \emph{query representation}.
The key difference is that GDS and GM are formulated inductively based on mappings onto operators of relational algebra. 
They thus mirror dataflow-type languages where visual symbols (directed hyperedges) represent operators like \emph{set difference} connecting two relational symbols, leading to a new third \add{symbol} as output.
We have proved that there are simple queries in relational calculus (recall \cref{ex:intro}) that cannot be represented in relational algebra with the same number of relational symbols. 
Thus any visual formalism based on relational algebra 
cannot represent the full range of relational query patterns.

\section{Conclusions and Future Work}

We motivated a criterion called \emph{pattern-preservation} that preserves logical query patterns across languages 
and gave evidence for its importance in designing diagrammatic representations.
To the best of our knowledge, our work is the the first to discuss and formalize the 
concepts of \emph{relational query patterns}
with a semantic definition that is applicable to any relational query language.
We also propose the ability to represent query patterns already used in existing query languages
as criteria for comparing and evaluating visual query representations.
As example, Jarke and Vassiliou's survey \cite{DBLP:journals/csur/JarkeV85} established various criteria for query language selection.
Those include usability and functional criteria.
No prior work we are aware of discusses diagrams 
under the view point of \emph{truthfully representing query patterns} (i.e.\ at least with the same number of extensional tables) 
used in relational query languages.

We formulated the non-disjunctive fragments of $\DatalogN$, $\RA$, safe $\TRC$, and corresponding $\SQL$ (interpreted under set semantics)
that naturally generalize conjunctive queries to nested queries with negation.
We prove that this important fragment 
allows a rather intuitive and, in hindsight, natural diagrammatic representation that can preserve the query pattern used across all 4 languages.
We call this representation \diagrams
and further prove that 
this formalism, extended with a representation of union, is complete for full safe relational calculus (though not pattern-preserving).
No prior diagrammatic representation we know is both relationally complete and representation-equivalent to relational calculus, at least for the non-disjunctive fragment.

Finding a \emph{pattern-preserving} diagrammatic representation for disjunction and even more general features of SQL 
(such as grouping and aggregates) is an open problem. 
For example, it is not clear how to achieve an intuitive representation for arbitrary nestings of disjunctions, such as
``$R.A < S.E \wedge (R.B < S.F \vee R.C < S.G)$''
or 
``$(R.A > 0 \wedge R.A < 10) \vee (R.A > 20 \wedge R.A < 30)$'' with \emph{minimal} additional notations.
Grounded in a long history of diagrammatic representations of logic,
we gave intuitive arguments for why visualizing disjunctions is inherently more difficult than conjunctions.
However, this is no proof it cannot be done with a novel approach.

\clearpage

\bibliographystyle{ACM-Reference-Format}

\bibliography{queryvis.bib}

\newpage
\appendix
\section{Nomenclature}
\label{sec:sec:appendix:nomenclature}

\begin{table}[h]
\centering
\small
\begin{tabularx}{\linewidth}{@{\hspace{0pt}} >{$}l<{$} @{\hspace{2mm}}X@{}} %
\hline
\textrm{Symbol}		& Definition 	\\
\hline
    \hline
	q(\mathcal{T})	& Query mapping a \emph{signature} $\mathcal{T}$ of tables to the query output.	\\
					& Example $q(R)$: $Q(x) \datarule R(x,y),R(\_,y).$ \\
	\overline q(\overline{\mathcal{T}})	
					& \emph{Query representation} with signature $\mathcal{T}$	of \emph{extensional tables}.\\
					& Example $\overline{q}(R, R)$: $Q(x) \datarule R(x,y),R(\_,y).$ \\
	q'(\overline{\mathcal{T}}')	
					& \emph{Shattered query} mapping signature $\overline{\mathcal{T}}'$ of tables to the query\\
					& output. Example $q'(R_1, R_2)$: $Q(x) \datarule R_1(x,y),R_2(\_,y).$ \\
	\equiv			& Logical equivalence between a query and its shattered queries with appropriate signature.
					Example $q'(R, R) \equiv q(R)$.	\\
	\L_1 \subseteq^\rep \L_2					
					& Language $\L_2$ can \emph{pattern-represent} language $\L_1$, i.e.\ 
					$\L_2$ can represent all relational query patterns of language $\L_1$. \\
	\L_1 \not \subseteq^\rep \L_2
					& Language $\L_2$ \emph{pattern-dominates} language $\L_1$, i.e.\ 
					$\L_2$ can represent all relational query patterns of language $\L_1$
					and $\L_2$ can represent relational query patterns that $\L_1$ cannot.\\
	\L_1 \equiv^\rep \L_2
					& Languages $\L_1$ and $\L_2$ are \emph{representation equivalent}, i.e.\
					they can express the identical set of relational query patterns.\\
\hline
\end{tabularx}
\end{table}

\section{Proofs \cref{sec:nondisjunctivefragment}}

\begin{proof}[Proof \cref{th:equivalence}]
	We prove each of the directions in turn.

\underline{$\symb \NDRA \rightarrow \DatalogND$}:
The proof for this direction is an easy induction on the size of the algebraic expression.
It is a minor adaptation of the translation from $\RA$ to $\DatalogN$ proposed in \cite{Ullman1988PrinceplesOfDatabase},
yet it also pays attention to the restricted fragment, 
and keeps the numbers of atoms constant during the translation, if possible.
Formally, we show that if an $\NDRA$ expression has $i$ occurrences of operators, 
then there is a $\DatalogND$ program that produces, 
as the relation for one of its IDB predicates, the value of the expression. 

The basis is $i = 0$, that is, a single operand. 
If this operand is a given relation $R$, then $R$ is an EDB relation and thus ``available" without the need for any rules. 
For the induction, consider an expression whose outermost operator is one of 6 operators: 
Cartesian product ($\times$), selection ($\sigma$), theta join ($\Join_{c}$), 
projection $\pi$, and difference ($-$).

Case 1: $Q = E_1 \times E_2$:
Let $\NDRA$ expressions $E_1$ and $E_2$ have $\DatalogND$ predicates $e_1$ and $e_2$ whose rules define their relations, and assume their relations are of arities $d$ and $m$, respectively. 
Then define $q$, the predicate for $Q$, by:
\begin{align*}
	q(x_1, \ldots , x_{d + m}) \datarule e_1(x_1, \ldots , x_d),  e_2(x_{d + i}, \ldots , x_{d + m}).
\end{align*}

Case 2: $Q = \sigma_c E$:
By restricting our language from $\RA$ to $\NDRA$, we only allow selections $\sigma_c(\varphi)$ where the condition $c$ is
a conjunction of simple selections $c = c_1 \wedge c_2 \wedge \cdots$, i.e.\
each selection $c_i$ is of the form 
$\sigma_{A_{i1} \theta A_{i2}}$ (join predicate) or
$\sigma_{A_{i1} \theta v}$ (selection predicate).
Let $e$ be a $\DatalogND$ predicate whose relation is the same as the relation for $E$, 
and suppose $e$ has arity $d$. Then the rule for $Q$ is:
\begin{align*}
	q(x_1, \cdots , x_d) \datarule e_1(x_1, \cdots , x_d), c_{\theta}.
\end{align*}
where $c_{\theta}$ is a conjunction of join predicates $x_i \theta x_j$
and selection predicates $x_i \theta v$
where $A_i$ and $A_j$ correspond to the attributes indexed by $x_i$ and $x_j$.

Case 3: $Q = E_1 \Join_{c} E_2$:
While the join operator is not a basic operator of relational algebra, 
built-in predicates are, in practice, commonly expressed directly through join conditions.
This case follows immediately from cases 1 and 2, and definition of joins as
$Q = E_1 \Join_{c} E_2 = \sigma_c (E_1 \times E_2)$.

Case 4: $Q = \pi_{i_1, \ldots, i_d}(E)$:
Let $E$ 's relation have arity $d$, and let $e$ be a predicate of arity $m$ whose rules produce the relation for $E$. 
Then the rule for $q$, the predicate corresponding to expression $Q$, is:
\begin{align*}
	q(x_{i_1}, \ldots, x_{i_d}) \datarule e(x_1, \ldots , x_m).
\end{align*}

Case 5: $Q = E_1 - E_2$: 
We know by definition of the set difference that $E_1$ and $E_2$ must have the same arities. Assume those to be $d$,
and that there are predicates $e_1$ and $e_2$ whose rules define their relations to be the same as the relations for $E_1$ and $E_2$, respectively.
Then we use rule:
\begin{align*}
	q(x_1, \cdots , x_n) \datarule e_1(x_1,  \ldots , x_d),  \neg e_2(x_1, \ldots, x_d).
\end{align*}
to define a predicate $q$ whose relation is the same as the relation for $Q$. 
We can easily check that safety for this rule is fulfilled as all variables appearing in the negated $e_2$ also appear in the positive $e_1$.

\begin{figure}[t]
\centering
	\hspace{5mm}		
    \includegraphics[scale=0.35]{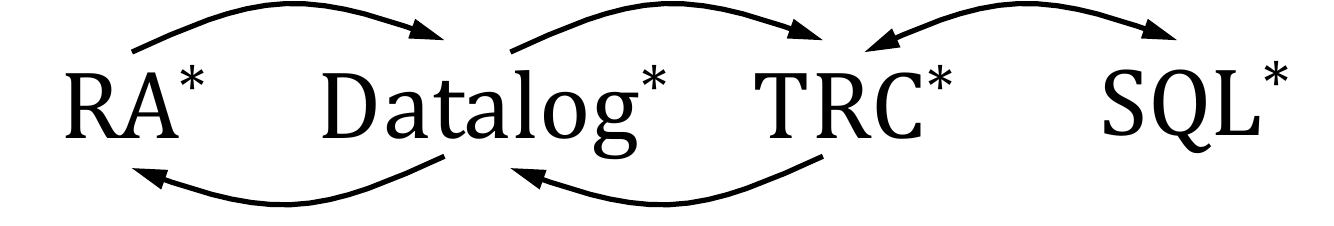}
\caption{Directions used in proof for \cref{th:equivalence} on logical expressiveness.}
\label{Fig_Equivalence}
\end{figure}

\begin{figure*}[t]
\centering
\begin{subfigure}[b]{.18\linewidth}		
\begin{align*}
	&	\{ q(A) \mid \exists r \in R, \exists s \in S [\\
	&	q.A = r.A \wedge r.B = s.B]\}	
\end{align*}
\vspace{-2mm}
    \caption{}
\end{subfigure}
\hspace{2mm}
\begin{subfigure}[b]{.136\linewidth}		
\begin{lstlisting}
SELECT DISTINCT R.A
FROM R, S
WHERE R.B = S.B
\end{lstlisting}
\vspace{0.5mm}
    \caption{}
\end{subfigure}
\hspace{3mm}
\begin{subfigure}[b]{.136\linewidth}
\begin{lstlisting}
SELECT DISTINCT R.A
FROM R 
WHERE exists
 (SELECT *
 FROM S
 WHERE R.B = S.B)
\end{lstlisting}
\vspace{-7mm}
\caption{}
\end{subfigure}
\hspace{3mm}	
\begin{subfigure}[b]{.136\linewidth}
\begin{lstlisting}
SELECT DISTINCT R.A
FROM R 
WHERE R.B in
  (SELECT S.B
  FROM S)
\end{lstlisting}
\vspace{-4.5mm}
\caption{}
\end{subfigure}
\hspace{3mm}
\begin{subfigure}[b]{.136\linewidth}		
\begin{lstlisting}
SELECT DISTINCT R.A
FROM R 
WHERE R.B = any
  (SELECT S.B
  FROM S )
\end{lstlisting}
\vspace{-4.5mm}
\caption{}
\end{subfigure}
\hspace{1mm}
\begin{subfigure}[b]{.14\linewidth}
    \includegraphics[scale=0.41]{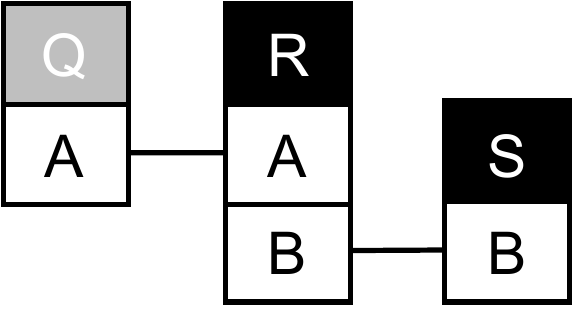}
	\vspace{1mm}
    \caption{}
    \label{Fig_R_S_join}
\end{subfigure}	
\hspace{0mm}
\begin{subfigure}[b]{.2\linewidth}		
\begin{align*}
	&	\{ q(A) \mid \exists r \in R[q.A = r.A \,\wedge \\
	&	 \neg (\exists s \in S [r.B = s.B])]\}	
\end{align*}
\vspace{-2mm}
    \caption{}
\end{subfigure}
\hspace{2mm}
\begin{subfigure}[b]{.116\linewidth}
	\phantom{test}
\end{subfigure}		
\hspace{3mm}	
\begin{subfigure}[b]{.136\linewidth}
\begin{lstlisting}
SELECT DISTINCT R.A
FROM R 
WHERE not exists
 (SELECT *
 FROM S
 WHERE R.B = S.B)
\end{lstlisting}
\vspace{-7mm}
\caption{}
\end{subfigure}
\hspace{3mm}	
\begin{subfigure}[b]{.136\linewidth}
\begin{lstlisting}
SELECT DISTINCT R.A
FROM R 
WHERE R.B not in
 (SELECT S.B
 FROM S)
\end{lstlisting}
\vspace{-4.5mm}
\caption{}
\end{subfigure}
\hspace{3mm}
\begin{subfigure}[b]{.136\linewidth}		
\begin{lstlisting}
SELECT DISTINCT R.A
FROM R 
WHERE R.B <> all
 (SELECT S.B
 FROM S)
\end{lstlisting}
\vspace{-4.5mm}
\caption{}
\end{subfigure}	
\hspace{1mm}
\begin{subfigure}[b]{.14\linewidth}
    \includegraphics[scale=0.41]{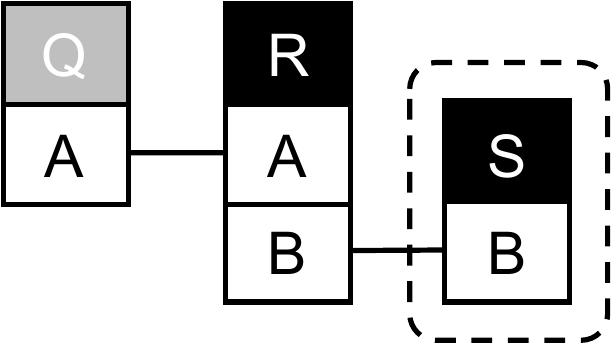}
	\vspace{-3.5mm}
    \caption{}
    \label{Fig_R_S_neg_join}
\end{subfigure}
\hspace{10mm}
\begin{subfigure}[b]{.2\linewidth}		
\begin{align*}
	&	\{ q(A) \mid \exists r \in R[q.A = r.A \,\wedge \\
	&	 \neg (\exists s \in S [r.B < s.B])]\}	
\end{align*}
\vspace{-2mm}
    \caption{}
\end{subfigure}
\hspace{2mm}
\begin{subfigure}[b]{.116\linewidth}
	\phantom{test}
\end{subfigure}		
\hspace{3mm}	
\begin{subfigure}[b]{.136\linewidth}
\begin{lstlisting}
SELECT DISTINCT R.A
FROM R 
WHERE not exists
 (SELECT *
 FROM S
 WHERE R.B < S.B)
\end{lstlisting}
\vspace{-7mm}
\caption{}
\end{subfigure}
\hspace{3mm}
\begin{subfigure}[b]{.136\linewidth}
	\phantom{test}
\end{subfigure}		
\hspace{3mm}
\begin{subfigure}[b]{.136\linewidth}		
\begin{lstlisting}
SELECT DISTINCT R.A
FROM R 
WHERE R.B >= all
 (SELECT S.B
 FROM S)
\end{lstlisting}
\vspace{-4.5mm}
\caption{}
\end{subfigure}	
\hspace{1mm}
\begin{subfigure}[b]{.14\linewidth}
    \includegraphics[scale=0.41]{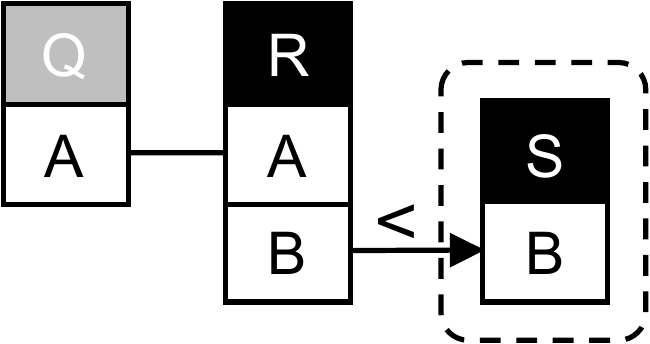}
	\vspace{-4mm}
    \caption{}
    \label{Fig_R_S_inequality_join}
\end{subfigure}
\caption{\Cref{ex:SQLvsTRC}: SQL has a redundant syntax, especially if interpreted under set semantics (``\sql{SELECT DISTINCT}''), 
binary logic (tables contain no null values) 
and compared with TRC.
Here, queries (a)--(e), queries (g)--(j), and queries (l)--(n) are equivalent.
On the right, (f), (k), and (o) show the three corresponding \diagrams (\cref{sec:QV}) that abstract away the syntactic variants 
and focus on the logical patterns of the queries.
}
\label{fig:SQLvariety}
\end{figure*}

\underline{$\symb \DatalogND \rightarrow \NDRA$}:
Typical textbook translations from $\DatalogN$ to $\RA$,
such as the one by Ullman \cite[Th 3.8, Alg 3.2, Alg 3.6]{Ullman1988PrinceplesOfDatabase}
need to compute the active domain by projecting all EDB relations onto each of their components 
and then taking the union of these projections and the set of constants appearing in the rules, if any. 
In contrast, we cannot create the active domain from a union of all constants used in the database
since $\NDRA$ lacks the union operator.

Let $\DatalogND$ program $\mathcal{P}$ be a collection of safe, nonrecursive Datalog rules, possibly with negated subgoals. 
By the safety condition, every variable that appears anywhere in the rule must appear in some nonnegated, relational subgoal of the body,
or must be bound by an equality (or a sequence of equalities) to a variable of such an ordinary predicate or to a constant~\cite{DBLP:journals/tkde/CeriGT89}.
From the definition of $\DatalogND$, each IDB appears in exactly one rule as head.
Then for each IDB predicate $q$ of $\mathcal{P}$ there is an expression $Q$ of relational algebra 
that computes the relation for $q$.
Since $\mathcal{P}$ is nonrecursive, we can order the predicates according to a topological sort of the dependency graph; 
that is, if $q_1$ appears as a subgoal in a rule for $q$, then $q_1$ precedes $q$ in the order. 

If a rule contains built-in predicates in the body (join predicates $x_i \theta x_j$ or selection predicates $x_i \theta v$), 
the translation first focuses on the body without predicates and then applies a selection $\sigma_c$ where the selection condition $c$
is a conjunction of the built-in predicates.

To express negated subgoals in the body, we need to use the set difference, and this requires us to complement negated subgoals with additional attributes.
Concretely, take a general Datalog rule without built-in predicates
\begin{align*}
	q(\vec x) \datarule p_1(\vec x_1), \ldots, p_k(\vec x_k),  \neg n_1(\vec y_1), \ldots,  \neg n_m(\vec y_m). 
\end{align*}
From the safety conditions of this rule we know that
all variables in negated atoms also need to appear in positive atoms:
$\bigcup \vec y_i \subseteq \bigcup \vec x_i$. 
Let $\vec z$ be the set of complementing attributes, i.e.\ the attributes that only appear in positive atoms:
$\vec z = \bigcup \vec x_i - \bigcup \vec y_i$.

Let $P_i$ and $N_i$ be the $\NDRA$ expressions corresponding to $\DatalogND$ predicates $p_i$ and $n_i$.
If $\vec z = \emptyset$, then 
\begin{align}
	Q &= (P_1 \Join \ldots \Join P_k) - (N_1 \Join \ldots \Join N_m) 	\notag \\
\intertext{Otherwise}
	Q &= (P_1 \Join \ldots \Join P_k) - \big((N_1 \Join \ldots \Join N_m) \times \pi_{\vec z}(P_1 \Join \ldots \Join P_k) \big)
	\label{eq:translation_Datalog_RA_problem}
\end{align}
This expression translates one single rule into a valid relational algebra expression without union or disjunction.
Since every IDB predicate $q$ appears in only one rule, we do not need union or disjunctions even if multiple rules are translated.
It then follows by induction on the order in which the IDB predicates are considered that each has a relation defined by some expression 
in $\NDRA$.

\underline{$\symb \NDTRC \rightarrow \DatalogND$}:
In this translation, 
we start from the canonical representation of $\NDTRC$
(\cref{sec:TRC})
where a set of existential quantifiers is always predated by the negation operator 
(except for the table variables at the root of the query).
This implies that we can decompose any query in $\NDTRC$ 
and write it as nested query components, each delimited by the scope of one negation operator.
Each query is then of the form:
\begin{align*}
	& \{ q(\vec A) \mid p_1 \in P_1, \ldots, p_k \in P_k[c_\textrm{out} \wedge c_{p} \\
	&\hspace{10mm}	\wedge \neg q_1(\vec A_1)
					\wedge \ldots \wedge \neg q_m(\vec A_m)] \}
\end{align*}
Here $\vec A$ is a set of attributes that correspond to the variables returned by the query
(or equivalently, variables that are passed to a nested query that determine whether that nested query is true or false),
$c_\textrm{out}$ is a conjunction of comparison predicates linking attributes from the output table $q$ 
to attributes from the local tables $P_i$,
$c_p$ is a conjunction of comparison predicates between the positive relations or constants, and
$\vec A_j$ are attributes from the output $q$ or input tables $P_i$ used in a nested query.

Notice that for safe queries, only attributes from the positive relations can be returned,
i.e.\ the output attributes need to be connected via \emph{equality predicates specified in $c_\textrm{out}$}.
However, nested queries do not need to be safe, 
which creates the one complication we need to take care of during the translation.
We proceed in two steps:

(1) First assume that each query is safe. 
Then each subquery can be immediately translated into a separate rule by induction on the nesting hierarchy from the inside out.
Basis of the induction for the leaf queries which are of the form:
\begin{align*}
	& \{ q(\vec A) \mid p_1 \in P_1, \ldots, p_k \in P_k[c_\textrm{out} \wedge c_{p}] \}
\end{align*}
A leaf query is translated into
\begin{align*}
	q(\vec x) \datarule P_1(\vec x_1), \ldots, P_k(\vec x_k), c_{\theta}.
\end{align*}
where $\vec x$ are attributes chosen from the relations $P_i$ as specified in $c_\textrm{out}$,
and $c_{\theta}$ is 
conjunction of comparison predicates between the positive relations $P_i$.

For the induction step, assume that each nested $q_i(\vec A_i)$ is safe and translated into a rule $q_i$
Then safe query $q$
\begin{align*}
	& \{ q(\vec A) \mid p_1 \in P_1, \ldots, p_k \in P_k[c_\textrm{out} \wedge c_{p} \\
	&\hspace{10mm}	\wedge \neg q_1(\vec A_1)
					\wedge \ldots \wedge \neg q_m(\vec A_m)] \}
\end{align*}
is translated into a rule
\begin{align*}
	q(\vec x) \datarule P_1(\vec x_1), \ldots, P_k(\vec x_k), \neg N_1(\vec y_1), \ldots, \neg N_m(\vec y_m), c_{\theta}.
\end{align*}
where $\vec x$ are attributes chosen from the positive relations $P_i$ as specified in $c_\textrm{out}$,
$c_{\theta}$ is conjunction of comparison predicates between the positive relations $P_i$,
and $\vec y_j$ are chosen from the variables used in the positive relations $P_k(\vec x_k)$.

(2) Next assume that a nested query is valid, yet not safe. 
This can happen because of two reasons:
($i$) Either some $q_j(\vec A_j)$ uses an attribute from the output $q(\vec A)$ directly;
or ($ii$) some predicate in $c_\textrm{out}$ connects an output predicate to a predicate from $P_j(\vec x_j)$ with an inequality predicate.
In both cases, we can make this query safe by adding one or more additional tables $P_{k+1}$
and adding appropriate predicates:
For case ($i$), we add an equality predicate to $c_\textrm{out}$ and replace the attribute specified in $q_j(\vec A_j)$.
For case ($ii$), we add an equality predicate to $c_p$.
We illustrate both cases of the translation with one example each.

\begin{example}[All quantification in $\DatalogND$]
We illustrate the translation with the help of the 
relational division example from \cref{ex:relationalDivision,fig:SQL_equivalence}:
\begin{align*}
	&\{ q(A) \mid \exists r \in R [r.A \equal q.A  \wedge \neg (\exists s \in S[	\\
	&\hspace{10mm}	
		\neg (\exists r_2 \in R 	
		[r_2.B \equal s.B \wedge r_2.A \equal r.A])])] \}
\end{align*}
Based on our extended safety condition for $\NDTRC$
\cref{def:anchor,def:NDTRC}, 
all predicates are anchored, i.e.\
they contain at least one attribute of a table that is existentially quantified 
inside the same negation scope as that predicate.
Those are 
$r.A$ in $r.A \equal q.A$,
$r_2.B$ in $r_2.B \equal s.B$, and
$r_2.A$ in $r_2.A \equal r.A$ (shown in red below):
\begin{align*}
	&\{ q(A) \mid \exists r \in R [\h{r.A} \equal q.A  \wedge \neg (\exists s \in S[	\\
	&\hspace{10mm}	
		\neg (\exists r_2 \in R 	
		[\h{r_2.B} \equal s.B \wedge \h{r_2.A} \equal r.A])])] \}
\end{align*}

Rewriting the query based on its recursive nested negation hierarchy allows us to identify 3 query components:
\begin{align*}
	\{ q(A)   &\mid  \exists r \in R [r.A \equal q.A \wedge \neg (q_1(r.A))] \}	\\
	\{ q_1(A) &\mid  \exists s \in S
		[\neg \h{q_2(q_1.A}, s.B\h{)}] \} \\
	\{q_2(A, B)  &\mid  \exists r_2 \in R 	
		[r_2.B \equal q_2.B \wedge r_2.A \equal q_2.A] \}
	\hspace{13mm}
\end{align*}
Now notice that the predicate $r_2.A=r.A$
(or $r_2.A=q_2.A$ in the recursive hierarchy)
is not limited: it references attribute $r.A$ that is outside the negation scope of the direct parent of the scope in which it appears.
This can also be seen from the recursive call $q_2(q_1.A, s.B)$ that ``passes through'' a predicate through the hierarchy.
It can also be seen from the fact that $q_1$ is not safe.

We can limit the predicate (or equivalently make $q_1$ safe) 
by adding another table $r_3 \in R$ in $q_1$ that accepts and hands over that attribute in the call hierarchy:
\begin{align*}
	\{ q(A)   &\mid  \exists r \in R [r.A \equal q.A \wedge \neg (q_1(r.A))] \}	\\
	\{ q_1(A) &\mid \exists s \in S, \h{\exists r_3 \in R}
		[\h{r_3.A = r.A} \wedge \neg q_2(\h{r_3.A}, s.B)] \} \\
	\{q_2(A, B)  &\mid \exists r_2 \in R 	
		[r_2.B \equal q_2.B \wedge r_2.A \equal q_2.A] \}	
\end{align*}
This rewritten query now allows a direct translation into $\DatalogND$ from the inside out:
\begin{align*}
	Q_2(x, y) 	& \datarule R(x,y). 		\\
	Q_1(x) 	& \datarule R(x,\_), S(y), \neg Q_2(x,y). 		\\
	Q(x) 	& \datarule R(x,\_), \neg Q_1(x).
\end{align*}
\end{example}

\begin{example}[Built-in predicates in $\DatalogND$]
We next illustrate the translation for a buit-in predicate with
$Q_3$
from
\cref{ex:DatalogLimits}
asking for values from $R$ for which no smaller value appears in $S$:
\begin{align*}
	& \{ q(A) \mid \exists r \in R[r.A \equal q.A
		\wedge \neg (\exists s \in S [	s.A < r.A])]\}	
\end{align*}

Rewriting the query based on its recursive nested negation hierarchy allows us to identify 2 query components:
\begin{align*}
	\{ q(A)   &\mid  \exists r \in R [r.A \equal q.A \wedge \neg (q_1(r.A))] \}	\\
	\{ q_1(A) &\mid  \exists s \in S [s.A < q_1.A] \} 
	\hspace{13mm}
\end{align*}
Now notice that the predicate $s.A < r.A$
is not limited: it references attribute $r.A$ that is outside the negation scope with an inequality instead of equality predicate.
This can also be seen from the fact that $q_1$ is not safe.

We can limit the predicate (or equivalently make $q_1$ safe) 
by adding another table $r_2 \in R$ in $q_1$:
\begin{align*}
	\{ q(A)   &\mid  \exists r \in R [r.A \equal q.A \wedge \neg (q_1(r.A))] \}	\\
	\{ q_1(A) &\mid  \exists s \in S, \h{r_2 \in R} [s.A < \h{r_2.A} \wedge \h{r_2.A} = q_1.A] \} 
	\hspace{13mm}
\end{align*}
This rewritten query now allows a direct translation into $\DatalogND$ from the inside out:
\begin{align*}
	Q_1(x)	& \datarule R(x),  S(y), x\!>\!y. 		\\
	Q(x) 	& \datarule R(x), \neg Q_1(x).
\end{align*}
\end{example}

It follows that every query in $\NDTRC$
can be translated into a logically equivalent query in $\DatalogND$.

\underline{$\symb \DatalogND \rightarrow \NDTRC$}:
We consider a general Datalog rule:
\begin{align*}
	q(\vec x) \datarule p_1(\vec x_1), \ldots, p_k(\vec x_k), \neg n_1(\vec y_1), \ldots, \neg n_m(\vec y_m), c_{\theta}.
\end{align*}
Here $c_{\theta}$ is a conjunction of built-in predicates that adhere to the standard safety conditions \cite{DBLP:journals/tkde/CeriGT89}.
We know the rule is safe and thus $\bigcup \vec y_i \subseteq \bigcup \vec x_i$. 
Let $\vec z$ again be the set of complementing attributes, i.e.\ the attributes that only appear in positive atoms:
$\vec z = \bigcup \vec x_i - \bigcup \vec y_i$.
The rule then translates into a $\TRC$ fragment
\begin{align*}
	\{q(\vec A) \mid 
		&p_1 \in P_1, \ldots, p_k \in P_k[c_\textrm{out} \wedge c_{p} \\
		&\wedge \neg(\exists n_1\in N_1, \ldots, n_m \in N_m[c_\textrm{in}])]\}
\end{align*}
Here $\vec A$ is a set of attributes that correspond to the variables returned by the Datalog rule 
(from safety conditions, only attributes from the positive relations can be returned),
$c_\textrm{out}$ is a conjunction of equality joins linking attributes from the output table $q$ to attributes from the input tables $P_i$,
$c_p$ is a conjunction of comparison predicates between the positive relations or constants, and
$c_\textrm{in}$ is a conjunction of equality predicates between exactly one negative relation
and either a positive relation or a constant.

\begin{figure}[t]
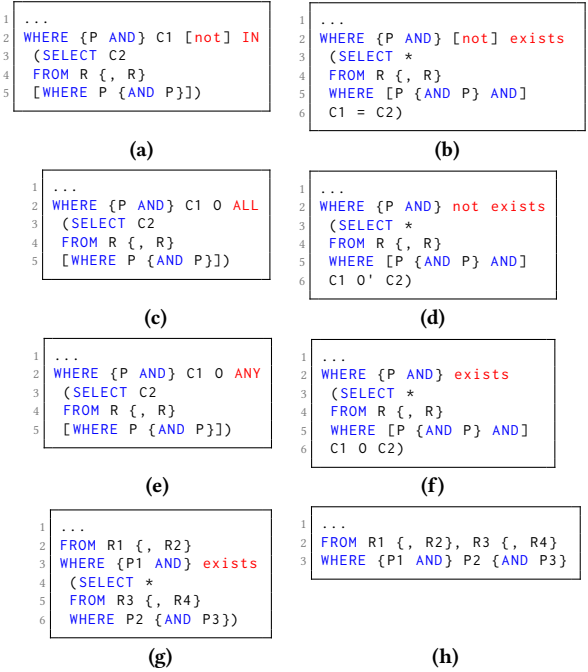

\centering
\begin{subfigure}[b]{.375\linewidth}		
\begin{lstlisting}
...
WHERE {P AND} C1 [not] IN
 (SELECT C2
 FROM R {, R}
 [WHERE P {AND P}])
\end{lstlisting}
\vspace{-4.4mm}
\caption{}
\label{fig:canonicalquery_a}
\end{subfigure}
\hspace{6mm}
\begin{subfigure}[b]{.405\linewidth}		
\begin{lstlisting}
...
WHERE {P AND} [not] exists
 (SELECT *
 FROM R {, R}
 WHERE [P {AND P} AND]
 C1 = C2)
\end{lstlisting}
\vspace{-7mm}
\caption{}
\label{fig:canonicalquery_b}
\end{subfigure}	
\hspace{1mm}
\begin{subfigure}[b]{.33\linewidth}		
\begin{lstlisting}
...
WHERE {P AND} C1 O ALL
 (SELECT C2
 FROM R {, R}
 [WHERE P {AND P}])
\end{lstlisting}
\vspace{-4.4mm}
\caption{}
\label{fig:canonicalquery_c}
\end{subfigure}
\hspace{6mm}
\begin{subfigure}[b]{.36\linewidth}		
\begin{lstlisting}
...
WHERE {P AND} not exists
 (SELECT *
 FROM R {, R}
 WHERE [P {AND P} AND]
 C1 O' C2)
\end{lstlisting}
\vspace{-7mm}
\caption{}
\label{fig:canonicalquery_d}
\end{subfigure}	
\hspace{6mm}
\begin{subfigure}[b]{.33\linewidth}		
\begin{lstlisting}
...
WHERE {P AND} C1 O ANY
 (SELECT C2
 FROM R {, R}
 [WHERE P {AND P}])
\end{lstlisting}
\vspace{-4.4mm}
\caption{}
\label{fig:canonicalquery_e}
\end{subfigure}
\hspace{6mm}
\begin{subfigure}[b]{.355\linewidth}		
\begin{lstlisting}
...
WHERE {P AND} exists
 (SELECT *
 FROM R {, R}
 WHERE [P {AND P} AND]
 C1 O C2)
\end{lstlisting}
\vspace{-7mm}
\caption{}
\label{fig:canonicalquery_f}
\end{subfigure}	
\hspace{6mm}
\begin{subfigure}[b]{.32\linewidth}		
\begin{lstlisting}
...
FROM R1 {, R2}
WHERE {P1 AND} exists
 (SELECT *
 FROM R3 {, R4}
 WHERE P2 {AND P3})
\end{lstlisting}
\vspace{-7mm}
\caption{}
\label{fig:canonicalquery_g}
\end{subfigure}	
\hspace{6mm}
\begin{subfigure}[b]{.395\linewidth}		
\begin{lstlisting}
...
FROM R1 {, R2}, R3 {, R4}
WHERE {P1 AND} P2 {AND P3}
\end{lstlisting}
\vspace{0.8mm}
\caption{}
\label{fig:canonicalquery_h}
\end{subfigure}	
\hspace{-5mm}
\caption{There is a natural structure isomorphism between the SQL variants defined by our grammar. 
The existential subqueries (right column) push any comparison or join predicate into the local scope of the nested query.}
\label{fig:canonicalquery}
\end{figure}

\underline{$\symb \NDTRC \leftrightarrow \NDSQL$}:
We prove equivalence in three steps:
We first reduce the syntactic variety of $\NDSQL$,
then define a canonical form,
and finally prove a one-to-one mapping between that canonical $\NDSQL$ and canonical $\NDTRC$.

1. Starting from \cref{table:supported_grammar}, 
we first transform ``membership subqueries'' and ``quantified subqueries'' into ``existential subqueries.''
We use the same grammar to describe this transformation.
Concretely, replace ``membership subqueries'' of the form
\cref{fig:canonicalquery_a}
with ``existential subqueries'' of the form
\cref{fig:canonicalquery_a},
and ``quantified subqueries''
\cref{fig:canonicalquery_c,fig:canonicalquery_e}
with
\cref{fig:canonicalquery_d,fig:canonicalquery_f}, respectively.
Here \sql{O'} is the complement operator of \sql{O} (for example ``<'' for ``>='')
and \sql{C1} and \sql{C2} represent different columns.

2. Similar to $\NDTRC$,
we pull existential quantifier of tables (table variables defined \sql{FROM} clauses)
as early as to either be in the root query, or
directly following a \sql{not exists}.
We show this recursive pulling out in the transition from 
\cref{fig:canonicalquery_g}
to
\cref{fig:canonicalquery_h}.

3. The resulting canonical $\NDSQL$ in now in a direct 1-to-1 correspondence to $\NDTRC$,
and the translation between $\NDSQL$ and $\NDTRC$ 
is then matter of translating the different syntactic expressions between the two languages:
The \sql{SELECT DISTINCT C \{, C\} } is equivalent to the output definition in $\NDTRC$,
each \sql{FROM R \{, R\} } defines the existentially quantified tuple variables $\exists \vec r \in \vec R[...]$,
each \sql{not exists(SELECT * FROM R \{, R\} ...)} corresponds to negated existentially quantified tuple variables,
$\neg(\exists \vec r \in \vec R [...])$,
and the syntax of predicates is identical.
\end{proof}

\begin{example}[$\NDSQL$ vs.\ $\NDTRC$]
\label{ex:SQLvsTRC}
\Cref{fig:SQLvariety} shows three different non-disjunctive queries in $\NDTRC$, various syntactic variants of $\NDSQL$, and $\diagrams$.
$\NDSQL$ queries (b), (h), and (m) are canonical and in a direct 1-to-1 relationship with $\NDTRC$.
\end{example}

\section{Proofs \cref{sec:structureisomorphism}}
\label{appendix:proofs4}

\begin{figure}[t]
\centering
\begin{subfigure}[b]{.58\linewidth}
	\centering
    \includegraphics[scale=0.35]{figs/Fig_Representation}
	\vspace{1mm}
    \caption{}
    \label{Fig_Representation}
\end{subfigure}	
\hspace{2mm}
\begin{subfigure}[b]{.35\linewidth}
	\centering
    \includegraphics[scale=0.35]{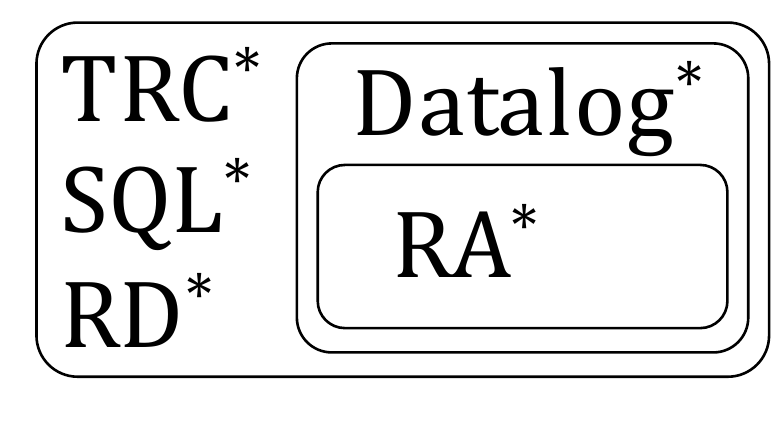}
	\vspace{-2mm}
    \caption{}
    \label{Fig_Representation_Hierarchy}
\end{subfigure}	
\caption{Directions used in proof for \cref{th:representations} (a) and resulting representation hierarchy (b).}
\end{figure}

\begin{proof}[Proof of \cref{th:representations}]
We prove each of the directions in turn. 
Notice that the logical equivalences already follow from the proof of \cref{th:equivalence}.
We need to only point out which directions are guaranteed to preserve the number of tables (and are thus representation-preserving).
For those directions that do not preserve the structure in general, we give a minimum counter example.

{$\symb \NDRA \subseteq^\rep \DatalogND$}: 
This direction follows immediately from the proof of \cref{th:equivalence} 
by observing each of the mappings in the 5 cases to be structure-preserving.

{$\symb \NDRA \not \supseteq^\rep \DatalogND$}:
We show that the set difference (or minus $-$) 
from $\NDRA$ cannot isomorphically represent negation from $\DatalogND$
if the complementing set of attributes is non-empty (see \cref{eq:translation_Datalog_RA_problem}).
We show that with our \cref{ex:intro} from the introduction:
\begin{align}
	Q(x,y)	 & \datarule R(x,y), \neg S(y)
	\label{eq:intro_appendix_example_Datalog}
\end{align}	
The binary \emph{minus operator from $\RA$ requires the same arity} of the two input relations.
Thus one cannot apply the minus operator directly to combine $R$ and $S$ as in $\DatalogN$. 
Any possible sequence that includes a minus thus either uses the minus on 1 attribute, or 2 attributes (or 3 or more attributes, but those require even more joins and thus more table instances).

Case 1: Minus on 2 (or more) attributes: 
Having 2 (or more) attributes for the minus requires us to increase the arity of the right side and thus $S$. 
This in turns requires a cross product with the domain from $R.A$ before the minus as in the following translation:
\begin{align*}
	R - (\pi_A R \times S) 		
\end{align*}
This in turn increases the number of input table instance used from 2 to at least 3, 
which prevents a structure-preserving representation.

Case 2: Minus on 1 attribute: 
Having 1 attribute on the minus requires us to increase the arity \emph{after applying the minus} (because our output has arity 2). 
This in turn unavoidably increase the number of input table instances to at least 3,
which again prevents a structure-preserving representation.
An example translation first uses a projection on $R$ before the minus (to reduce the left input to arity 1) and then a subsequent join again with $R$ after the minus:
\begin{align*}
	R \Join_B (\pi_B R - S)		
\end{align*}

It follows that in whatever way the $\DatalogN$ expression 
\cref{eq:intro_appendix_example_Datalog} is expressed in $\RA$, 
the expression will have at least 3 references to input tables.
Thus $\RA$ cannot preserve the representation from \cref{eq:intro_appendix_example_Datalog}.

{$\symb \DatalogND \subseteq^\rep \NDTRC$}: 
This direction follows immediately from the proof of \cref{th:equivalence} 
by observing the mappings of each Datalog rule to be structure-preserving.

\begin{figure*}[t]
\centering
\hspace{1mm}
\begin{subfigure}[b]{.14\linewidth}
\begin{lstlisting}
SELECT DISTINCT R.A
FROM R
WHERE not exists
 (SELECT *
 FROM S
 WHERE not exists
  (SELECT *
  FROM R AS R2
  WHERE R2.B = S.B
  AND R2.A = R.A))
\end{lstlisting}
\vspace{-7mm}
    \caption{}
    \label{fig:SQL_equivalence1}
\end{subfigure}
\hspace{1mm}
\begin{subfigure}[b]{.18\linewidth}
    \includegraphics[scale=0.35]{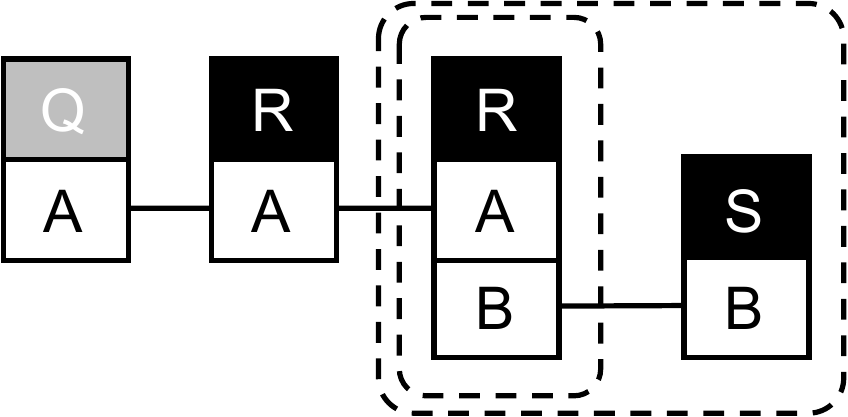}
	\vspace{2mm}
    \caption{}
    \label{Fig_relational_division_1}
\end{subfigure}
\hspace{12mm}
\begin{subfigure}[b]{.14\linewidth}
\begin{lstlisting}
SELECT DISTINCT R.A
FROM R
WHERE not exists
 (SELECT *
 FROM S, R AS R3
 WHERE R3.A = R.A
 AND not exists
  (SELECT *
  FROM R AS R2
  WHERE R2.B = S.B
  AND R2.A = R3.A))
\end{lstlisting}
\vspace{-7mm}
    \caption{}
    \label{fig:SQL_equivalence2}
\end{subfigure}
\hspace{1mm}
\begin{subfigure}[b]{.20\linewidth}
    \includegraphics[scale=0.35]{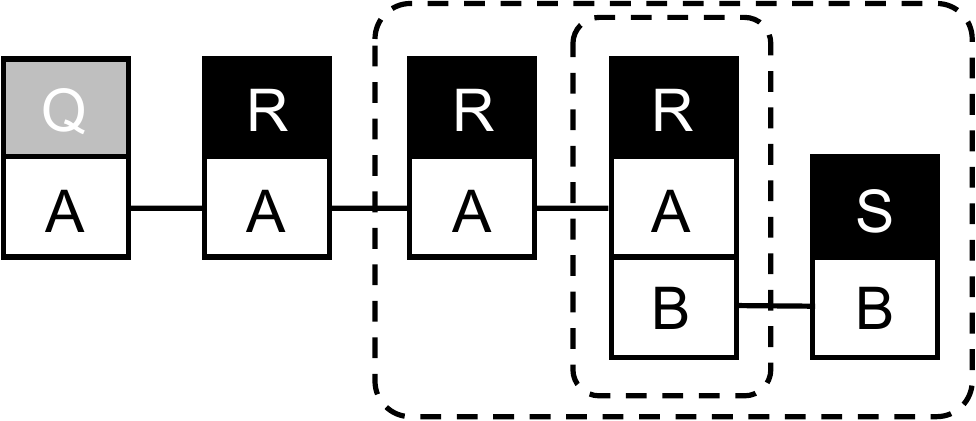}
	\vspace{2mm}
    \caption{}
    \label{Fig_relational_division_2}
\end{subfigure}
\hspace{2mm}
\begin{subfigure}[b]{.18\linewidth}
\begin{lstlisting}
SELECT DISTINCT R.A
FROM R
WHERE R.A not in
 (SELECT R3.A
 FROM S, R AS R3
 WHERE (R3.A, S.B) not in
  (SELECT R2.A, R2.B
  FROM R AS R2))
\end{lstlisting}
\vspace{-7mm}
    \caption{}
    \label{fig:SQL_equivalence3}
\end{subfigure}
\hspace{35.5mm}
\vspace{-4mm}
\caption{\Cref{ex:relationalDivision}: Relational division in SQL (a)(c)(e) and as \diagrams (b)(d).
All 5 representations are \emph{logically equivalent}, but only the partitions
\{(a), (b)\}
and
\{(c), (d), (e)\} are also
\emph{pattern-isomorph} (which is what we expect).
}
\label{fig:SQL_equivalence}
\end{figure*}

{$\symb \DatalogND \not \supseteq^\rep \NDTRC$}:
We show with our \cref{ex:relationalDivision} that $\RA$ cannot isomorphically represent relational division from $\NDTRC$.

Consider a schema $R(A,B), S(B)$ and the query asking for attribute values from $R.A$ that co-occur in $R$ with all attribute values from $S.B$.
Relational division can be written in $\TRC$ as:
\begin{align}
\begin{aligned}
\{ q(A) \mid 
	&\exists r \in R 	[q.A \equal r.A \wedge \neg (\exists s \in S[	\\
	&\neg (\exists r_2 \in R[r_2.B \equal s.B \wedge r_2.A \equal r.A])])] \}
\end{aligned}
\label{TRC:division_appendix}
\end{align}
Notice that \cref{TRC:division_appendix} uses 2 occurrences of $R$ and 1 occurence of $S$. 
Further notice that the predicate ``$r_2.A \equal r.A$'' joins two $R$ tables across two different negation scopes.
\Cref{Fig_relational_division_1} shows that query pattern as \diagram with a join between the two corresponding $R$ tables
\emph{across two negation boxes}.

We now show that there is no way to represent relational division in $\DatalogN$ 
with this pattern (and thus with only 2 occurrences of the $R$ symbol).
The key ingredient for this proof is the fact that the safety condition of $\DatalogND$ 
requires that each variable occurring in a negated atom also needs to be occur in at least one non-negated atom \emph{of the same rule}~\cite{DBLP:journals/tkde/CeriGT89}.
As such, it can model negation \emph{only one rule at a time}
(each rule only allows application of one negation).

As a consequence, it cannot model a query pattern 
with a join predicate across \emph{two} negations that is needed
for the $\NDTRC$ expressions 
from \cref{TRC:division_appendix}.
Instead, it needs to use another occurrence of $R$ as ``guard'' for each negation.
Thus $\DatalogN$ cannot preserve the pattern from \cref{TRC:division_appendix}.

This is achieved by the standard translation into $\DatalogN$ with 3 occurrences of $R$ and two rules:
The first rule finds all the $R.A$ that do not co-occur with all $S.B$ values. 
The second rule then finds the complement against the domain from $R.A$:
\begin{align}
\begin{aligned}
	I(x) 	& \datarule R(x,\_), S(y), \neg R(x,y) 		\\
	Q(x) 	& \datarule R(x,\_), \neg I(x) 		
\end{aligned}
\label{relational_division_datalog_appendix}
\end{align}	
Notice that the first ``extra'' atom $R(x,\_)$ 
is needed for the aforementioned safety condition of $\DatalogN$.
\Cref{Fig_equivalence1} shows that logical pattern 
with an extra repeated table $R$.

{$\symb \NDSQL \equiv^\rep \NDTRC$}:
This also follows immediately from the proof of \cref{th:equivalence} by observing the 1-to-1 correspondences of the mappings in both directions.
\end{proof}

\begin{figure*}[t]
\centering	
\hspace{-10mm}
	\begin{subfigure}[b]{.35\linewidth}
	    \includegraphics[scale=0.34]{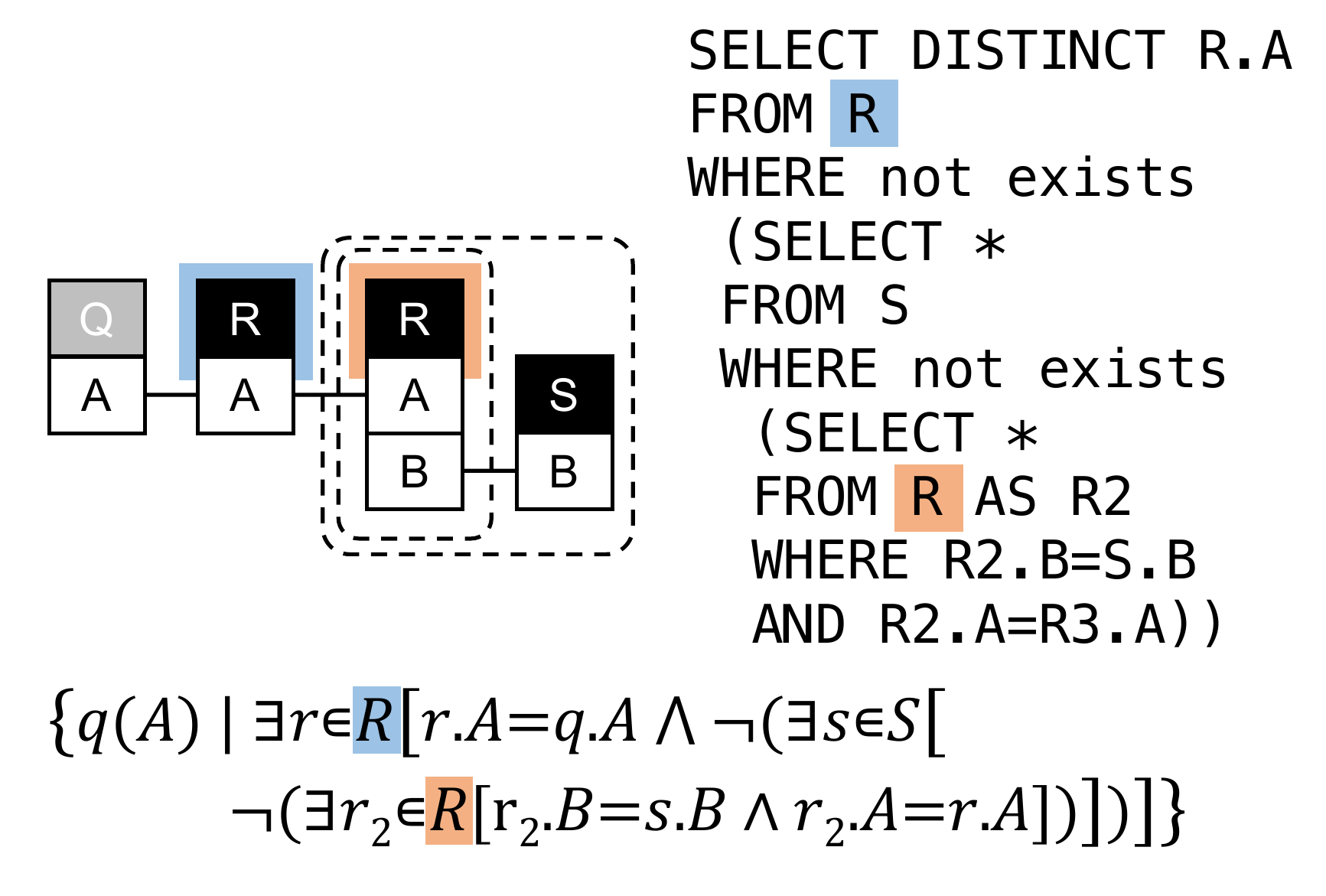}
		\vspace{-1mm}
	    \caption{}
	    \label{Fig_equivalence2}
	\end{subfigure}
	\hspace{10mm}
	\begin{subfigure}[b]{.6\linewidth}
	    \includegraphics[scale=0.34]{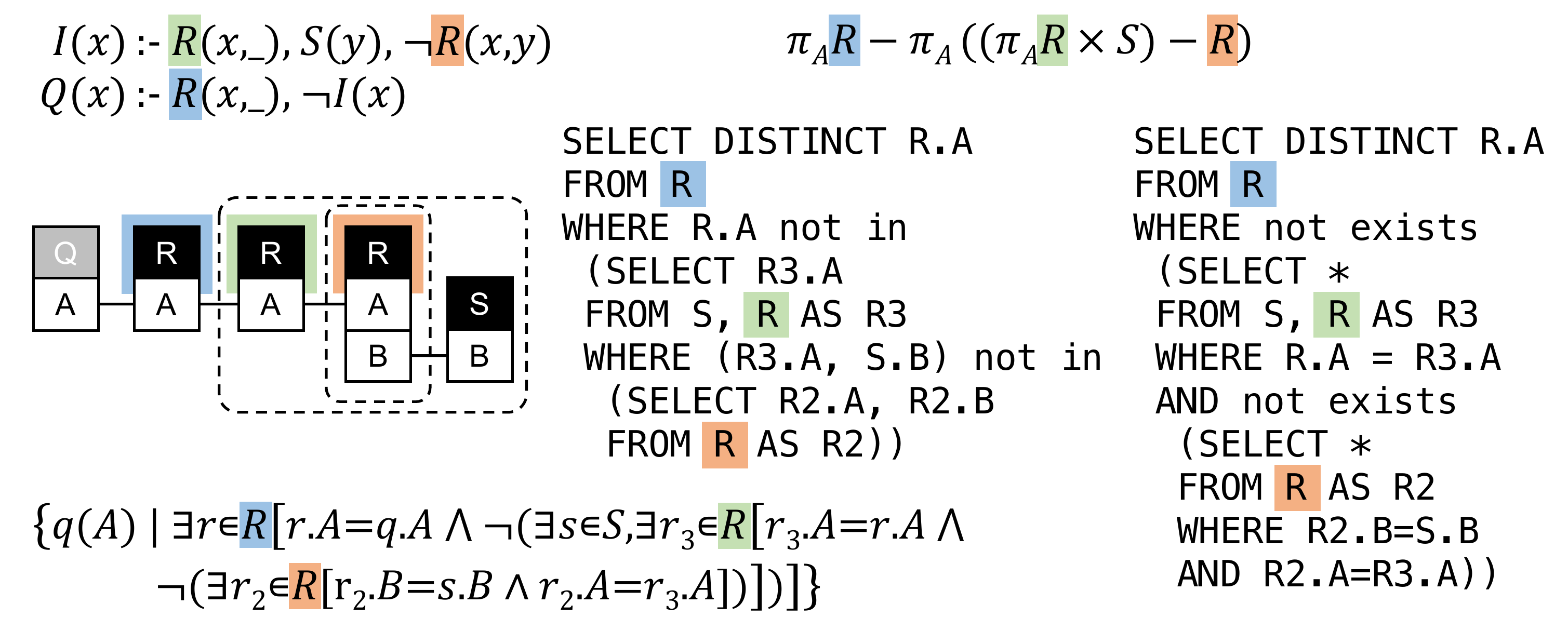}
		\vspace{-1mm}
	    \caption{}
	    \label{Fig_equivalence1}
	\end{subfigure}	
	\hspace{-13mm}
\caption{\Cref{ex:relationalDivision} and \cref{fig:SQL_equivalence} continued:
Two logically-equivalent sets (a) and (b) of relational division in 5 query languages 
(\diagrams, $\NDSQL$, $\NDRA$, $\DatalogND$, $\NDTRC$).
The queries in (a) use 2 occurrences of $R$, whereas the ones in (b) use 3 occurrences of $R$.
We highlight the two or three occurrences of $R$ across the different languages that can mapped to each other 
according to our pattern isomorphism defined in \cref{def:isomorphism}.
We prove in \cref{appendix:proofs4} that for (a), there is no pattern-isomorph representation in $\DatalogN$ (thus neither in $\RA$).
}
\label{Fig_equivalence}
\end{figure*}

\section{More Illustrations \cref{sec:structureisomorphism}}

We next illustrate 
with the help of the more complicated example of relational division
that there is a structure-preserving mapping from $\RA$ to $\TRC$, but not in the other direction.

\begin{example}[$\TRC$ and $\RA$ are not representation-equivalent]\label{ex:relationalDivision}
Assume a schema $R(A,B), S(B)$.
Consider the relational division asking for attribute values from $R.A$ that co-occur in $R$ with all attribute values from $S.B$.
The translation
into $\TRC$  is
\begin{align}
\begin{aligned}
\{ q(A) \mid 
	&\exists r \in R 	[q.A \equal r.A \wedge \neg (\exists s \in S[	\\
	&\neg (\exists r_2 \in R[r_2.B \equal s.B \wedge r_2.A \equal r.A])])] \}	
\end{aligned}
\label{TRC:division1}
\end{align}
The corresponding canonical $\SQL$ statement is shown in \cref{fig:SQL_equivalence1}.
Relational division expressed in primitive $\RA$ 
is
\begin{align}
	\pi_A R - \pi_A\big( (\pi_A R \times S) - R \big)
	\label{RA:division}
\end{align}
The translation into $\DatalogN$ uses two rules:
\begin{align}
\begin{aligned}
	I(x) 	& \datarule R(x,\_), S(y), \neg R(x,y). 		\\
	Q(x) 	& \datarule R(x,\_), \neg I(x). 		
\end{aligned}
\label{datalog}
\end{align}	
The atoms $R(x,\_)$ are needed for the safety condition of $\DatalogN$.
This translation is part of a standard proof for equivalence of expressiveness between $\RA$ and safe $\TRC$ in textbooks such as 
\cite{Ullman1988PrinceplesOfDatabase,DBLP:books/cs/Maier83,DBLP:books/aw/AbiteboulHV95}.

Now notice an arguably important difference between the three expressions:
$\TRC$ \cref{TRC:division1} uses the atom $R$ \emph{two} times, whereas
$\RA$ \cref{RA:division} and $\DatalogN$ \cref{datalog} use $R$ \emph{three} times.
It turns out that there is no way to represent relational division in primitive $\RA$ or $\DatalogN$ with only two occurrences of the $R$ symbol 
(see \cref{th:representations}).

There is, however, an alternative representation in $\TRC$ that preserves the $\RA$ structure with three occurrences of $R$:
\begin{align}
\begin{aligned}	
	\{ q(A) \mid \exists r \in R &[q.A \equal r.A \wedge 
		\neg (\exists s \in S, \exists r_3 \in R [r_3.A = r.A  		\\							
		&\hspace{0mm} \wedge \neg (\exists r_2 \in R[r_2.B \equal s.B \wedge r_2.A \equal r_3.A])])]  \}		\label{TRC:division2}
\end{aligned}
\end{align}
Notice that now there a natural 1-to-1 correspondence between the atoms in 
$\TRC$ \cref{TRC:division2} and the atoms in $\RA$ \cref{RA:division}.
This correspondence is even more intuitive by mapping the correspondence between two logically-equivalent $\SQL$ statements
(\cref{fig:SQL_equivalence2,fig:SQL_equivalence3}) and $\RA$ \cref{RA:division}:
for example, lines 4--8 in \cref{fig:SQL_equivalence3} translate into 
the $\RA$ fragment ``$\pi_A\big( (\pi_A R \times S) - R \big)$'', 
which corresponds to the IDB predicate $\sql{Temp}(x)$ in $\DatalogN$ \cref{datalog}.

In other words, while all of these 7 queries are \emph{logically equivalent},
they partition into two disjoint sets that are ``\emph{pattern-isomorph}'':
\begin{align*}
\textrm{Set 1} =
		& \{\RA \textrm{ \cref{RA:division}}, 
			\TRC \textrm{ \cref{TRC:division2}}, 
			\SQL \textrm{ \cref{fig:SQL_equivalence2}}, 
			\SQL \textrm{ \cref{fig:SQL_equivalence3}}, 
			\DatalogN \textrm{ \cref{datalog}} \} \\
\textrm{Set 2} =
		& \{\TRC \textrm{ \cref{TRC:division1}}, 
			\SQL \textrm{ \cref{fig:SQL_equivalence1}} \} 
\end{align*}	
This suggests that $\TRC$ and $\RA$ are not \emph{representation equivalent}.

We next prove the pattern-isomorphism between 
RA query 
\cref{RA:division}
and TRC query
\cref{TRC:division2} with our formalism.
First, write their shattered queries
$q'_{\textrm{RA}}(R_1, R_2, S, R_3)$
and
$q'_{\textrm{TRC}}(R_1, S, R_2, R_3)$
with
\begin{align*}
	q'_{\textrm{RA}} = 
	\pi_A R_1 - \pi_A\big( (\pi_A R_2 \times S) - R_3 \big)
\end{align*}
\begin{align*}
	\{ q'_{\textrm{TRC}}(A) \mid \exists r \in R_1 &[q.A \equal r.A \wedge 
		\neg (\exists s \in S, \exists r_3 \in R_2 [r_3.A = r.A  		\\							
		&\hspace{0mm} \wedge \neg (\exists r_2 \in R_3[r_2.B \equal s.B \wedge r_2.A \equal r_3.A])])]  \}		
\end{align*}
Second, define the
homomorphism $h(R_1, R_2, S, R_3) = (R_1, S, R_2, R_3)$, which is injective and surjective and thus an isomorphism between the signature of the shattered queries.
We can now easily verify that the shattered queries are logically equivalent 
after composition with $h$:
$q'_{\textrm{RA}} \equiv q'_{\textrm{TRC}} \circ h$.
In other words:
\begin{align*}
q'_{\textrm{RA}}(R_1, R_2, S, R_3) \equiv
q'_{\textrm{TRC}}(h(R_1, R_2, S, R_3)) =
q'_{\textrm{TRC}}(R_1, S, R_2, R_3)
\end{align*}

\Cref{Fig_equivalence} illustrates the pattern isomorphism within two sets of queries with the color-highlighted $R$ atoms. 
Notice in 
\cref{Fig_equivalence2} the correspondences between the blue and orange highlighted tables $R$ across the SQL and the TRC statements and the \diagram.
Notice that the constraints between their ``A'' attributes (``R2.A=R.A'') make use of the nesting hierarchy: two levels of the ``not exists'' nesting hierarchy in SQL and two levels in the negation hierarchy in TRC.
Similarly, in \cref{Fig_equivalence1}
	the two SQL variants use different syntactic constructs to represent the single negation hierarchy between R3 and R2. 
	Then, see how Datalog represents this constraint without referring to the explicit attributes ``A'' but by positional reference and using the repeated variable $x$ together with ``not'' to represent the same logical constraint.
	RA represents the same logical constraint by projecting attribute ``A'' from the green instance R on the left side of a cross product, before the set difference with the yellow instance $R$.
	Despite this extreme syntactic variants and logical equivalence of all these queries, 
	by defining individual pairwise isomorphisms between extensional tables, our
	formalisms allows to partition these queries into two sets within which the queries are pattern-isomorph. 
\end{example}

\section{More illustrations for \cref{sec:QV}}

A table can be represented by any visual grouping of its attributes (see \cref{Fig_Table_Attributes_Set_Visualization} for examples).
Our choice in \diagrams is to use the typical UML convention of representing tables as rectangular boxes
with the table name on top and attribute names below in separate rows
(\cref{Fig_Table_Attributes_Set_Visualization_a}).
This choice may affect the readability and usability of \diagrams,
yet does not affect their semantics and pattern expressiveness.

\begin{figure}
\centering
\begin{subfigure}[b]{0.15\linewidth}
	\centering
    \includegraphics[scale=0.42]{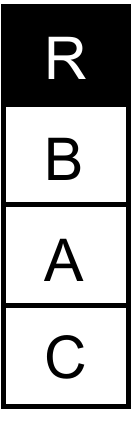}
	\vspace{-1mm}
    \caption{}
    \label{Fig_Table_Attributes_Set_Visualization_a}
\end{subfigure}
\begin{subfigure}[b]{0.18\linewidth}
	\centering	
    \includegraphics[scale=0.42]{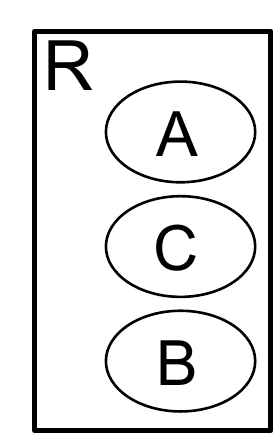}
    \caption{}
    \label{Fig_Table_Attributes_Set_Visualization_b}
\end{subfigure}
\begin{subfigure}[b]{0.23\linewidth}
	\centering	
    \includegraphics[scale=0.42]{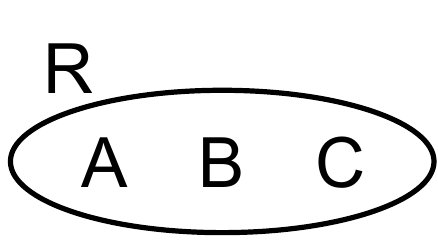}
    \caption{}
    \label{Fig_Table_Attributes_Set_Visualization_c}
\end{subfigure}
\begin{subfigure}[b]{0.17\linewidth}
	\centering	
    \includegraphics[scale=0.42]{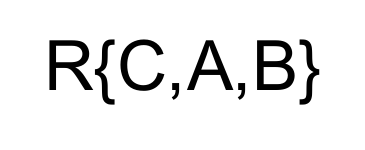}
    \caption{}
    \label{Fig_Table_Attributes_Set_Visualization_d}
\end{subfigure}
\begin{subfigure}[b]{0.21\linewidth}
	\centering	
    \includegraphics[scale=0.42]{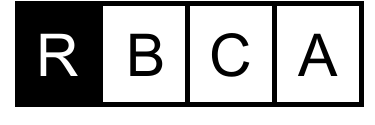}
    \caption{}
    \label{Fig_Table_Attributes_Set_Visualization_e}
\end{subfigure}
\caption{A few ways to visualize a table and its set of attributes as a group of nodes.
We use (a) inspired by UML conventions.
}
\label{Fig_Table_Attributes_Set_Visualization}
\end{figure}

\section{More illustrations for \cref{sec:completeness}}

\begin{figure}[t]
\centering	
\begin{subfigure}[b]{.375\linewidth}		
\begin{lstlisting}
SELECT DISTINCT S.sname
FROM Sailor S, Reserves R
WHERE S.sid = R.sid
AND not 
 (not exists
  (SELECT *
  FROM Boat B
  WHERE color='red'
  AND R.bid = B.bid)
 AND not exists
  (SELECT *
  FROM Boat B
  WHERE color='blue'
  AND R.bid = B.bid)  
\end{lstlisting}
\vspace{-6mm}
\caption{}
\label{fig:Sailor_disjunction_a}
\end{subfigure}	
\hspace{1mm}
\begin{subfigure}[b]{.51\linewidth}
    \includegraphics[scale=0.34]{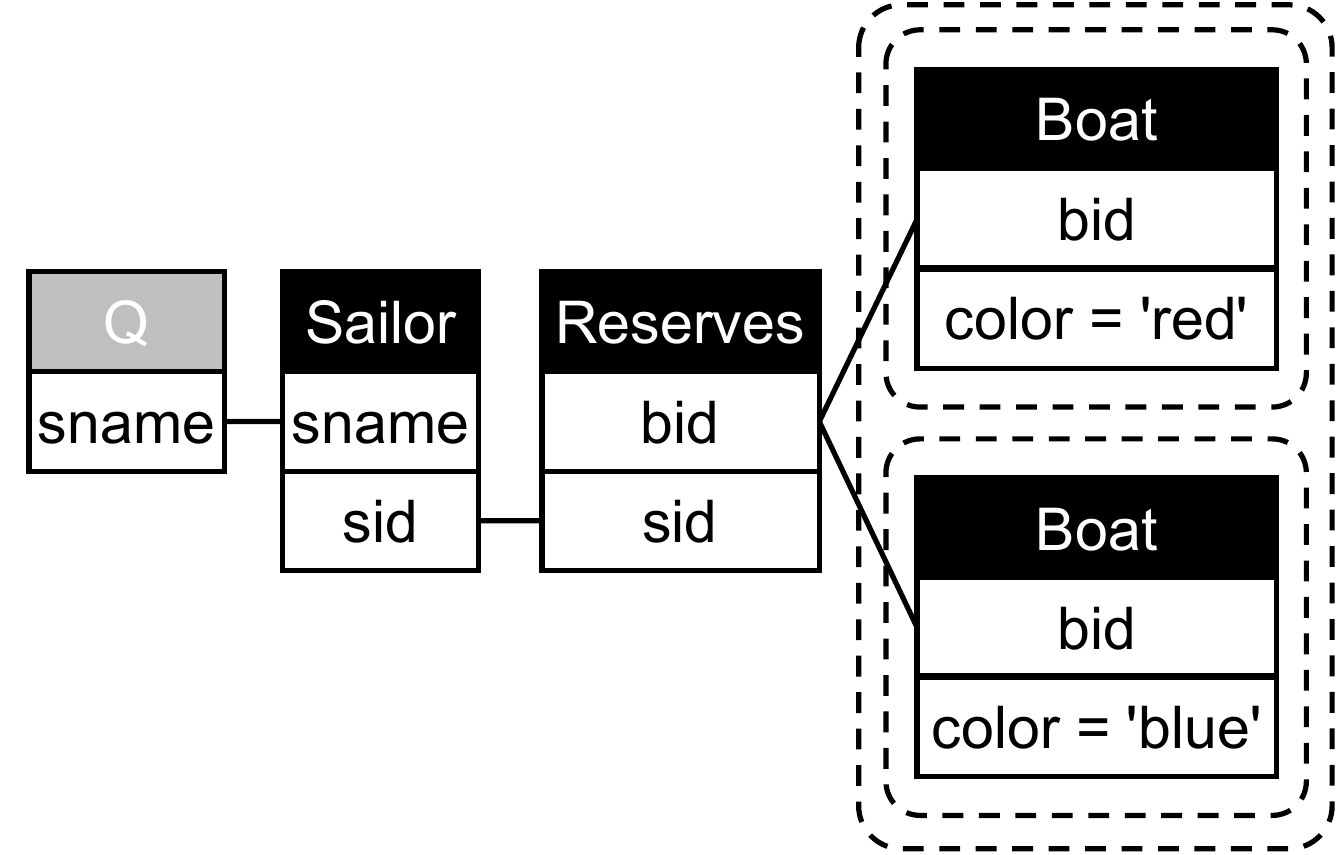}
	\vspace{-1mm}
    \caption{}
    \label{fig:Sailor_disjunction_b}
\end{subfigure}	
\caption{\Cref{ex:red_or_blue}:
A query with simple disjunction ``Find sailors who reserve a red or a blue boat'' 
can be represented with double negation 
in the non-disjunctive fragment of SQL (a) and whereas \diagram.}
\label{fig:Sailor_disjunction}
\end{figure}

\begin{example}[Red or blue]
Consider the following $\NDTRC$ query asking for sailors 
who have reserved a red or a blue boat:\label{ex:red_or_blue}		%
\begin{align*}
	\{ 
	& q(\sql{sname}) \mid \exists s \in \sql{Sailor}, r \in \sql{Reserves} [q.\sql{sname} = s.\sql{sname} \,\wedge  	\notag\\
	& s.\sql{sid} = r.\sql{sid} \wedge \h{\exists b \in \sql{Boat}[}										\\	%
	& \hspace{2mm}	b.\sql{bid} = r.\sql{bid} \wedge 
		(b.\sql{color} = \sql{`red'} \;\h{\vee}\; b2.\sql{color} = \sql{`blue'} ) \h{]} ] \}						\notag
\end{align*}
Using De Morgan's Law $(A \vee B) = \neg(\neg A \wedge \neg B)$ applied to quantifiers, 
we can transform the disjunction into double-negation with conjunction.
This transformation comes at the cost of repeated uses of extensional tables and is thus \emph{not pattern-preserving}:
\begin{align}
	\{ 
	& q(\sql{sname}) \mid \exists s \in \sql{Sailor}, r \in \sql{Reserves} [q.\sql{sname} = s.\sql{sname} \,\wedge  				\notag\\
	& s.\sql{sid} = r.\sql{sid} \wedge \h{\neg \big(}																\label{eq:red_or_blue_NDTRC}\\
	& \hspace{2mm}	\h{\neg(\exists b1 \in \sql{Boat}[}b1.\sql{bid} = r.\sql{bid} \wedge b1.\sql{color} = \sql{`red'} \h{])} \,\h{\wedge} 		\notag\\
	& \hspace{2mm}	\h{\neg(\exists b2 \in \sql{Boat}[}b2.\sql{bid} = r.\sql{bid} \wedge b2.\sql{color} = \sql{`blue'} \h{]) \big)} ] \}	 	\notag
\end{align}
\Cref{fig:Sailor_disjunction_a} shows 
\cref{eq:red_or_blue_NDTRC}
translated into canoical
$\NDSQL$
and 
\cref{fig:Sailor_disjunction_b}
its translation into a \diagram.
Notice how the non-disjunctive fragment repeats the boats table twice.
\end{example}

\section{Proofs \cref{sec:completeness}}

\begin{proof}[Proof \cref{th:completeness}]
	Given a safe $\TRC$ expression.
	We pull any existential quantifier as early as to either be at the start of the query, or directly following a negation operator. 

	First, consider a nested query with disjunctions in the $\sql{WHERE}$ conditions,
	possibly nested with conjunctions.
	Rewrite the conditions as DNF, i.e.\ as
	\begin{align*}
		&\neg (\exists \vec r \in \vec R[f_1(\vec r') \vee f_2(\vec r') \cdots \vee f_k(\vec r')]) \\
	\intertext{Here,
	$\vec r$ is a set of table variables, $\vec R$ a set of tables, 
	and each $f_i$ is a conjunction of predicates in free and/or quantified variables $\vec r' \supseteq \vec r$.
	Next rewrite it as:}
		&\neg (\exists \vec r_1 \in \vec R[f_1(\vec r')]) \wedge 
			\neg (\exists \vec r_2 \in \vec R[f_2(\vec r')]) \wedge \cdots \wedge
			\neg (\exists \vec r_k \in \vec R[f_k(\vec r')])	
	\end{align*}
	This fragment is in $\NDTRC$ and can be visualized by $\diagrams$.

	Second, for remaining disjunctions in the top query $q_0$, rewrite the query as union over queries without disjunction:
	\begin{align*}
	\phantom{=} &\{ q(\vec A) \mid
		\exists \vec r \in \vec R
		[f_0 \wedge (f_1(\vec r) \vee f_2(\vec r) \cdots \vee f_k(\vec r)] \}	\\
	= &\{ q(\vec A) \mid 
		\exists \vec r \in \vec R
		[f_0 \wedge f_1(\vec r)] \}  \,\cup \\
	  &\{ q(\vec A) \mid 
		\exists \vec r \in \vec R
		[f_0 \wedge f_2(\vec r)] \}	\cup \cdots \cup \\
	  &\{ q(\vec A) \mid 
		\exists \vec r \in \vec R
		[f_0 \wedge f_k(\vec r)] \}	
	\end{align*}
	Here, $f_0$ is a conjunction of attribute assignments to the output table $q$ and nested subqueries.
\end{proof}

\section{More Illustrations for \cref{sec:sentences}}

\begin{figure}[t]
\centering	
\begin{subfigure}[b]{.35\linewidth}		
\begin{lstlisting}
SELECT DISTINCT S.sname
FROM Sailor S
WHERE not exists
 (SELECT *
 FROM Boat B
 WHERE B.color = 'red'
 AND not exists
  (SELECT *
  FROM RESERVES R
  WHERE R.bid = B.bid
  AND R.sid = S.sid))
\end{lstlisting}
\vspace{-6mm}
\caption{}
\label{SQL_Sailor_exists_Sailor_all_red_boats_text}
\end{subfigure}	
\hspace{13mm}
\begin{subfigure}[b]{.37\linewidth}		
\begin{lstlisting}
SELECT exists
 (SELECT *
 FROM Sailor S
 WHERE not exists
  (SELECT *
  FROM Boat B
  WHERE B.color = 'red'
  AND not exists
   (SELECT *
   FROM RESERVES R
   WHERE R.bid = B.bid
   AND R.sid = S.sid)))
\end{lstlisting}
\vspace{-6mm}
\caption{}
\label{SQL_Sailor_exists_Sailor_all_red_boats}
\end{subfigure}	
\hspace{1mm}
\begin{subfigure}[b]{.56\linewidth}
    \includegraphics[scale=0.37]{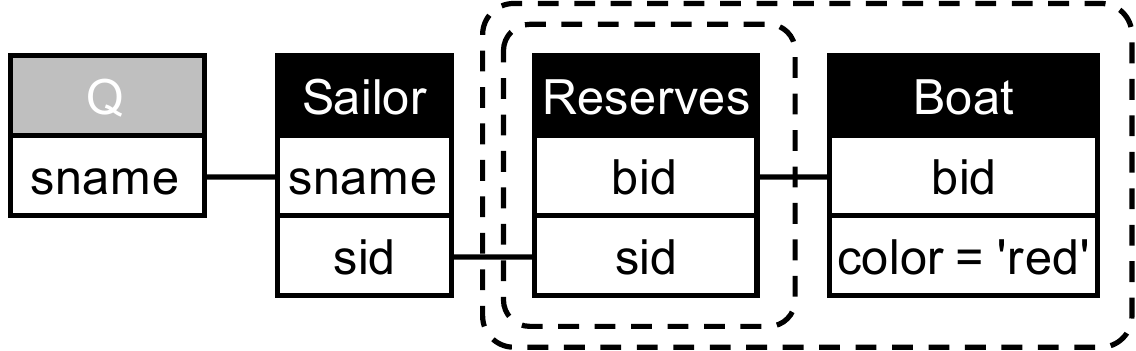}
    \caption{}
    \label{Fig_Sailor_Sailors_all_red_boats}
\end{subfigure}	
\begin{subfigure}[b]{.43\linewidth}
    \includegraphics[scale=0.37]{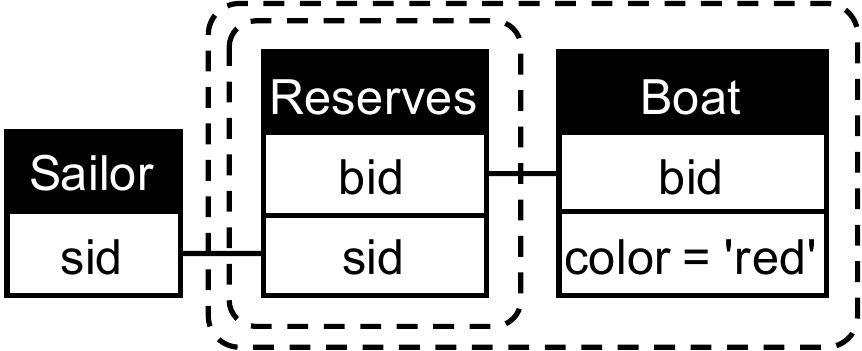}
	\vspace{0.5mm}
    \caption{}
    \label{Fig_Sailor_exists_Sailor_all_red_boats}
\end{subfigure}	
\caption{\Cref{ex:sailorsallredboats}: Sailors reserving all red boats.}
\label{fig:Sailor_sentences}
\end{figure}

\begin{example}[Sailors reserving all red boats]
	\label{ex:sailorsallredboats}
Consider the sailor database \cite{RamakrishnanGehrke:DBMS2000} that models sailors reserving boats:
Sailor(sid, sname, rating, age),
Reserves(sid, bid, day),
Boat(bid,bname,color),
and the query
``Find sailors who reserved all red boats:''
\begin{align}
\begin{aligned}
\!\!\!\!\!\!\{q(\sql{sname}) \mid 
	& \exists s \in \sql{Sailor} [q.\sql{sname}=s.\sql{sname} \ \wedge \\
	&\neg (\exists b \in \sql{Boat} [b.\sql{color} = \sql{'red'} \wedge \\
	&\neg (\exists r \in \sql{Reserves} [r.\sql{bid} = b.\sql{bid} \wedge r.\sql{sid} = s.\sql{sid}])]) \}
\end{aligned}
\end{align}

Contrast it with the logical statement
``There is a sailor who reserved all red boats.''
In $\NDTRC$, the difference is achieved 
by leaving away curly brackets and any mentions of the output table
(highlighted for a different example in green color in \cref{Fig_TRC_vs_RD_a}):
\begin{align}
\begin{aligned}
	&\exists s \in \sql{Sailor}[ \\
	&\neg (\exists b \in \sql{Boat} [b.\sql{color} = \sql{'red'} \wedge 
	\\
	& \neg (\exists r \in \sql{Reserves} [r.\sql{bid} = b.\sql{bid} \wedge r.\sql{sid} = s.\sql{sid}])])
\end{aligned}
\label{trc:existssailorsallredboats}
\end{align}
Similarly, \diagrams
loose the output table
(contrast \cref{Fig_Sailor_Sailors_all_red_boats}
with \cref{Fig_Sailor_exists_Sailor_all_red_boats} and their respective SQL statements).
\end{example}

\begin{figure}[t]
\centering	
\begin{subfigure}[b]{.28\linewidth}		
\begin{lstlisting}
SELECT not
 (not exists
  (SELECT *
  FROM R
  WHERE R.A=1)
 AND not exists
  (SELECT *
  FROM R R2
  WHERE R2.A=2)))
\end{lstlisting}
\vspace{-6mm}
\caption{}
\label{SQL_Disjunction}
\end{subfigure}	
\hspace{5mm}
\begin{subfigure}[b]{.3\linewidth}
    \includegraphics[scale=0.4]{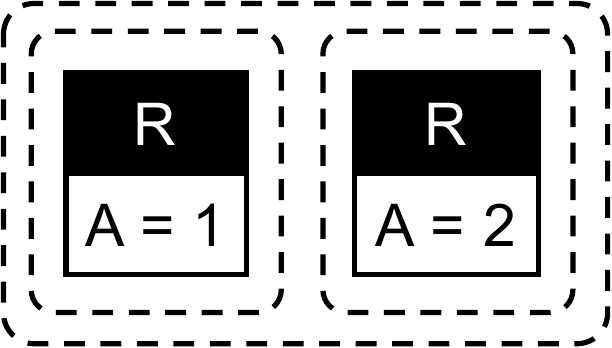}
\vspace{4mm}
\caption{}
\label{Fig_Disjunction}
\end{subfigure}	
\hspace{3mm}
\caption{\Cref{ex:disjunction}: $\exists r \in R [R.A = 1 \vee R.A = 2]$.}
\label{fig:Disjunction}
\end{figure}

We give an example that shows that in order to express sentences (instead of queries), 
and to be relationally complete (in that we would like to be able to express all logical sentences), 
we actually would not have to introduce the visual union.
This is in stark contrast to the union at the root being \emph{necessary} for queries.

\begin{example}[Disjunctions]
	\label{ex:disjunction}
Consider the simplest disjunction
\begin{align*}
	& \exists r \in R [R.A = 1 \;\h{\vee}\; R.A = 2]
\end{align*}
We can remove the disjunction with a double negation:
\begin{align*}
	& \exists r \in R [R.A = 1] \;\h{\vee}\; \exists r \in R[R.A = 2] \\
	& \h{\neg}(\h{\neg}(\exists r \in R [R.A = 1]  \;\h{\vee}\; \exists r \in R[R.A = 2])) \\	
	& \h{\neg}(\h{\neg}(\exists r \in R [R.A = 1]) \;\h{\wedge}\; \h{\neg}(\exists r \in R[R.A = 2])) 
\end{align*}

The first 4 steps of the translation in \cref{sec:fromTRCtoRD} still work
and leads to \cref{Fig_Disjunction}.
For $\SQL$, the query uses the second new rule to express double negation before the first \sql{FROM} clause.
\end{example}

\section{More detailed Related work \cref{SEC:RELATEDWORK}}

\subsection{Peirce's existential graphs (\cref{sec:Peirce})}

We mentioned in \cref{sec:Peirce} the complications arising from LI's (Lines of Identities) being overloaded.
Any LI can branch into multiple endings (also called \emph{ligatures}), and may have \emph{loose endings}, and may represent multiple existentially-quantified variables, together with cuts being applied to such LI's can quickly lead to hard-to-interpret diagrams (see e.g., the increasingly-unreadable figures in \cite[pp. 42-49]{Shin:2002}).
This led to several attempts in the literature to provide ``reading algorithms'' of those graphs (e.g., \cite{Zeman:1964,Roberts:1973,Shin:2002}) and rather complicated proofs of the expressiveness of beta graphs~\cite{Zeman:1964}, assuming a correct reading.
As example, the paper by Dau \cite{Dau:2006} points out an error in Shin's reading algorithm \cite{Shin:2002}.
However, Dau's correction to Shin~\cite{Dau:2006} itself also has errors.
For example, the interpretation of the right-most diagram in \cite[Fig 2]{Dau:2006} 
(reproduced as \cref{fig:Fig_Dau}) is wrong and misses one equality.
The given interpretation is
\begin{align*}
&\exists x. \exists y. \exists z [S(x) \wedge P(y) \wedge T(z) \wedge \neg(x = y \wedge y = z)]
\intertext
{
whereas it should be
}
&\exists x. \exists y. \exists z [S(x) \wedge P(y) \wedge T(z) \wedge \neg(x = y \wedge y = z \wedge x = z)]
\end{align*}
This is just an intuitive example how difficult beta graphs are in practice to interpret, even by the experts, 
and even by experts pointing out errors from other experts.

\begin{figure}[t]
\centering
\includegraphics[scale=0.33]{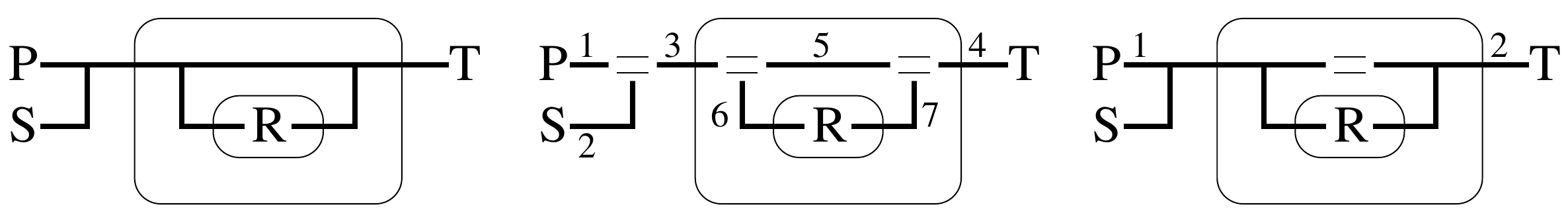}
\vspace{0mm}
\caption{
Figure copied from Dau \cite{Dau:2006} discussing an example beta graph whose interpretation provided by Shin \cite{Shin:2002} is incorrect, 
together with two alternative ways of splitting the LI's in order to interpret the graph correctly.
The details of the arguments are intricate and not important here. 
What matters is that a lot of disagreement exists as to how interpret
LI's correctly.
\diagrams avoid this problem entirely by using lines only as comparison predicates.
}
\label{fig:EG_comparison_Dau}
\end{figure}

\begin{figure}[t]
\centering
\includegraphics[scale=0.19]{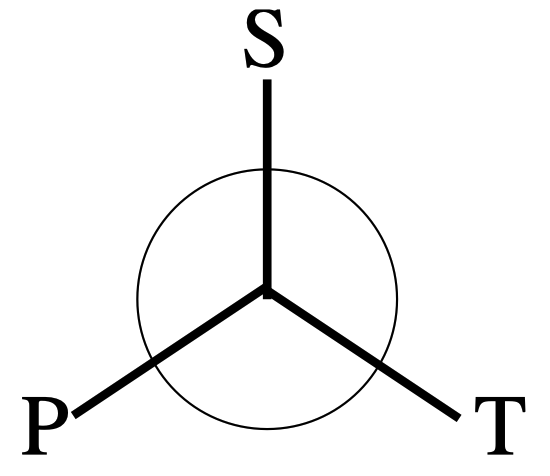}
\vspace{0mm}
\caption{
Right-most diagram of Figure 2 in Dau \cite{Dau:2006}.
}
\label{fig:Fig_Dau}
\end{figure}

\begin{figure*}[t]
\centering
\hspace{-1mm}
\begin{subfigure}[b]{.25\linewidth}
    \includegraphics[scale=0.35]{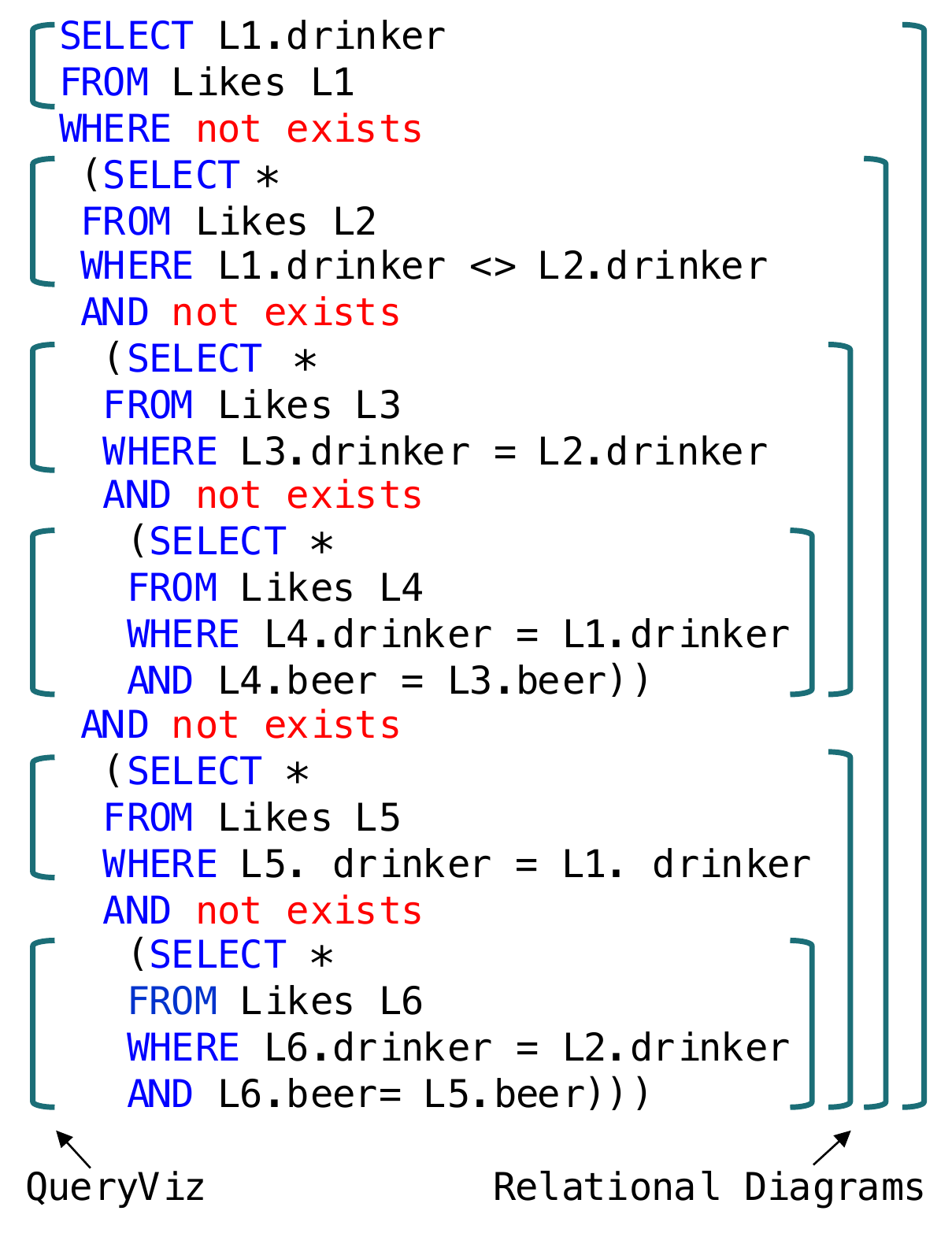}
	\vspace{-3mm}
    \caption{}
    \label{Fig_long_nested_query}
\end{subfigure}	
\hspace{3mm}
\begin{subfigure}[b]{.355\linewidth}
    \includegraphics[scale=0.35]{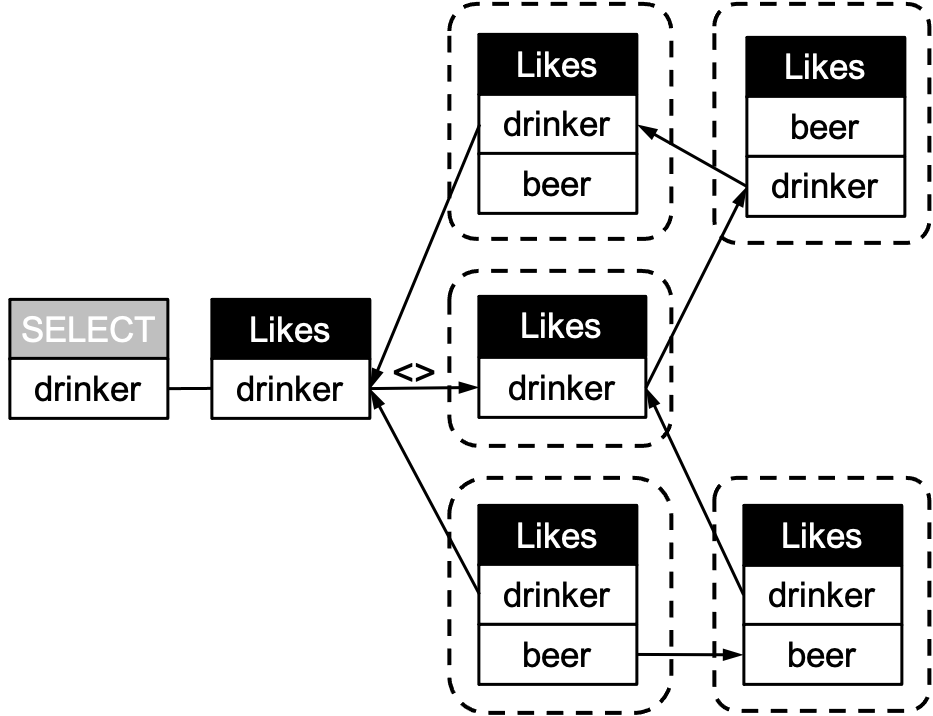}
	\vspace{3mm}
    \caption{}
    \label{Fig_QV_unique_beer_tastes2}
\end{subfigure}	
\begin{subfigure}[b]{.355\linewidth}
    \includegraphics[scale=0.35]{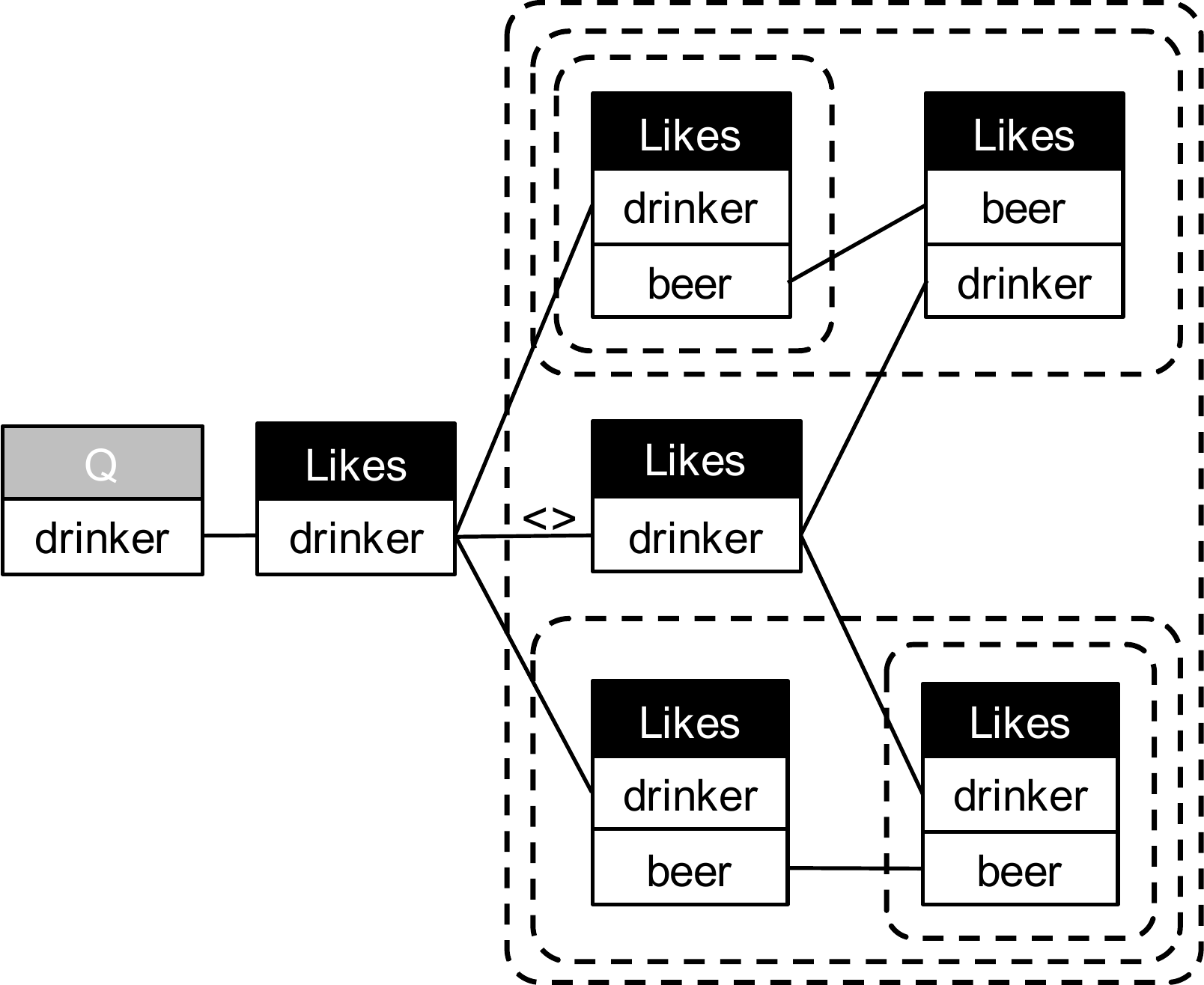}
	\vspace{1.3mm}
    \caption{}
    \label{Fig_QV_unique_beer_tastes1}
\end{subfigure}	
\hspace{-1mm}
\hspace{-2mm}
\caption{Illustrations for \Cref{ex:uniquebeers}:
(a) Unique-set-query ``\emph{Find drinkers with a unique beer taste}'' used by \cite{DBLP:conf/sigmod/LeventidisZDGJR20}, 
(b) $\queryviz$ diagram with reading order encoded by arrows (redrawn according to \cite{DBLP:conf/sigmod/LeventidisZDGJR20}),
(c) \diagrams with a nested scoping and no need for arrows.
}
\label{fig:beerquery}
\end{figure*}

\subsection{\queryviz~(\cref{sec:queryvis1_vs_2})}

We use the ``unique beer taste'' query \cite{DBLP:conf/sigmod/LeventidisZDGJR20} to show the difference in design decisions.

\begin{example}[Unique-set-query]
\label{ex:uniquebeers}
Consider the $\SQL$ query from \cref{Fig_long_nested_query}
asking to find ``\emph{drinkers who like a unique set of beers},'' i.e.\ no other drinker
likes the exact same set of beers.
The scoping brackets to the left of the query in  \cref{Fig_long_nested_query} 
show the content of boxes used by $\queryviz$, which include all tables \emph{from each individual query block}.
Without the additional visual symbol of arrows, 
this diagram becomes ambiguous to interpret.
To mitigate this problem,  
the design of \queryviz \cite{DBLP:conf/sigmod/LeventidisZDGJR20,DanaparamitaG2011:QueryViz}
uses
directed arrows with an implied \emph{reading order}~(\cref{Fig_QV_unique_beer_tastes2}).

The scoping brackets to the right in \cref{Fig_long_nested_query} 
show the nesting of the variables scopes in queries, which are also reflected in
the dashed bounding boxes in \diagrams 
(\cref{Fig_QV_unique_beer_tastes1}).
\end{example}

The design decision
by \cite{DBLP:conf/sigmod/LeventidisZDGJR20} 
are justified in terms of usability 
(for ``most'' queries the diagrams are not ambiguous and the reduction in nesting simplifies their interpretation), 
yet requires overloading of the meaning of arrows.
Two conceptual problem with these diagrams are:
1. $\queryviz$ requires each partition of the canvas to contain a relation from the relational schema.
Our earlier examples from \cref{fig:Sailor_disjunction,fig:Disjunction} show examples that can thus not be handled.
2. $\queryviz$ does not guarantee to unambiguously visualize nested queries with nesting depth $\geq 4$. 
This was alluded to already in \cite{DBLP:conf/sigmod/LeventidisZDGJR20}, and we next give an example to illustrate:

\begin{example}[Ambiguous $\queryviz$]
	\label{ex:ambiguousQV}
We next give a minimum example for when $\queryviz$ becomes ambiguous.
Consider the two different SQL queries~\cref{Fig_ambiguous_QV_1,Fig_ambiguous_QV_2}. 
Following the algorithm given in \cite{DBLP:conf/sigmod/LeventidisZDGJR20}, 
both lead to the same visual representation \cref{Fig_ambiguous_QV}.
In other words, it is is not possible to uniquely interpret the diagram in \cref{Fig_ambiguous_QV}.
\end{example}

\begin{figure}[t]
\centering	
\begin{subfigure}[b]{.32\linewidth}		
\begin{lstlisting}
SELECT DISTINCT *
FROM R
WHERE not exists
 (SELECT *
 FROM S
 WHERE not exists
  (SELECT *
  FROM T
  WHERE T.A = R.A
  AND T.B = S.B
  AND not exists
   (SELECT *
   FROM U
   WHERE U.C = S.C
   AND not exists
    (SELECT *
    FROM V
    WHERE V.D = T.D
    AND V.E = U.E))))
\end{lstlisting}
\vspace{-6mm}
\caption{}
\label{Fig_ambiguous_QV_1}
\end{subfigure}	
\hspace{10mm}
\begin{subfigure}[b]{.32\linewidth}		
\begin{lstlisting}
SELECT DISTINCT *
FROM R
WHERE not exists
 (SELECT *
 FROM V
 WHERE not exists
  (SELECT *
  FROM T
  WHERE T.A = R.A
  AND T.D = V.D
  AND not exists
   (SELECT *
   FROM U
   WHERE U.E = V.E
   AND not exists
    (SELECT *
    FROM S
    WHERE S.B = T.B
    AND S.C = U.C))))
\end{lstlisting}
\vspace{-6mm}
\caption{}
\label{Fig_ambiguous_QV_2}
\end{subfigure}	
\begin{subfigure}[b]{0.65\linewidth}
    \includegraphics[scale=0.35]{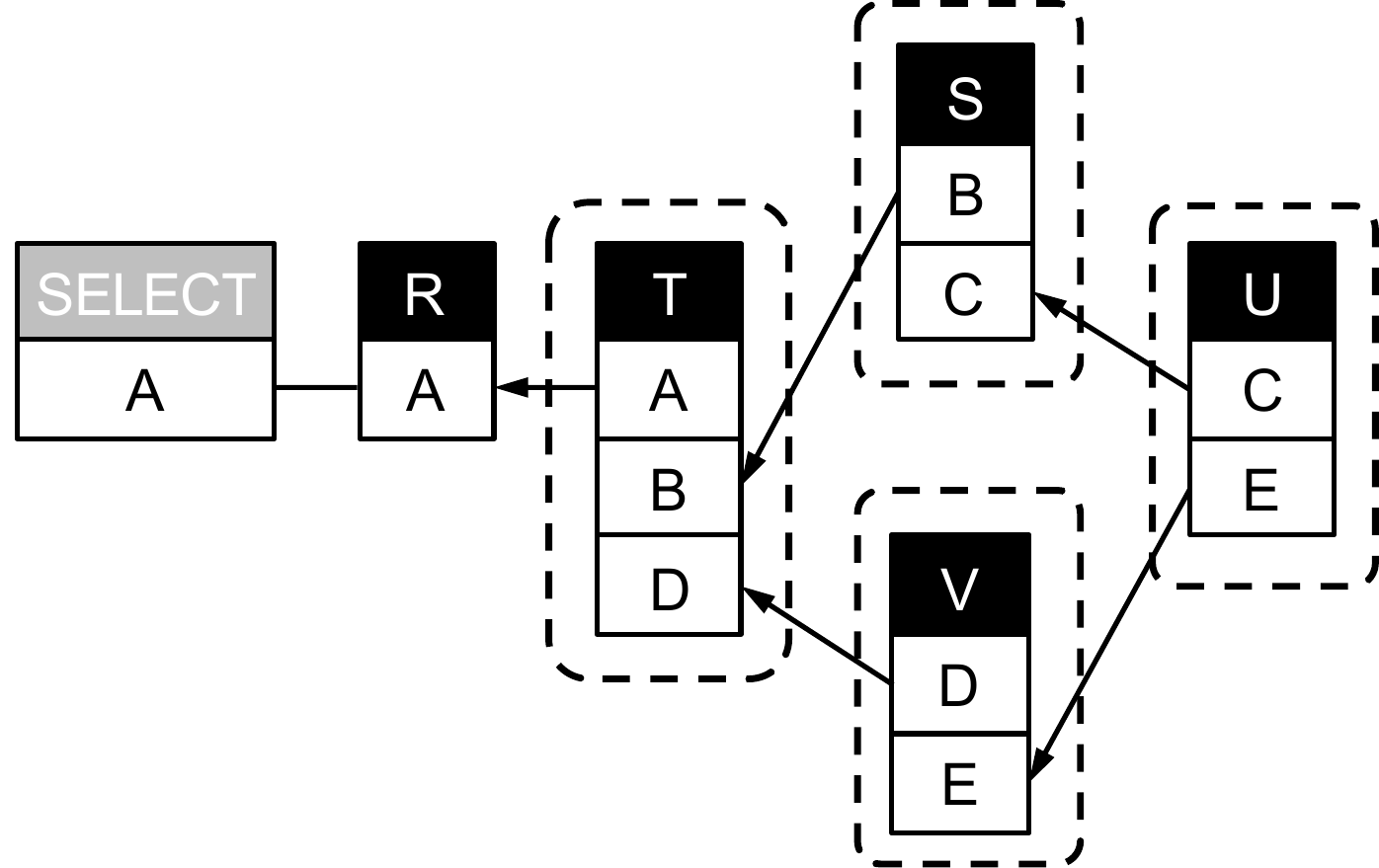}
	\vspace{-2mm}
    \caption{}
    \label{Fig_ambiguous_QV}
\end{subfigure}	
\hspace{-1mm}
\caption{
\Cref{ex:ambiguousQV}:
Minimal example showing that $\queryviz$ is not sound for nested queries with 4 levels: 
Two \emph{different queries} (a), (b) that are translated into the same $\queryviz$ diagram (c).}
\label{fig:QV_ambiguousness}
\end{figure}

\subsection{DFQL}

DFQL (Dataflow Query Language) is an example visual representation that is relationally complete \cite{DBLP:journals/iam/ClarkW94,DBLP:journals/vlc/CatarciCLB97}
by mapping its visual symbols to the operators of relational algebra.
Aside from providing basic set of operators derived from the requirements for being as expressive as first-order predicate calculus, 
DFQL also provides diagrammatic representation of grouping operators in both comparison functions and aggregations.
Following the same procedurality as $\RA$, DFQL expresses the dataflow in a top-down tree-like structure. 
However, since DFQL focuses on the 1-to-1 correspondence to relational algebra, 
it also can not generate a representation-equivalent diagram for query
\cref{fig:SQL_equivalence1}
which has no structure-isomorph representation in $\RA$. 
See the following example for details:

\begin{example}[Sailors reserving all red boats in DFQL]
\label{DFQL_redboats}
Representing the query ``Find sailors who have reserved all red boats'' 
(recall \cref{ex:sailorsallredboats,Fig_Sailor_Sailors_all_red_boats}) in the formalism of DFQL, 
the entire query can be visualized in one single connected tree-like diagram (unlike QBE which needs to visualize a temporary table to hold the intermediary values).
However, since the language is based on the operators of RA, 
there is no pattern-isomorph expression of the query (\cref{Fig_Sailor_Sailors_all_red_boats}) in relational algebra.
Instead, the logically-equivalent representation in $\RA$ is as follows:
	\begin{align}
	\begin{aligned}	
		\mathit{Q} =
		\pi_{\sql{sname}} \big(
			&\sql{Sailor}\Join \big(
				\pi_{\sql{sid}} \sql{Sailor} - \pi_{\sql{sid}}\big(
				(\pi_{\sql{sid}} \sql{Sailor} \\
			&\times \pi_{bid}\sigma_{color='red'} \mathit{Boat}) - \pi_{\sql{sid,bid}} \sql{Reserves}\big) \big) \big)
	\end{aligned}
	\end{align}
	The join between \sql{Sailor S} and \sql{Sailor S2} is necessary to project column \sql{sname} from the table.
	This later query can be visualized by DFQL in a pattern-preserving way as \cref{Fig_DFQL_red_boats}. 
	One can easily find a 1-to-1 mapping between DFQL operators and this $\RA$ expression.
	
	Notice that for the same arguments, there is also no pattern-isomorph expression of the query shown in 
	\cref{Fig_RA_vs_Datalog_g} and DFQL needs two extensional tables for input table $R$ to represent that query.
\end{example}

\begin{figure}[t]
\centering
\vspace*{-4mm}
\hspace*{-8mm}
\includegraphics[scale=0.35]{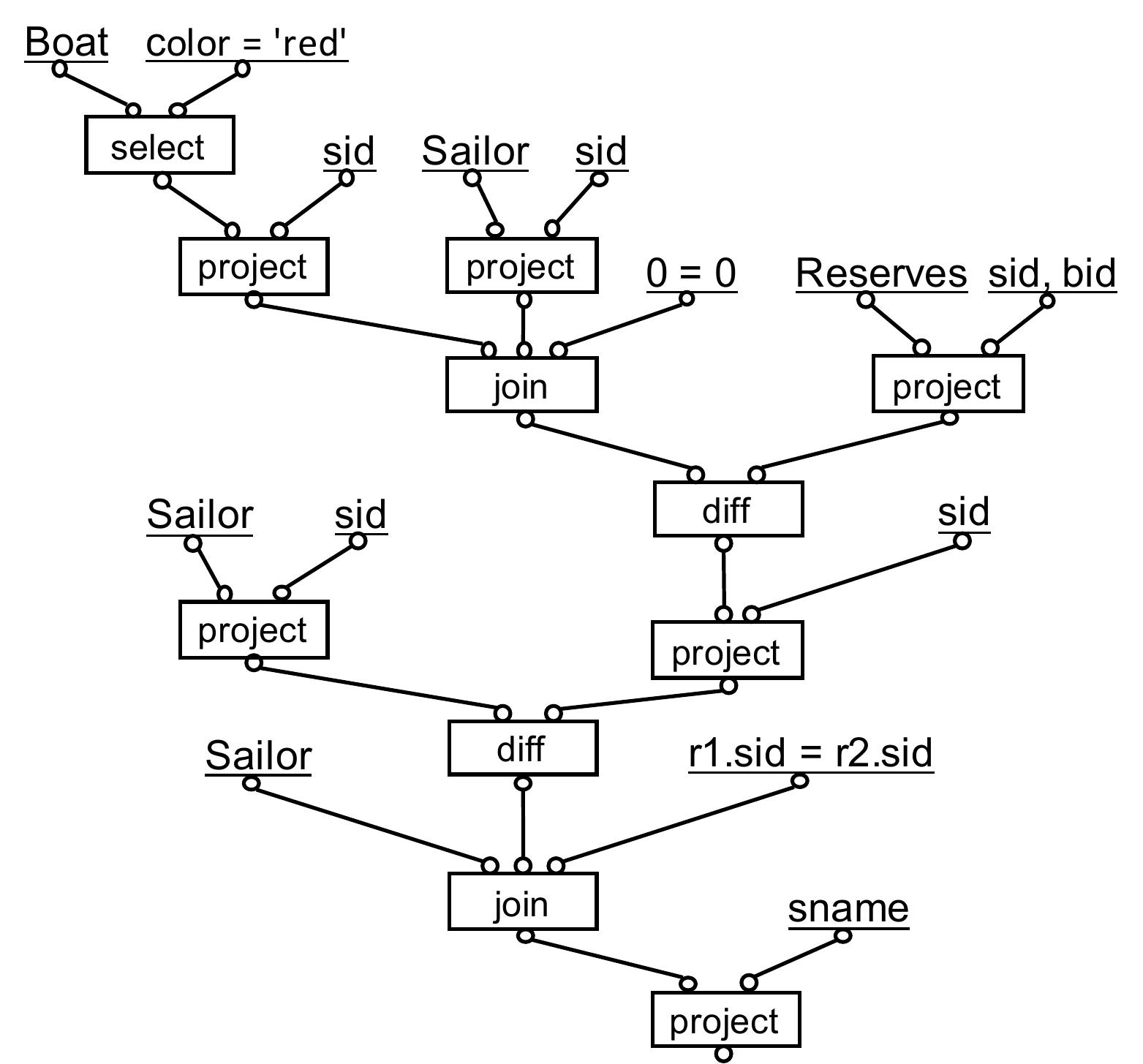}
\caption{\Cref{DFQL_redboats}: DFQL visualization of the query ``Find sailors who have reserved all red boats'' 
that is structurally equivalent to the pattern used by relational algebra. 
The \emph{diff} operator is equivalent to binary \emph{$-$} (minus) in RA and the tautology ``0 = 0'' in \emph{join} operator is required to create a Cartesian Join in DFQL \cite{DBLP:journals/iam/ClarkW94}.
Compare the difficulty in perceiving a logical pattern in this visualization against the one from \diagrams in \cref{Fig_Sailor_Sailors_all_red_boats}.
}
\label{Fig_DFQL_red_boats}
\end{figure}

\subsection{Tools for Query Visualisations}
The four projects that we know of that focus on the problem of visualizing existing relational queries are
QueryVis~\cite{DanaparamitaG2011:QueryViz,DBLP:conf/sigmod/LeventidisZDGJR20,gatterbauer2011databases}
(which we showed is not relationally complete, yet which inspired a lot of our work),
\mbox{GraphSQL}~\cite{DBLP:conf/dexaw/CerulloP07},
Visual SQL~\cite{DBLP:conf/er/JaakkolaT03},
(both of which maintain the 1-to-1 correspondence to SQL,
and syntactic variants of the same query like \cref{fig:SQLvariety} lead to different representations),
and Snowflake join \cite{snowflake} (which is a pure query visualization approach that focuses on join queries with optional grouping, but does not support any nested queries with negation).
Compared to all these visual representations, ours is the only one that is relationally complete and that can preserve and represent all logical patterns in the non-disjunctive fragment of relational query languages.

\subsection{Applications for Query interpretation}
\emph{Query Interpretation} is the problem of reading and understanding an existing query. 
It is often as hard as {query composition}, i.e., creating a new query~\cite{DBLP:journals/csur/Reisner81}. 
In the past, several projects have focused on building Query Management Systems 
that help users issue queries by leveraging an existing log of queries. Known systems to date include CQMS~\cite{DBLP:conf/cidr/KhoussainovaBGKS09,KhoussainovaKBS:2011}, SQL QuerIE~\cite{DBLP:conf/ssdbm/ChatzopoulouEP09,QueRIERecommendations:2010}, DBease~\cite{LiFWWF2011:DBease}, and SQLShare~\cite{HoweC2010:SQLshare}.
All of those are motivated by making SQL composition easier and thus databases more usable~\cite{DBLP:conf/sigmod/JagadishCEJLNY07}, especially for non-sophisticated database users.
An essential ingredient of such systems is a \emph{query browse} facility, i.e., 
a way
that allows the user to browse and quickly choose between several queries proposed by the system. 
This, in turn, requires a user to \emph{quickly understand} existing queries.
Whereas visual systems for \emph{specifying} queries have been studied
extensively
(a 1997 survey by Catarci et al.~\cite{DBLP:journals/vlc/CatarciCLB97} 
cites over 150 references), 
the explicit reverse problem
of visualizing and thereby helping \emph{interpret a relational query that has already been written}
has not drawn 
much attention, despite 
very early \cite{Reisner1975:HumanFactors,DBLP:journals/csur/Reisner81}
and very recent work
\cite{DBLP:conf/sigmod/LeventidisZDGJR20}
repeatedly showing
that visualizations of relational queries can help users understand them faster than SQL text.

\begin{figure*}[t]
\centering
\begin{subfigure}[b]{.136\linewidth}
\begin{lstlisting}
SELECT DISTINCT R.A
FROM R 
WHERE exists
 (SELECT *
 FROM S
 WHERE R.B = S.B)
\end{lstlisting}
\vspace{-7mm}
\caption{$\NDSQL$}
\end{subfigure}
\hspace{3mm}	
\begin{subfigure}[b]{.2\linewidth}		
\begin{align*}
	&	\{ q(A) \mid \exists r \in R, \exists s \in S [\\
	&	q.A = r.A \wedge r.B = s.B]\}	
\end{align*}
\vspace{-4mm}
\caption{$\NDTRC$}
\end{subfigure}
\begin{subfigure}[b]{.18\linewidth}
\begin{align*}
	R \Join S
\end{align*}
\vspace{-4mm}
\caption{$\NDRA$}
\label{}
\end{subfigure}
\begin{subfigure}[b]{.22\linewidth}
\begin{align*}
	Q(x) 	& \datarule R(x), \neg S(x).
\end{align*}
\vspace{-4mm}
\caption{$\DatalogND$}
\label{}
\end{subfigure}
\begin{subfigure}[b]{.14\linewidth}
    \includegraphics[scale=0.41]{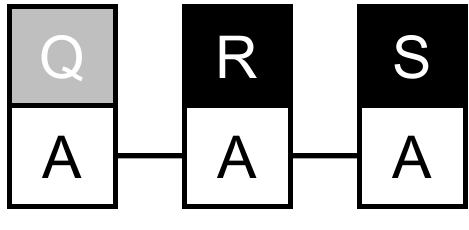}
	\vspace{1mm}
    \caption{}
    \label{Fig_DatalogLimits_1}
\end{subfigure}	
\hspace{10mm}
\begin{subfigure}[b]{.136\linewidth}
\begin{lstlisting}
SELECT DISTINCT R.A
FROM R 
WHERE not exists
 (SELECT *
 FROM S
 WHERE R.B = S.B)
\end{lstlisting}
\vspace{-7mm}
\caption{$\NDSQL$}
\end{subfigure}
\hspace{3mm}	
\begin{subfigure}[b]{.2\linewidth}		
\begin{align*}
	&	\{ q(A) \mid \exists r \in R[q.A = r.A  \\
	&	\wedge \neg (\exists s \in S [r.B = s.B])]\}	
\end{align*}
\vspace{-4mm}
\caption{$\NDTRC$}
\end{subfigure}
\begin{subfigure}[b]{.18\linewidth}
\begin{align*}
	R - S
\end{align*}
\vspace{-4mm}
\caption{$\NDRA$}
\label{}
\end{subfigure}
\begin{subfigure}[b]{.22\linewidth}
\begin{align*}
	Q(x) 	& \datarule R(x), \neg S(x).
\end{align*}
\vspace{-4mm}
\caption{$\DatalogND$}
\label{}
\end{subfigure}
\begin{subfigure}[b]{.14\linewidth}
\includegraphics[scale=0.41]{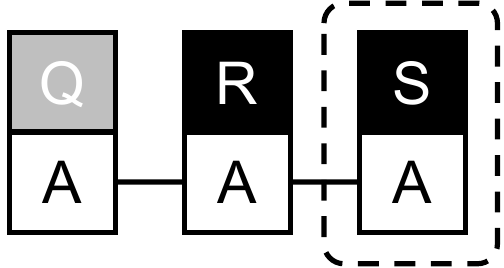}
\caption{}
\label{Fig_DatalogLimits_2}
\end{subfigure}
\hspace{10mm}
\begin{subfigure}[b]{.136\linewidth}
\begin{lstlisting}
SELECT DISTINCT R.A
FROM R 
WHERE not exists
 (SELECT *
 FROM S
 WHERE R.B > S.B)
\end{lstlisting}
\vspace{-7mm}
\caption{$\NDSQL$}
\end{subfigure}
\hspace{3mm}
\begin{subfigure}[b]{.2\linewidth}		
\begin{align*}
	&	\{ q(A) \mid \exists r \in R[q.A = r.A  \\
	&	\wedge \neg (\exists s \in S [r.A > s.A])]\}	
\end{align*}
\vspace{-4mm}
\caption{$\NDTRC$}
\end{subfigure}
\begin{subfigure}[b]{.18\linewidth}
\centering
\h{\frownie{}}
\vspace{1mm}
\caption{$\NDRA$}
\end{subfigure}
\begin{subfigure}[b]{.22\linewidth}
\centering
\h{\frownie{}}
\vspace{1mm}
\caption{$\DatalogND$}
\label{}
\end{subfigure}
\begin{subfigure}[b]{.14\linewidth}
\includegraphics[scale=0.41]{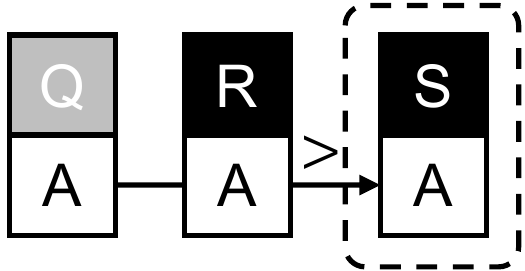}
\caption{}
\label{Fig_DatalogLimits_3}
\end{subfigure}
\hspace{10mm}
\begin{subfigure}[b]{.136\linewidth}
\begin{lstlisting}
SELECT DISTINCT R.A
FROM R 
WHERE not exists
 (SELECT *
 FROM S, R as R2
 WHERE R2.A > S.A
 AND R2.A = R.A)
\end{lstlisting}
\vspace{-7mm}
\caption{$\NDSQL$}
\end{subfigure}
\hspace{3mm}
\begin{subfigure}[b]{.2\linewidth}		
\begin{align*}
	& \{ q(A) \mid \exists r \in R[q.A \!=\! r.A  \\
	& \wedge \neg (\exists s \in S, r_2 \in R [\\
	& r_2.A \!>\! s.A \wedge r_2.A \!=\! r.A])]\}	
\end{align*}
\vspace{-4mm}
\caption{$\NDTRC$}
\end{subfigure}
\begin{subfigure}[b]{.18\linewidth}
\begin{align*}
	&R - \\
	&\big( \pi_{R.A} \sigma_{R.A>R.A} (R \times S) \big)
\end{align*}
\vspace{-4mm}
\caption{$\NDRA$}
\label{}
\end{subfigure}
\begin{subfigure}[b]{.22\linewidth}
\begin{align*}
	I(x)	& \datarule R(x),  S(y), x\!>\!y. 		\\[-1mm]
	Q(x) 	& \datarule R(x), \neg I(x).
\end{align*}
\vspace{-4mm}
\caption{$\DatalogND$}
\label{Fig_RA_vs_Datalog_a}
\end{subfigure}
\begin{subfigure}[b]{.14\linewidth}
\includegraphics[scale=0.41]{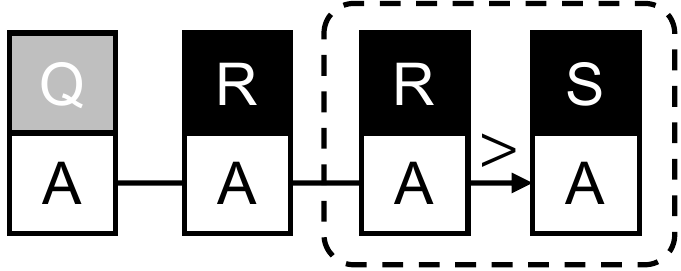}
\vspace{-2mm}
\caption{}
\label{Fig_DatalogLimits_4}
\end{subfigure}	
\caption{\Cref{ex:DatalogLimits}: 
The first two rows show $Q_1$ and $Q_2$, respectively. The last two rows show $Q_3$.
Notice that the $\NDSQL$, $\NDTRC$, and \diagram
queries from the third row have no pattern-isomorph query
in $\NDRA$ or $\DatalogND$.
To express $Q_3$ instead $\NDRA$ and $\DatalogND$ require an additional cross-join with $R.A$,
which is shown in the forth row.
}
\label{fig:DatalogLimits}
\end{figure*}

\subsection{Limits of Datalog for representing patterns}

We have shown earlier that $\DatalogND$ cannot represent all Query patterns from $\NDTRC$.
We next use another example to illustrates that
this limit of $\Datalog$ does not only appear with deeply nested queries;
it already appears for simply nested queries and is an immediate consequence of $\Datalog$'s safety conditions for built-in predicates.

\begin{example}[Limits of Datalog]
Consider two unary tables $R(A)$ and $S(A)$ and three questions:
\label{ex:DatalogLimits}
\begin{align*}	
	&Q_1: \textrm{Find values from R that also appear in S.} \\
	&Q_2: \textrm{Find values from R that do not appear in S} \\
	&Q_3: \textrm{Find values from R for which no smaller value appears in S.} 
\end{align*}	
The first three lines  of \cref{fig:DatalogLimits} show these queries expressed in 
$\NDSQL$,
$\NDTRC$, 
$\NDRA$, 
$\DatalogND$, 
and
\diagrams.

Notice that the $\NDSQL$, $\NDTRC$, and \diagram
queries from the third row have no pattern-isomorph query
in $\NDRA$ or $\DatalogND$.
The safety condition of $\Datalog$ requires each variable to appear in a non-negated atom. 
This criterion requires a cross-join with the domain of $R.A$ in a separate rule before the negation can be applied on an equality predicate.
For the same reason, $\RA$ cannot apply the set difference directly and also requires an additional cross-join with $R.A$.
The forth row shows the resulting resulting $\NDRA$ and $\DatalogND$ 
queries together with their pattern-isomorph queries in 
 $\NDSQL$, $\NDTRC$, and \diagrams.
\end{example}

\end{document}